\documentclass{jpp}

\usepackage{graphicx}
\usepackage[makeroom]{cancel}

\usepackage{natbib}
\usepackage{amsmath}
\usepackage{amssymb}
\usepackage{xcolor}
\usepackage{bm}
\usepackage{authblk}
\usepackage{xcolor}
\usepackage{comment}
\usepackage{soul}

\ifCUPmtlplainloaded \else
  \checkfont{eurm10}
  \iffontfound
    \IfFileExists{upmath.sty}
      {\typeout{^^JFound AMS Euler Roman fonts on the system,
                   using the 'upmath' package.^^J}%
       \usepackage{upmath}}
      {\typeout{^^JFound AMS Euler Roman fonts on the system, but you
                   dont seem to have the}%
       \typeout{'upmath' package installed. JPP.cls can take advantage
                 of these fonts, if you use 'upmath' package.^^J}%
      }
  \else
  \fi
\fi

\ifCUPmtlplainloaded \else
  \checkfont{msam10}
  \iffontfound
    \IfFileExists{amssymb.sty}
      {\typeout{^^JFound AMS Symbol fonts on the system, using the
                'amssymb' package.^^J}%
       \usepackage{amssymb}%
       \let\le=\leqslant  \let\leq=\leqslant
       \let\ge=\geqslant  \let\geq=\geqslant
      }{}
  \fi
\fi

\ifCUPmtlplainloaded \else
  \checkfont{msam10}
  \iffontfound
    \IfFileExists{amssymb.sty}
      {\typeout{^^JFound AMS Symbol fonts on the system, using the
                'amssymb' package.^^J}%
       \usepackage{amssymb}%
       \let\le=\leqslant  \let\leq=\leqslant
       \let\ge=\geqslant  \let\geq=\geqslant
      }{}
  \fi
\fi

\ifCUPmtlplainloaded \else
\IfFileExists{amsbsy.sty}
             {\typeout{^^JFound the 'amsbsy' package on the system, using it.^^J}%
     \usepackage{amsbsy}}
    {\providecommand\boldsymbol[1]{\mbox{\boldmath $##1$}}}
\fi

\newsavebox{\astrutbox}
\sbox{\astrutbox}{\rule[-5pt]{0pt}{20pt}}

\newcommand{\Alfven}{Alfv\'{e}n }
\newcommand{\Alfvenic}{Alfv\'{e}nic }
\newcommand{\V}[1]{\boldsymbol{#1}} 
\newcommand{\T}[1]{{\tt #1}}
\newcommand{\xhat}{\mbox{$\hat{\V{x}}$}}

\newcommand{\zhat}{\mbox{$\hat{\V{z}}$}}
\newcommand{\bhat}{\mbox{$\hat{\V{b}}$}}
\newcommand{\figref}[1]{Figure~\ref{#1}}
\newcommand{\secref}[1]{\S\ref{#1}}
\newcommand{\appref}[1]{Appendix~\ref{#1}}

\title[Transit-Time Damping]{The Velocity-Space Signature of Transit-Time Damping}

\author[1]{Rui Huang\thanks{Email address for correspondence: rui-huang@uiowa.edu}}
\author[1]{Gregory G. Howes}
\author[2]{Andrew J. McCubbin}

\affil[1]{Department of Physics and Astronomy, University of Iowa, IA 52242, USA}
\affil[2]{Applied Physics Laboratory, Johns Hopkins University, MD 20723, USA}

\pubyear{2024}
\volume{?}
\pagerange{?}
\date{?; revised ?; accepted ?. - To be entered by editorial office}

\begin{document}
\maketitle

\begin{abstract}
Transit-time damping (TTD) is a process in which the magnetic mirror force---induced by the parallel gradient of magnetic field strength---interacts with resonant plasma particles in a time-varying magnetic field, leading to the collisionless damping of electromagnetic waves and the resulting energization of those particles through the perpendicular component of the electric field, $E_\perp$. In this study, we utilize the recently developed field-particle correlation technique to analyze gyrokinetic simulation data. This method enables the identification of the velocity-space structure of the TTD energy transfer rate between waves and particles during the damping of plasma turbulence. Our analysis reveals a unique bipolar pattern of energy transfer in velocity space characteristic of TTD. By identifying this pattern, we provide clear evidence of TTD's significant role in the damping of strong plasma turbulence. Additionally, we compare the TTD signature with that of Landau damping (LD). Although they both produce a bipolar pattern of phase-space energy density loss and gain about the parallel resonant velocity of the \Alfvenic waves, they are mediated by different forces and exhibit different behaviors as $v_\perp \to 0$. We also explore how the dominant damping mechanism varies with ion plasma beta $\beta_i$, showing that TTD dominates over LD for $\beta_i > 1$. This work deepens our understanding of the role of TTD in the damping of weakly collisional plasma turbulence and paves the way to seek the signature of TTD using \emph{in situ} spacecraft observations of turbulence in space plasmas.

\end{abstract}


\section{Introduction}

A key area of research in the study of turbulence in weakly collisional plasmas is understanding how energy from the fluctuating plasma flows and electromagnetic fields is converted into plasma particle energy. This phenomenon is especially relevant in heliospheric plasmas like the solar wind, where the characteristic low density and high temperature lead to weakly collisional plasma dynamics. The dissipation of turbulence in such space and astrophysical plasmas is likely mediated by three categories of mechanisms: (i) resonant wave-particle interactions, such as Landau damping \citep{Landau:1946, Chen:2019}, transit-time damping \citep{stix:1992, Barnes:1966}, and cyclotron damping \citep{Isenberg:1983, Isenberg:2019}; (ii) non-resonant wave-particle interactions, including stochastic heating \citep{Chandran:2010, Chandran:2013, Martinovic:2020, Cerri:2021}, magnetic pumping \citep{Lichko:2020,montag:2022}, and ``viscous'' damping mediated by kinetic temperature anisotropy instabilities \citep{Arzamasskiy:2023}; and (iii) dissipation within coherent structures, in particular collisionless magnetic reconnection that may occur in current sheets that are found to arise naturally in plasma turbulence \citep{Osman:2011, Zhdankin:2015, Mallet:2017, Loureiro:2017a}.

Given the low collisionality of these plasma environments, the six-dimensional (3D-3V) kinetic plasma theory is essential for analyzing the evolution of the turbulence and its dissipation through collisionless interactions between electromagnetic fields and plasma particles \citep{Howes:2017c}. Although \emph{in situ} spacecraft measurements in the solar wind provide invaluable data, they are often limited to a single point, or a few points, in space, 
which presents a significant challenge for the investigation of the physical mechanisms that remove energy from the turbulent fluctuations and consequently energize the plasma particles.  The recently developed field–particle correlation technique \citep{Klein:2016a, Howes:2017a, Klein:2017b} enables direct measurements of the electromagnetic fields and particle velocity distributions at a single point in space to be combined to create a velocity-space signature of particle energization that can be used to identify the physical mechanisms responsible for damping the turbulence and to estimate the resulting rate of the change of particle energy density. Consequently, this technique provides an innovative means to utilize \emph{in situ} spacecraft observations to identify specific collisionless damping mechanisms and determine particle heating rates.

This technique has shown success in identifying several damping mechanisms in weakly collisional turbulent plasmas, such as ion Landau damping \citep{Klein:2017b, TCLi:2019}, ion cyclotron damping \citep{Klein:2020, Afshari:2023}, electron Landau damping \citep{Chen:2019, TCLi:2019, Afshari:2021, Conley:2023}, and magnetic pumping \citep{montag:2022}. However, the role of transit-time damping, a resonant wave-particle interaction, in the damping of plasma turbulence remains unconfirmed. The focus of this paper is to employ the field-particle correlation technique to identify the velocity-space signature of ion energization through transit-time damping and to recover this signature from simulations of strong plasma turbulence.

The structure of this paper is laid out as follows. We derive the specific form of the field-particle correlation for transit-time damping in \secref{sec:ttd_fpc_formula}. This is followed by an exploration of the expected transit-time damping signature in \secref{sec:predict_ttd_sig}. In \secref{sec:singleKAW}, we conduct single kinetic \Alfven wave simulations to investigate the velocity-space signature characteristic of transit-time damping. Subsequently, in \secref{sec:turb}, we delve into turbulence simulations, presenting details for distinguishing transit-time damping from turbulence damping process. \secref{sec:conclusion} summarizes our findings and outlines potential future applications for further research.

\section{Transit-Time Damping}
\label{sec:transittimedamping}

The idea of transit-time damping (TTD) had its origins in mid-20th-century plasma physics when transit-time magnetic pumping was proposed as a means to heat confined plasma \citep{spitzer:1953}. This method is characterized by a modulation of the magnetic field magnitude at a frequency considerably lower than the ion cyclotron frequency; the evolution of the parallel and perpendicular particle velocities in the time-varying magnetic field, combined with a weak collisionality, leads to a net transfer of energy to the plasma particles. The term ``transit-time" refers to the duration necessary for an ion to traverse from one side to the other across the confined region.

\begin{figure}
\begin{center}
            \includegraphics[width=0.75 \textwidth]{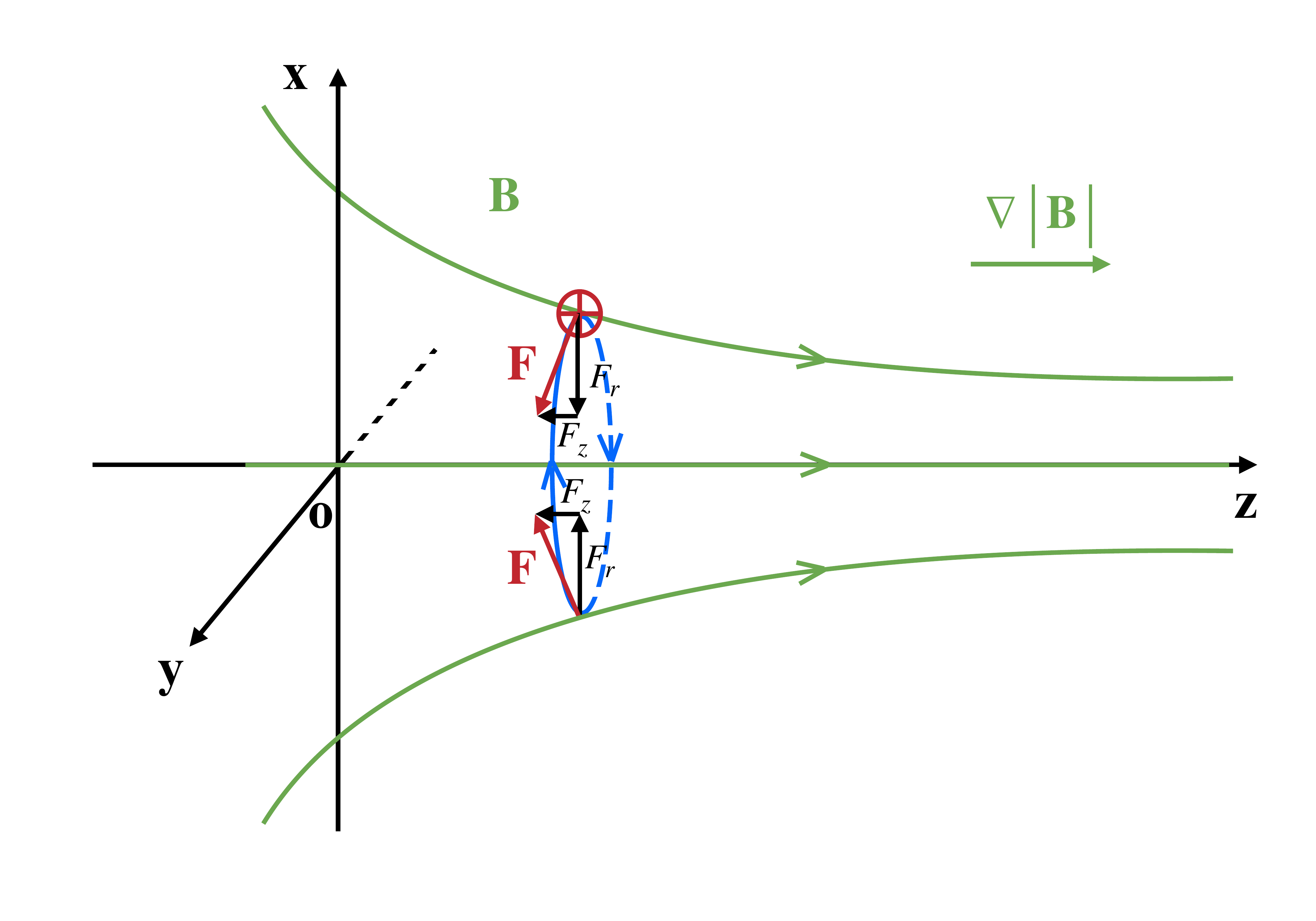} 
\end{center}   
\caption{Diagram of the radial component $F_r$ and axial component $F_z$ of the Lorentz force of the magnetic field (red) on a positively charged particle (red $+$) in a converging magnetic field (green) with increasing magnitude in the $+z$ direction.  Averaged over the Larmor orbit of the particle (blue), the net magnetic mirror force is in the direction of decreasing magnetic field magnitude, here the $-z$ direction.} 
    \label{fig:mirror_force}
\end{figure}

The magnetic mirror force plays a key role in the dynamics of TTD. In a static magnetic field with a spatial variation of the magnetic field magnitude along the direction parallel to the field, the mirror force accelerates charged particles in the direction of decreasing field magnitude.  In a cylindrical coordinate system aligned with the magnetic field direction, the condition  $\nabla \cdot \V{B}=0$  implies that an increase of the magnetic field along the axial direction must be accompanied by the convergence of the field in the radial direction, as shown in \figref{fig:mirror_force}. For a particle with a guiding center on the axis, the particle will experience an inward radial field throughout its Larmor orbit.  The Lorentz force, which acts perpendicularly to the magnetic field direction at the particle position, will therefore have both a large radial and a small axial component.  Averaged over the full Larmor orbit, the net nonzero axial component accelerates the particle in the direction of the decreasing magnetic field magnitude. 
Because the magnetic field can do no work, the total energy of the particle remains constant---the change in the parallel velocity is accompanied by a small change in the perpendicular velocity governed by the average of the radial component of the Lorentz force.  The net effect of the magnetic mirror force is that, as a particle moves in the direction of the increasing magnetic field, the mirror force reduces the velocity $v_\parallel$ parallel to the mean magnetic field over the Larmor orbit and increases perpendicular velocity $v_\perp$ to maintain a constant total velocity $v=(v^2_\perp+ v^2_\parallel)^{1/2}$.

Although the mirror force in a static magnetic field cannot change the energy of particles, any changes of the magnetic field in time will induce an electric field according to Faraday's Law.  Work done by that induced electric field can do work on the particles, providing the key element underlying the physics of TTD.  In a collisionless plasma, collisionless wave-particle interactions are governed by the resonance condition, given by $\omega - k_\parallel v_\parallel = n \Omega_s$, where  $n\Omega_s$ for $n=0,\pm 1, \pm 2, \ldots$ incorporates the cyclotron harmonics of the particle motion in a magnetic field \citep{melrose:1980}.
Here we adopt the convention that the value of $\omega$ is positive, so that the sign of $\V{k}$ indicates the 
direction of the phase velocity.
The $n=0$ resonance, known as the Landau resonance, describes resonant interactions with particles that have parallel velocities near the phase velocity of the wave, $v_\parallel \simeq \omega/k_\parallel$,
enabling energy exchange between the particles and the wave through two mechanisms: (i) the electrostatic force due to the parallel component of the electric field governing the phenomenon of Landau damping (LD) \citep{Landau:1946, Villani:2014}; (ii) the magnetic mirror force governing the phenomenon of TTD \citep{stix:1992}, also known as Barnes damping \citep{Barnes:1966}. In the case of TTD, the perpendicular component of the electric field, induced by the change in the magnetic field magnitude along the parallel direction, accelerates the particle by changing the perpendicular velocity $v_\perp$; the mirror force effectively converts this perpendicular velocity into parallel velocity, leading to a net acceleration of the particle along the axial direction parallel to the magnetic field \citep{Howes:2024}.  For a Maxwellian distribution of particles, there are more particles with parallel velocities $|v_\parallel| < \omega/|k_\parallel|$ than particles with $|v_\parallel| > \omega/|k_\parallel|$, so the net effect on the distribution is an increase of the particle energy, leading to damping of the wave. 
A detailed demonstration of this phenomenon for a model moving magnetic mirror field is presented in  \secref{sec:predict_ttd_sig}.

\subsection{Field-Particle Correlation for Transit-Time Damping \label{sec:ttd_fpc_formula}}

To determine the appropriate form of the field-particle correlation to diagnose TTD via the magnetic mirror force, we start with the Vlasov equation for a species $s$ being acted upon by a general force $\V{F}_s$,
\begin{equation}
\frac{\partial f_s}{\partial t} + \V{v}\cdot \nabla f_s + \frac
     {\V{F}_s}{m_s} \cdot \frac{\partial f_s}{\partial \V{v}} =0.
\label{eq:Vlasov}
\end{equation}
Multiplying the Vlasov equation by $m_s v^2/2$, we obtain an expression for the rate of change of the \emph{phase-space energy density},
$w_s(\V{r},\V{v},t) \equiv m_s v^2 f_s(\V{r},\V{v},t)/2 $, 
\begin{equation}
  \frac{\partial w_s(\V{r},\V{v},t)}{\partial t} = - \V{v}\cdot \nabla  w_s  -
\frac{v^2}{2}  \V{F}_s \cdot \frac{\partial f_s}{\partial \V{v}}.
  \label{eq:dws}
\end{equation}
Previous analysis of this equation \citep{Klein:2016a, Howes:2017a} has shown that, if integrated over space (with appropriate infinite or periodic boundary conditions), the change in the total kinetic energy of the particles  $\mathfrak{W}_s (t)= \int d^3\V{r} \int d^3\V{v} \ w_s(\V{r},\V{v},t)$ is due to work done on the particle species by the force  $\V{F}_s$.  Therefore, the field-particle correlation due to a general force  $\V{F}_s$ at spatial position $\V{r}_0$ is defined as a time-average over a correlation interval $\tau$ of the last term on the right hand side of \eqref{eq:dws},
\begin{equation}
  C_{\V{F}_s}(\V{r}_0,\V{v},t;\tau)  \equiv \frac{1}{\tau}\int^{t+\tau/2}_{t-\tau/2}\frac{-v^2}{2}  \V{F}_s(\V{r}_0,\V{v},t') \cdot \frac{\partial f_s(\V{r}_0,\V{v},t')}{\partial \V{v}} dt'.
  \label{eq:corrf}
\end{equation}
Note here that the correlation interval $\tau$ is a parameter of the field-particle correlation analysis, and so is included as a secondary argument, separated by a semicolon from the primary arguments that define the dimensions of the 3D-3V phase space in position, velocity, and time.

If we consider the force due to an electric field $\V{F}_s=q_s \V{E}$, we obtain the established field-particle correlation  due to the electric field \citep{Klein:2016a, Howes:2017a, Klein:2017b},
\begin{equation}
  C_{\V{E}, s}(\V{r}_0,\V{v},t;\tau)  = \frac{1}{\tau}\int^{t+\tau/2}_{t-\tau/2}\frac{-q_s v^2}{2}  \V{E}(\V{r}_0,t') \cdot \frac{\partial f_s(\V{r}_0,\V{v},t')}{\partial \V{v}} dt'.
  \label{eq:corrld}
\end{equation}

The collisionless transfer of energy between electromagnetic waves and particles in TTD is mediated by the magnetic mirror force,  $\V{F}_s = - \mu_s \bhat \cdot \nabla \V{B}$, where the magnetic moment for a particle of species $s$ is given by 
$\mu_s=m_s v_\perp^2/(2 B)$, the unit vector in the direction of the magnetic field is given by $\bhat \equiv \V{B}/B$, and the magnitude of the magnetic field is $B=|\V{B}|$.  Substituting the magnetic mirror force into \eqref{eq:corrf}, we obtain 
\begin{equation}
  C_{\V{B}, s}(\V{r}_0,\V{v},t;\tau)  = \frac{1}{\tau}\int^{t+\tau/2}_{t-\tau/2}\frac{m_s v^2v_\perp^2}{4 B}   (\bhat \cdot \nabla) \V{B}(\V{r}_0,t') \cdot \frac{\partial f_s(\V{r}_0,\V{v},t')}{\partial \V{v}} dt'.
  \label{eq:corrttd}
\end{equation}

A few modifications of the form of the field-particle correlation for TTD given in \eqref{eq:corrttd} are helpful
for its application to the gyrokinetic simulations presented here.  First, we exploit two important characteristics of TTD and turbulence: (i) TTD is most effective in damping the dominant \Alfvenic fluctuations\footnote{This peak in TTD damping rates is clearly shown in the linear dispersion relation plots for the \Alfven wave mode shown in \figref{fig:ldr_mr1836}.} in turbulence with perpendicular wavelengths at the ion scales, $k_\perp \rho_i \sim k \rho_i \sim 1$ where $k_\parallel/k_\perp \ll 1$; and (ii) for most turbulent space and astrophysical plasmas of interest, the amplitude of the magnetic fluctuations $\delta \V{B}$ at ion scales $k \rho_i \sim 1$ is much smaller than the magnitude of the mean magnetic field $\V{B}_0$.  Therefore, if we separate the magnetic field into its mean plus the fluctuations, $\V{B}=\V{B}_0+\delta \V{B}$, where $|\delta \V{B}| \ll |\V{B}_0| $, the change in the magnetic field magnitude $\delta | \V{B}|$ (which is the key ingredient for the magnetic mirror force) can be expressed as $\delta B_\parallel$ by recognizing
\begin{equation}
\delta | \V{B}| = |\V{B}|-|\V{B}_0| = \sqrt{(\V{B}_0 + \delta \V{B})^2} - B_0 = \sqrt{B_0^2 + 2 \delta \V{B} \cdot\V{B}_0 + |\delta \V{B}|^2 } -B_0 \simeq \delta B_\parallel
\label{eq:deltabpar}
\end{equation}
where we use a binomial expansion to eliminate the square root, neglect the small $|\delta \V{B}|^2$ term, and write $\delta B_\parallel = \delta \V{B} \cdot (\V{B}_0/B_0)$ as the variation in the component of the perturbed magnetic field parallel to the mean magnetic field.  Furthermore, separating term $v^2=v_\perp^2+v_\parallel^2$ in the correlation \eqref{eq:corrttd}, it is easy to show that the $v_\perp^2$ contribution yields a perfect differential in $f_s$ when integrated over all parallel velocity \citep{Howes:2017a}, so we choose to omit this term since it leads to zero net change in the particle energy.  Finally, we write the gradient along the magnetic field direction by $\nabla_\parallel \equiv \bhat \cdot \nabla$, leading to our preferred form of the field-particle correlation for TTD,
\begin{equation}
  C_{\delta B_\parallel, s}(\V{r}_0,\V{v},t;\tau)  = \frac{1}{\tau}\int^{t+\tau/2}_{t-\tau/2}\frac{m_s v_\parallel^2v_\perp^2}{4 B}   \nabla_\parallel \delta B_\parallel  \frac{\partial f_s(\V{r}_0,\V{v},t')}{\partial v_\parallel} dt'.
  \label{eq:corrttd_final}
\end{equation}

In the gyrokinetic system of equations \citep{antonsen:1980,frieman:1982,howes:2006,schekochihin:2009}, the gyro-averaged effect of parallel magnetic field gradients leads to the magnetic mirror force, which can accelerate particles in the direction parallel to the magnetic field via the Landau resonance, and therefore can lead to collisionless TTD of electromagnetic fluctuations.  A rigorous derivation of the gyrokinetic equation in \appref{sec:gk_theory} shows explicitly the two collisionless wave-particle interactions via the Landau resonance---specifically, LD and TTD.  The natural form of the field-particle correlation arising from the gyrokinetic version of the generalized energy density equation is slightly different from the field-particle correlation for TTD given in \eqref{eq:corrttd_final}, but the gyrokinetic form requires the gyroaveraged distribution function, which is not accessible through single-point spacecraft measurements and is not a natural quantity that can easily be derived from other kinetic simulation approaches, such as particle-in-cell or Vlasov simulations. Therefore, we choose here to use the perturbed distribution functions and electromagnetic fields generated by our gyrokinetic simulations, but we analyze them using \eqref{eq:corrttd_final} which is more directly applicable to spacecraft measurements or alternative kinetic simulation approaches, such as particle-in-cell codes.

Utilizing the gyroaveraged distribution function and corresponding fields, the quantity $C_{\delta B_\parallel, s}(\V{r}_0, \V{v}, t; \tau)$  reveals the velocity-space signature of TTD on the gyrotropic velocity space $(v_\perp, v_\parallel)$. To simplify the notation, henceforth we shall employ $C_{\delta B_\parallel, s} (v_\parallel, v_\perp)$ to signify the gyrotropic correlation, explicitly noting the associated spatial position $\V{r}_0$, time $t$, and the correlation interval $\tau$ only when necessary.
The resonant structure of the velocity-space signature of TTD is primarily a function of $v_\parallel$, so it is often useful to define the \emph{reduced parallel field-particle correlation} by integrating $C_{\delta B_\parallel, s} (v_\parallel, v_\perp,t)$  over $v_\perp$, given by $C_{\delta B_\parallel, s}(v_\parallel,t) \equiv 2 \pi \int C_{\delta B_\parallel, s}(v_\parallel, v_\perp,t) v_\perp d v_\perp$, where the extra $2 \pi v_\perp$ factor arises from the integration over the gyrophase in 3V phase space.
A \emph{timestack plot} of the reduced parallel correlation $C_{\delta B_\parallel, s}(v_\parallel,t)$ reveals the persistence in time of any resonant velocity-space signatures in $v_\parallel$.  We can also consider the rate of change of the kinetic energy density of species $s$ due to TTD by integrating the gyrotropic correlation over all velocity space: $(\partial W_s/\partial t)_{\mbox{TTD}} = \int C_{\delta B_\parallel, s}(v_\parallel, v_\perp) d^3 \V{v}$.

In closing, note that the gyroaveraging procedure employed in the derivation of the system of gyrokinetics enables the variations in the perpendicular components of the electric field $E_\perp$ to be expressed in terms of changes in the parallel component of the magnetic field $\delta B_\parallel$, as shown by \eqref{eq:ex_bz_conversion}. In a system where the gyroaverage has not been performed, the work done by TTD is actually mediated (at the position of the particle) by the perpendicular component of the electric field $E_\perp$ \citep{Howes:2024}.  Therefore, the perpendicular electric field correlation, given by summing the two perpendicular contributions to the electric field correlation,  $C_{E_\perp}(\V{r}_0,\V{v},t;\tau)$ \citep{Klein:2020, Afshari:2023}, can be used to seek the velocity-space signature of TTD at the parallel resonant phase velocity, as seen recently in hybrid particle-in-cell simulations of plasma turbulence \citep{Cerri:2021}.

\subsection{Prediction of the Velocity-Space Signature of Transit-Time Damping \label{sec:predict_ttd_sig}}

To predict the velocity-space signature of TTD, we begin with a simple model of a magnetic field with an amplitude variation that varies along the mean field direction $z$, given in cylindrical coordinates $(r,\phi,z)$ by
\begin{equation}
    \label{eq:toymodelmirror}
    \V{B}(r,\phi,z) = - \frac{\delta B_z }{4} k r\sin(k z') \hat{\V{r}} + \left \{ B_0 + \frac{\delta B_z}{2} [1 - \cos (k z')] \right \} \hat{\V z} \\
\end{equation}
where the wavenumber $k$ of the spatial variation of the magnetic field magnitude is along the mean field direction $z$, and $z'=z-U t$, such that that pattern moves in the $+z$ direction with a phase speed $U \geq 0$. The corresponding electric field variation can be determined by the Lorentz transform\footnote{The equations for spatial and temporal variations of the magnetic and electric field  given by \eqref{eq:toymodelmirror} and \eqref{eq:toymodelmirror_e} satisfy Faraday's Law, indicating explicitly that the electric field arises from the time variation of the magnetic field, but mathematically solving for \eqref{eq:toymodelmirror_e} is more easily done using the Lorentz transform.} from the primed (wave) frame $K'$ in which the magnetic field pattern is stationary (and therefore $\V{E}'=0$) to the unprimed (lab) frame $K$.  This Lorentz transformation in the non-relativistic limit $U/c \ll 1$ is given by 
$\V{E}=\V{E}' - \V{U} \times \V{B}$ and $\V{B}=\V{B}'$  \citep{Howes:2014a}, where the transformation velocity is just $\V{U} =U \zhat$.  The resulting induced electric field in the lab frame $K$ is given by 
\begin{equation}
    \label{eq:toymodelmirror_e}
    \V{E}(r,\phi,z) = U\frac{\delta B_z}{4} kr  \sin (k z')  \hat{\boldsymbol{\phi}}. \\
\end{equation}
With this simple model, we can illustrate how a single particle is accelerated into different regions of velocity space by the electromagnetic fields. Extending this approach to consider a distribution of particles will enable us to predict the qualitative and quantitative features of the velocity-space signature of TTD.

\begin{figure}

\begin{center}
      \includegraphics[width=0.48\textwidth]{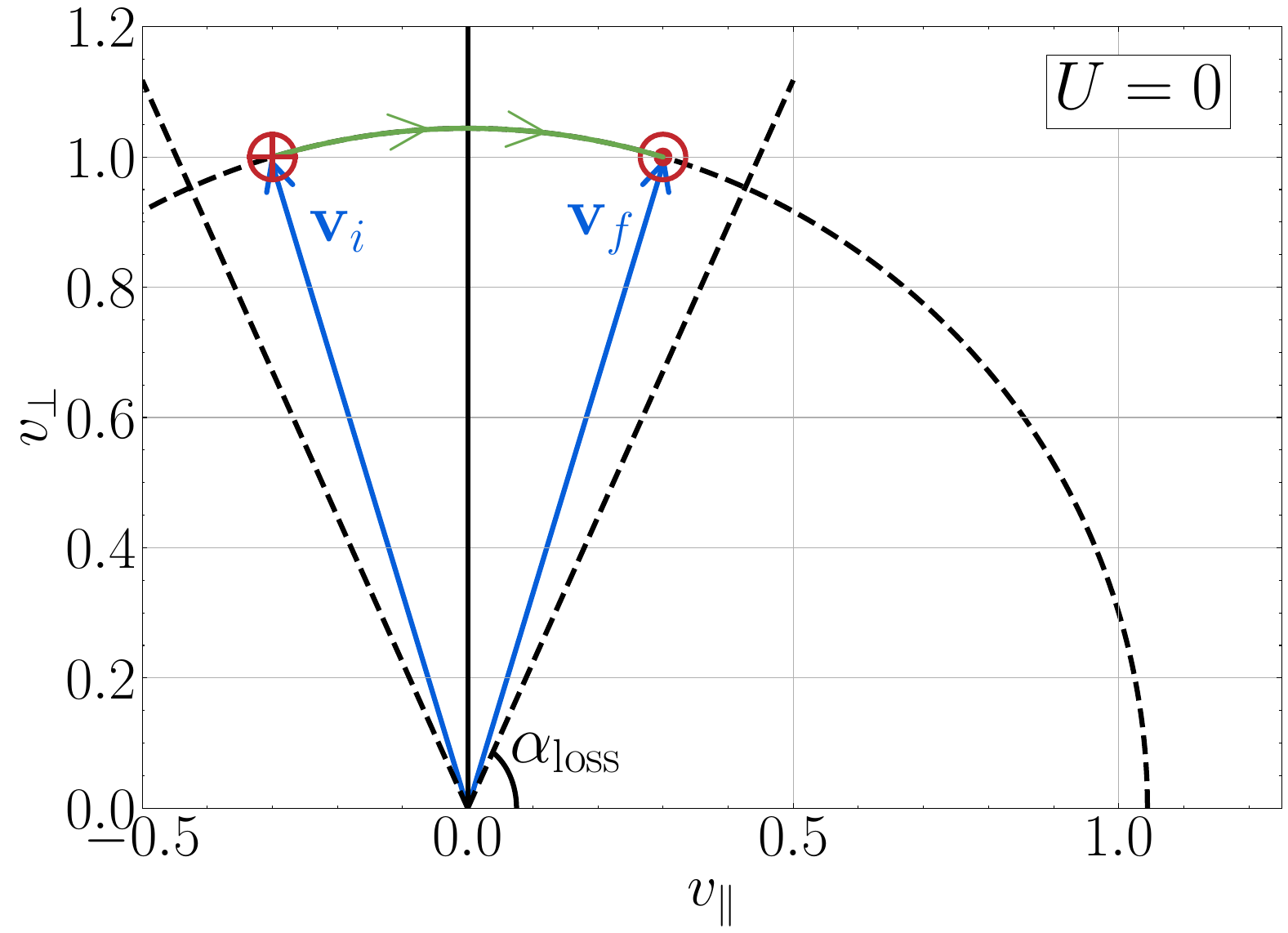}
      \includegraphics[width=0.48\textwidth]{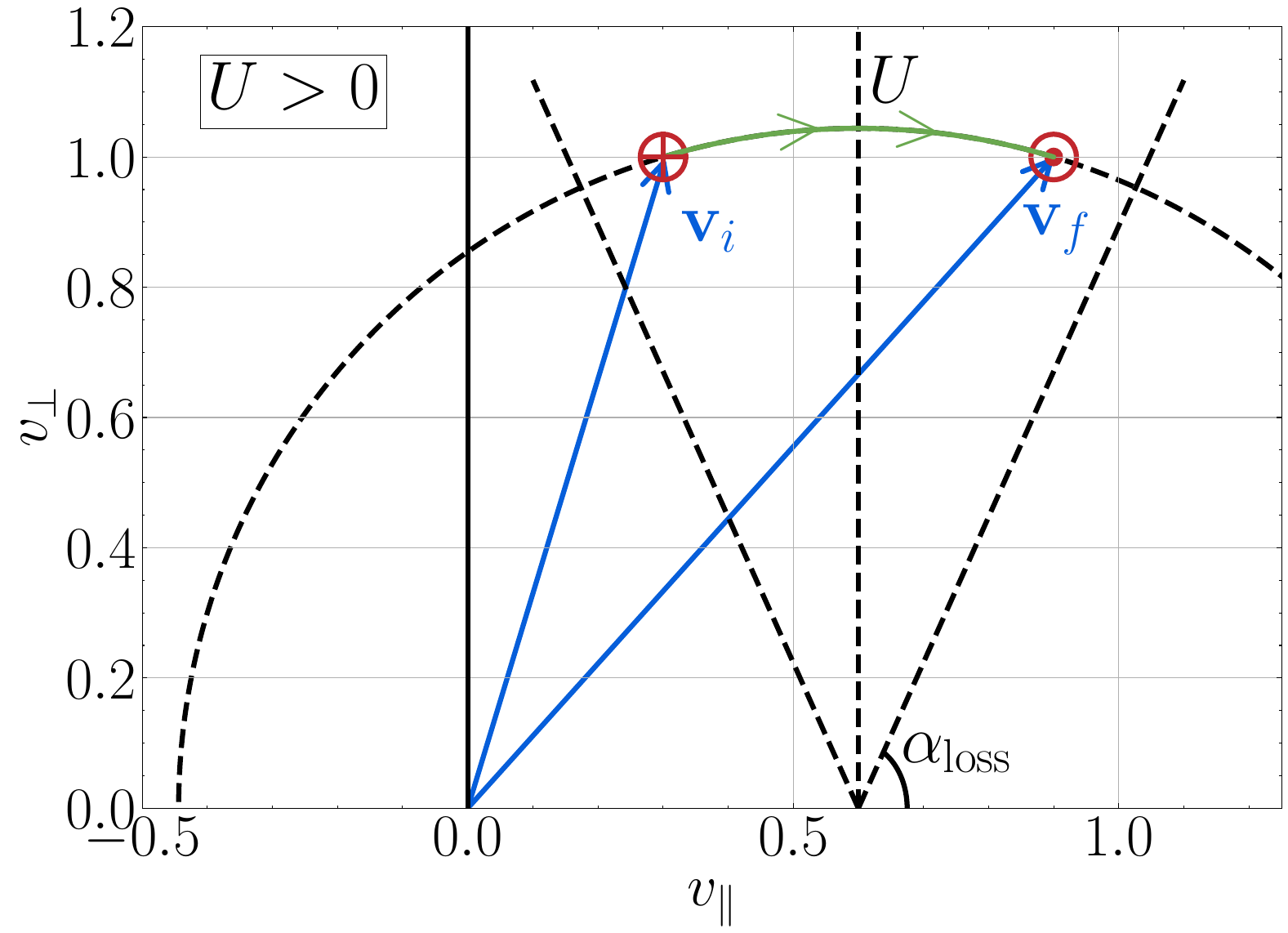}
   \end{center}
 \vskip -2in
\hspace*{0.05in} (a)\hspace*{2.3in} (b)
\vskip +2.2in
\begin{center}
        \includegraphics[width=0.6 \textwidth]{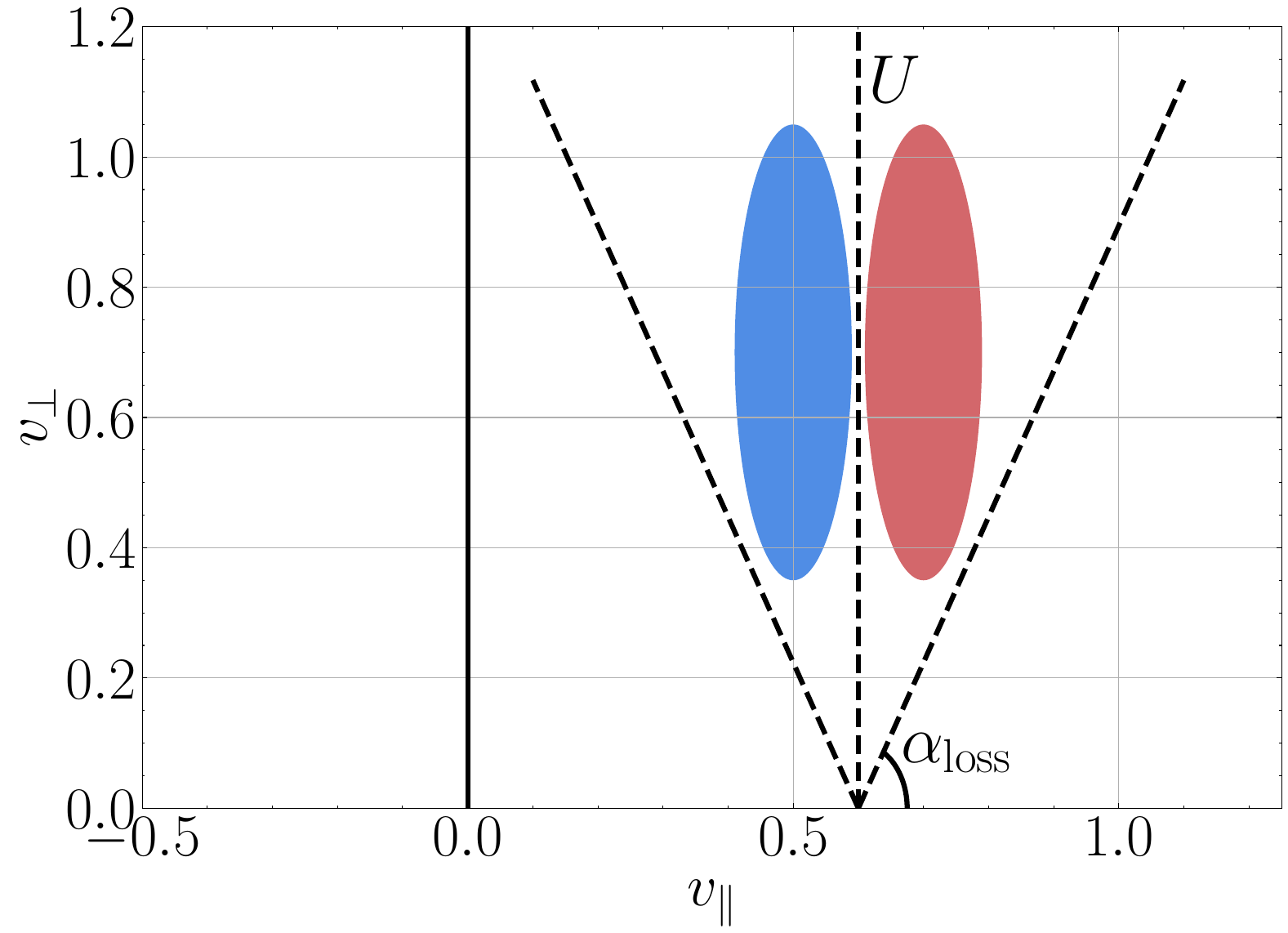}
        \includegraphics[width=0.24 \textwidth]{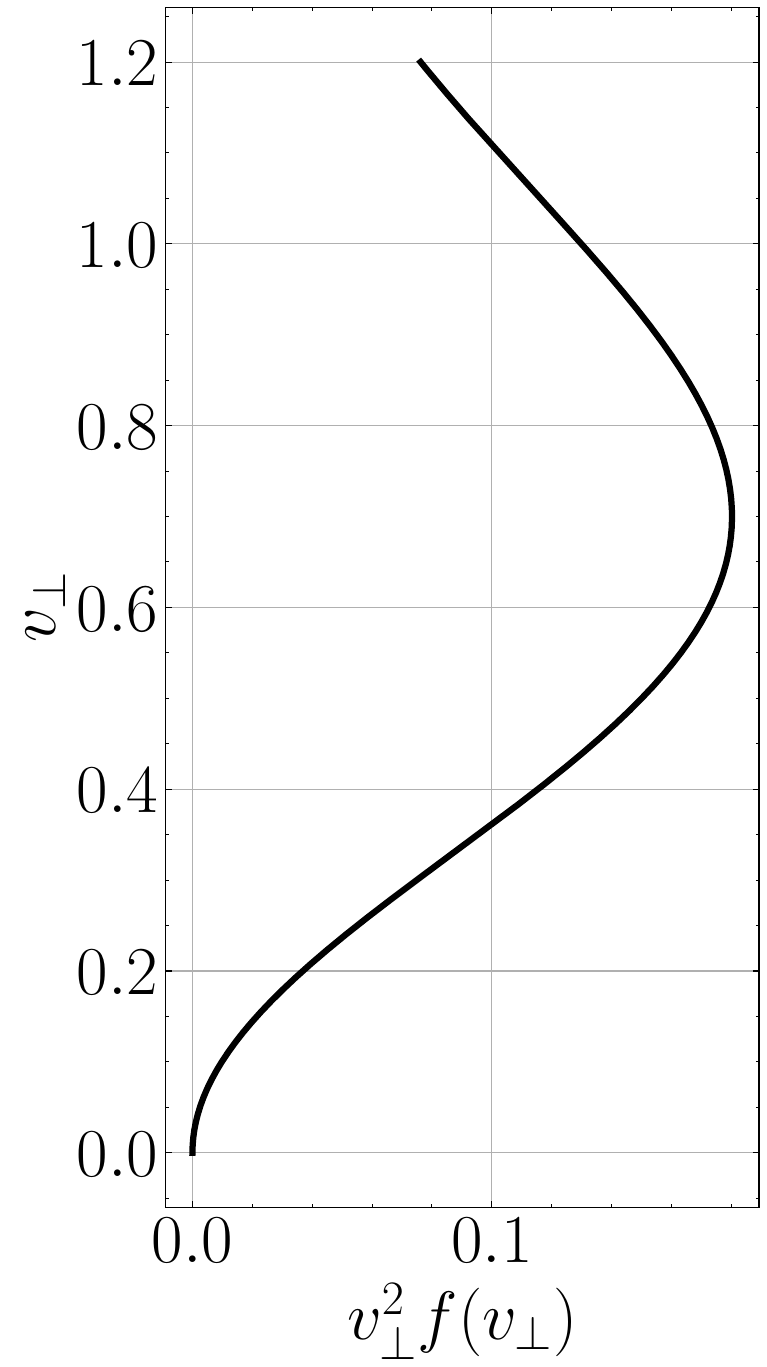}
    \end{center}
    \vskip -2.4in
\hspace*{0.3in} (c)\hspace*{3.1in} (d)
\vskip +2.2in
\caption{Diagram of the magnetic mirror force and prediction for the velocity-space signature of TTD: (a) $v_\perp$ versus $v_\parallel$ for the single particle motion in a static magnetic mirror field; (b) $v_\perp$ versus $v_\parallel$ for the single particle motion in a moving magnetic mirror field, where the vertical black dashed line denotes the wave phase velocity $U$; (c) the predicted velocity-space signature for a Maxwellian velocity distribution function, where the phase-space energy density decreases at $v_\parallel < U$ (blue) and increases at $v_\parallel > U$ (red); (d) effective $v_\perp$ weighting of correlation  $v_\perp^2f(v_\perp)$, which constrains the velocity-space signature in the $v_\perp$ direction.} 
    \label{fig:ttd_cartoon}
\end{figure}

Consider first the acceleration of a single particle in a stationary mirror field with $U=0$, as shown in \figref{fig:ttd_cartoon}(a), for a ``wave'' amplitude of $\delta B_z/B_0=0.2$, giving a mirror ratio of $B_\text{max}/B_\text{min} = 1.2$.
The particle begins at the minimum in the magnetic field at $z=0$ with an initial perpendicular velocity $v_\perp$ and an initial parallel velocity $v_\parallel<0$, given by the red $+$ in the figure at the tip of the initial velocity vector $\V{v}_i$ (blue).  As the particle moves into the increasing magnetic field at $z<0$, the mirror force increases $v_\perp$ and decreases $v_\parallel$ such that the particle moves through velocity space (green arrow) on a circle of constant total velocity $v = \sqrt{v_\parallel^2+v_\perp^2}$ (black dashed circle).  The particle will be reflected by the mirror field if the particle has an initial pitch angle $\alpha = \tan^{-1}(v_\perp/v_\parallel)$ larger than the loss cone angle $\alpha_{\mbox{loss}} = \sin^{-1}\left( \sqrt{B_\text{min}/B_\text{max}}\right)$.  For $B_\text{max}/B_\text{min} = 1.2$, the loss cone angle is $\alpha_{\mbox{loss}} = 66^\circ$; the particle depicted in \figref{fig:ttd_cartoon}(a) has an initial pitch angle $\alpha > \alpha_{\mbox{loss}}$ and is therefore reflected by the mirror field. The particle follows this circular trajectory in $(v_\parallel,v_\perp)$ velocity space until it returns to its initial axial position $z=0$, ending up with a final velocity $\V{v}_f$ (blue) with the same perpendicular component but an equal and opposite parallel component. Thus, the kinetic energy of the particle does not change, consistent with the fact that magnetic fields do no work on particles. The particle has simply been reflected by the mirror, reversing the sign of its parallel velocity.

Next, we consider the case for a magnetic mirror field moving with velocity  $\V{U} =U \zhat$, where $U>0$.  In the wave frame, moving at velocity $\V{U} =U \zhat$ in which the magnetic field is stationary, the acceleration of the particle by the magnetic field must be the same as the stationary case in \figref{fig:ttd_cartoon}(a).   But, in the lab frame, depicted in \figref{fig:ttd_cartoon}(b), the particle now moves on a circular trajectory in velocity space centered about the mirror velocity $U$, given by a constant magnitude of velocity in the wave frame $v_0=\sqrt{(v_\parallel-U)^2+v_\perp^2}$.  Here the particle is initially moving in the same direction as the mirror field but with a slower initial parallel velocity $0 \le v_\parallel \le U$, given by the red $+$ in the figure at the tip of the initial velocity vector $\V{v}_i$ (blue). If the pitch angle \emph{in the wave frame}  $\alpha_w = \tan^{-1}[v_\perp/(v_\parallel-U)]$ is greater than the loss-cone angle,  $\alpha_w > \alpha_{\mbox{loss}}$, the particle will be reflected by the moving mirror field, leading to a net acceleration in the axial direction, ultimately ending up with a parallel velocity greater than the mirror velocity $v_\parallel > U$, with a final velocity vector $\V{v}_f$ (blue).  In this case, the induced electric field given by \eqref{eq:toymodelmirror_e} has done work on the particle \citep{Howes:2024}, ultimately leading to a net acceleration in the axial direction.  This process is the fundamental energy transfer underlying the physics of TTD.

Finally, we consider how this understanding of the single particle motion and acceleration can be combined with a distribution of initial particle velocities to predict the velocity-space signature of TTD.  Note that for the sinusoidally oscillating magnetic field magnitude given by \eqref{eq:toymodelmirror}, the long time evolution of the particle in velocity space would oscillate back and forth between $\V{v}_i$ and  $\V{v}_f$ along the green trajectory shown in \figref{fig:ttd_cartoon}(a) and (b); for example, if the particle started with initial velocity $\V{v}_f$, the other side of the mirror field would lead to a reflection in the opposite direction, ultimately resulting in the particle ending up with a final velocity $\V{v}_i$.   In the case of a moving mirror field, for a Maxwellian velocity distribution there will be more particles with parallel velocities $v_\parallel < U$ than with $v_\parallel > U$, so the net effect is that more particles will gain energy than lose energy, leading to a net energization of the particles and consequent damping of the electromagnetic wave.  Only particles with pitch angles
 \emph{in the wave frame} larger than the loss cone angle, $\alpha_w > \alpha_{\mbox{loss}}$,  will undergo the mirror reflection, so that net effect on the distribution is an acceleration of particles from  $v_\parallel < U$ to $v_\parallel > U$.  The resulting change in the phase-space energy density leads to the \emph{prediction of the velocity-space signature of TTD} depicted in \figref{fig:ttd_cartoon}(c):  a loss of phase-space energy density (blue) in the region $v_\parallel < U$, and a gain of phase-space energy density (red) in the region $v_\parallel > U$. The extent of this velocity-space signature in $(v_\parallel,v_\perp)$ velocity space is confined by two effects: (i) only particles outside of the loss cone will experience a net acceleration; and (ii) the signature is weighted by the $v_\perp^2$ weighting in 
 \eqref{eq:corrttd_final} for the rate of change of phase-space energy density by TTD (which arises from the magnetic moment $\mu=m v_\perp^2/(2B)$ dependence of the mirror force) combined with the reduced perpendicular velocity distribution $f(v_\perp)$, where this net weighting of $v_\perp^2 f(v_\perp)$ is shown in \figref{fig:ttd_cartoon}(d).  Thus, the velocity-space signature of TTD in \figref{fig:ttd_cartoon}(c) is restricted to ``Landau resonant'' particles with parallel velocities near the velocity of the magnetic field pattern $v_\parallel \sim U$ and to a region away from the $v_\perp=0$ axis, unlike the velocity-space signature of LD \citep{Klein:2016a, Howes:2017a, Klein:2017b} which extends down to $v_\perp=0$.

\section{Single Kinetic \Alfven Wave (KAW) Simulations \label{sec:singleKAW}}

Here we perform numerical simulations of single kinetic \Alfven waves to determine the velocity-space signature of TTD using the Astrophysical Gyrokinetics Code, \T{AstroGK} \citep{numata:2010}.
\T{AstroGK} evolves the gyroaveraged scalar potential
$\phi(\V{r})$, parallel vector potential $A_\parallel(\V{r})$, and the
parallel magnetic field fluctuation $\delta B_\parallel(\V{r})$, as well as the 
gyrokinetic distribution function $h_s(\V{r},v_\perp,v_\parallel)$, in
a triply-periodic slab geometry of size $L_\perp^2\times L_\parallel$
elongated along the straight, uniform mean magnetic field $\V{B}_0=B_0 \zhat$. The domain-scale wavenumbers are defined by $k_{\parallel 0} = 2 \pi/L_\parallel$ and $k_{\perp 0} = 2 \pi/L_\perp$.  The gyrokinetic expansion
parameter is defined by $\epsilon \sim k_{\parallel 0}/k_{\perp 0} \ll 1$ \citep{howes:2006}, and all quantities are scaled to accommodate an arbitrary value of $\epsilon$. The gyrokinetic distribution function is related to the total distribution function $f_s$ via
\begin{equation}
f_s(\V{r}, \V{v}, t) = 
F_{0s}(v)\left( 1 - \frac{q_s \phi(\V{r},t)}{T_{s}} \right)
+ {h_s}(\V{R}_s, v_\perp, v_\parallel, t) + \delta f_{2s} + ...
\label{eqn:fullF}
\end{equation}
where $F_{0s}$ is the equilibrium distribution, $\V{r}$ is the spatial
position, $\V{R}_s$ is the associated species gyrocenter related to $\V{r}$ by $\V{r}=\V{R}_s - \V{v} \times \zhat/\Omega_s$, and $\delta
f_{2s}$ are corrections second-order in the gyrokinetic expansion
parameter $\epsilon$ which are not retained \citep{howes:2006}. 
The code employs
a pseudospectral method in the $(x,y)$ (perpendicular) plane and finite differencing in the
$z$-direction. The velocity distribution is resolved on a grid in energy $E=v_\parallel^2 + v_\perp^2$ and pitch
angle $\lambda=v^2_\perp/v^2$ space, with the points selected on a Legendre polynomial basis. A
fully conservative, linearized, gyroaveraged collision operator is employed \citep{Abel:2008, Barnes:2009} to ensure velocity-space structure in $h_s$ remains resolved throughout the simulation evolution.
We normalize time using the domain-scale \Alfven wave frequency $\omega_A \equiv k_{\parallel 0} v_A$, and particle velocity is normalized to the ion thermal velocity $v_\text{ti} = \sqrt{2 T_{i}/m_i}$, where the Boltzmann constant has been absorbed to yield temperature in units of energy.

\subsection{Single Wave Simulation Set Up \label{sec:singleKAW_sim}}

We first determine the velocity-space signature of TTD by performing simulations of single kinetic \Alfven waves (KAWs) for a fully ionized proton-electron plasma with Maxwellian equilibrium velocity distributions with a temperature ratio $T_i / T_e = 1$ and a realistic ion-to-electron mass ratio of $m_i/m_e = 1836$.  We perform three simulations with ion plasma beta $\beta_i=0.3, 1, 3$ and sample the time evolution of the electromagnetic fields and gyrokinetic distribution function at discrete single points throughout the simulation domain with dimensions $L_\perp = 2 \pi \rho_i$ and $L_\parallel = 2 \pi a_0$, yielding an arbitrary expansion parameter $\epsilon = \rho_i/a_0 \ll 1$.  
The ion plasma beta is defined by $\beta_i=8 \pi n_i T_i/B^2=v_{ti}^2/v_A^2$, where the ion thermal velocity is $v_{ti}= \sqrt{2 Ti/m_i}$, the \Alfven velocity is $v_A=B/\sqrt{4 \pi n_i m_i}$, and the Boltzmann constant is absorbed to give temperature in units of energy. Here $\rho_i\equiv v_{ti}/\Omega_i$ is the ion Larmor radius, where the angular ion (proton) cyclotron frequency is $\Omega_i=q_i B_0/m_i$. The dimensions of these single-wave simulations are  $(n_x, n_y, n_z, n_{\lambda}, n_E, n_s) = (10, 10, 32, 128, 64, 2)$, where $n_s$ denotes the number of species. Using the solutions to the linear gyrokinetic dispersion relation \citep{howes:2006}, a single plane-wave KAW with wavevector $(k_x \rho_i,k_y \rho_i, k_\parallel a_0) = (1,0,1)$ and amplitude\footnote{Note that the gyrokinetic simulation results can be scaled to any value of the gyrokinetic expansion parameter, $\epsilon \ll 1$.} $\delta B_\perp/B_0=0.125 \epsilon$ is initialized throughout the domain and allowed to evolve linearly for 5 wave periods with enhanced collisionality to eliminate any transients associated with the initialization, yielding a clean, single KAW with $k_\perp \rho_i =1$.   The simulation is then restarted with lowered collisionalities $\nu_s/(k_{\parallel 0} v_{ti}) = 2 \times 10^{-3}$ and evolved to allow collisionless wave-particle interactions to damp the wave. We have verified that the collisionless damping rates of the initialized KAWs agree with the analytical predictions from Vlasov-Maxwell and gyrokinetic linear dispersion relations.

\subsection{The Velocity-Space Signature of Transit-Time Damping}
\label{sec:iTTD}

\begin{figure}
    \begin{center}
        \includegraphics[width=0.45 \textwidth]{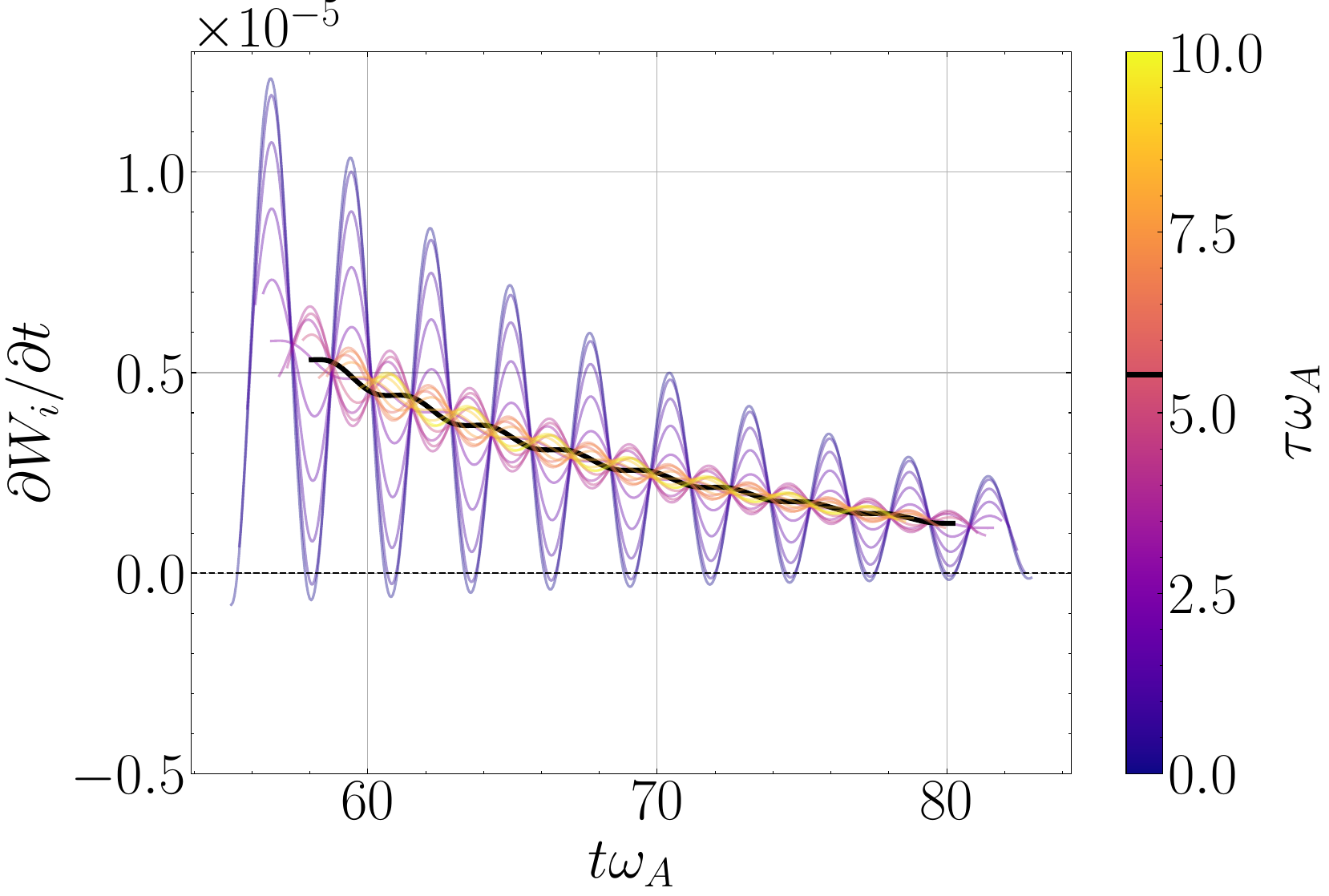}
        \includegraphics[width=0.45 \textwidth]{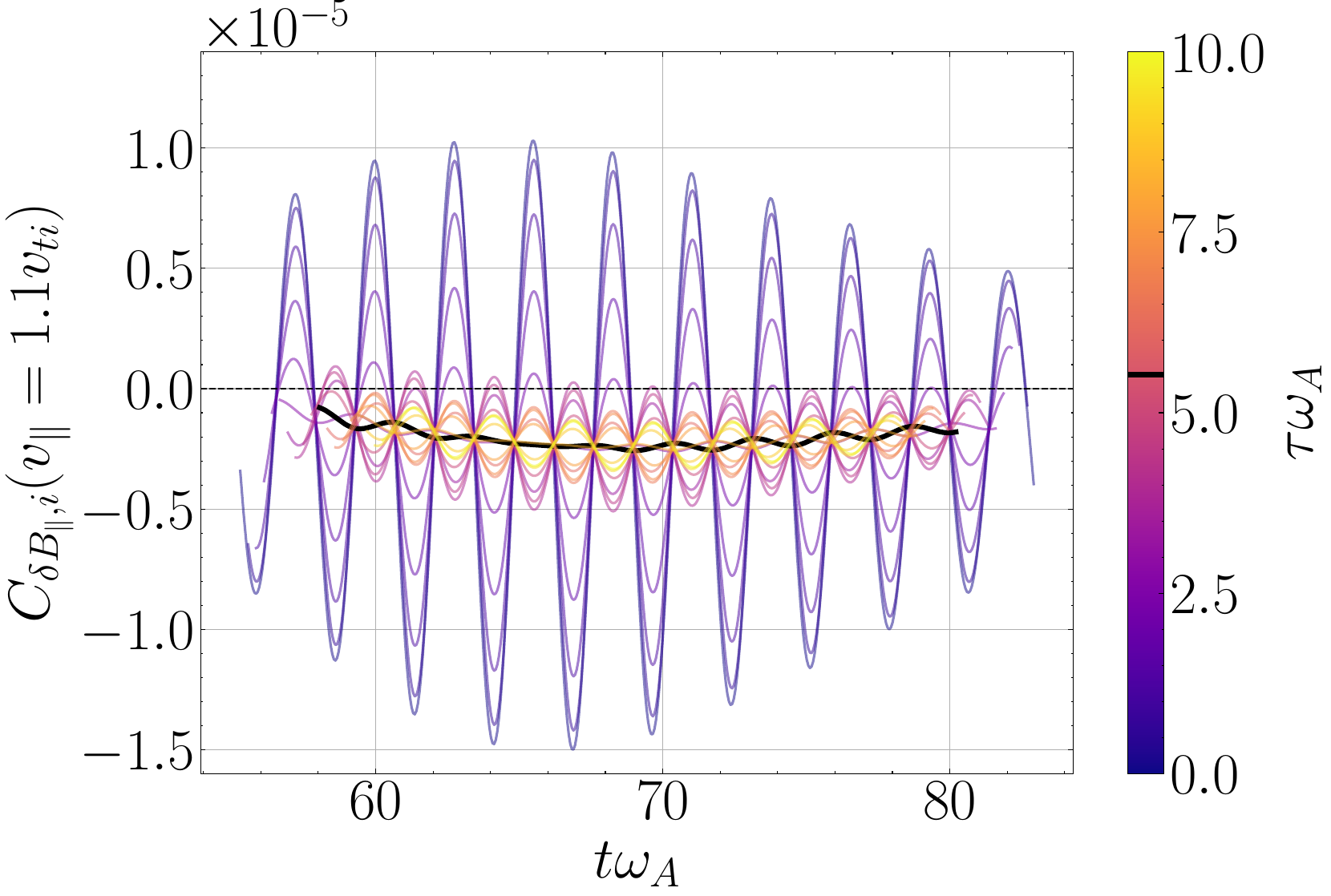}
   \end{center}
 \vskip -1.7in
\hspace*{0.15in} (a)\hspace*{2.3in} (b)
\vskip +1.8in
\begin{center}
      \includegraphics[width=0.45\textwidth]{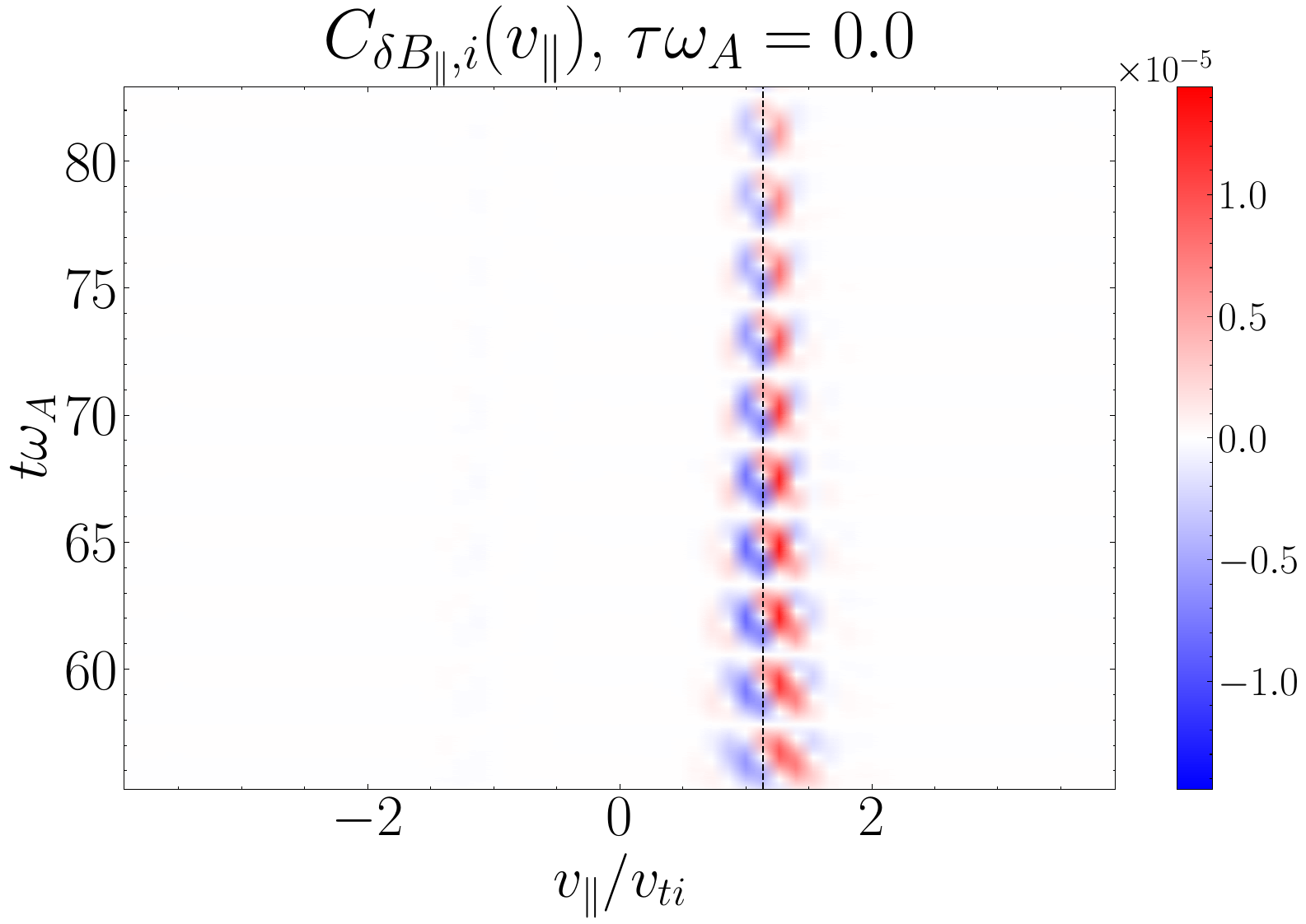}
      \includegraphics[width=0.45\textwidth]{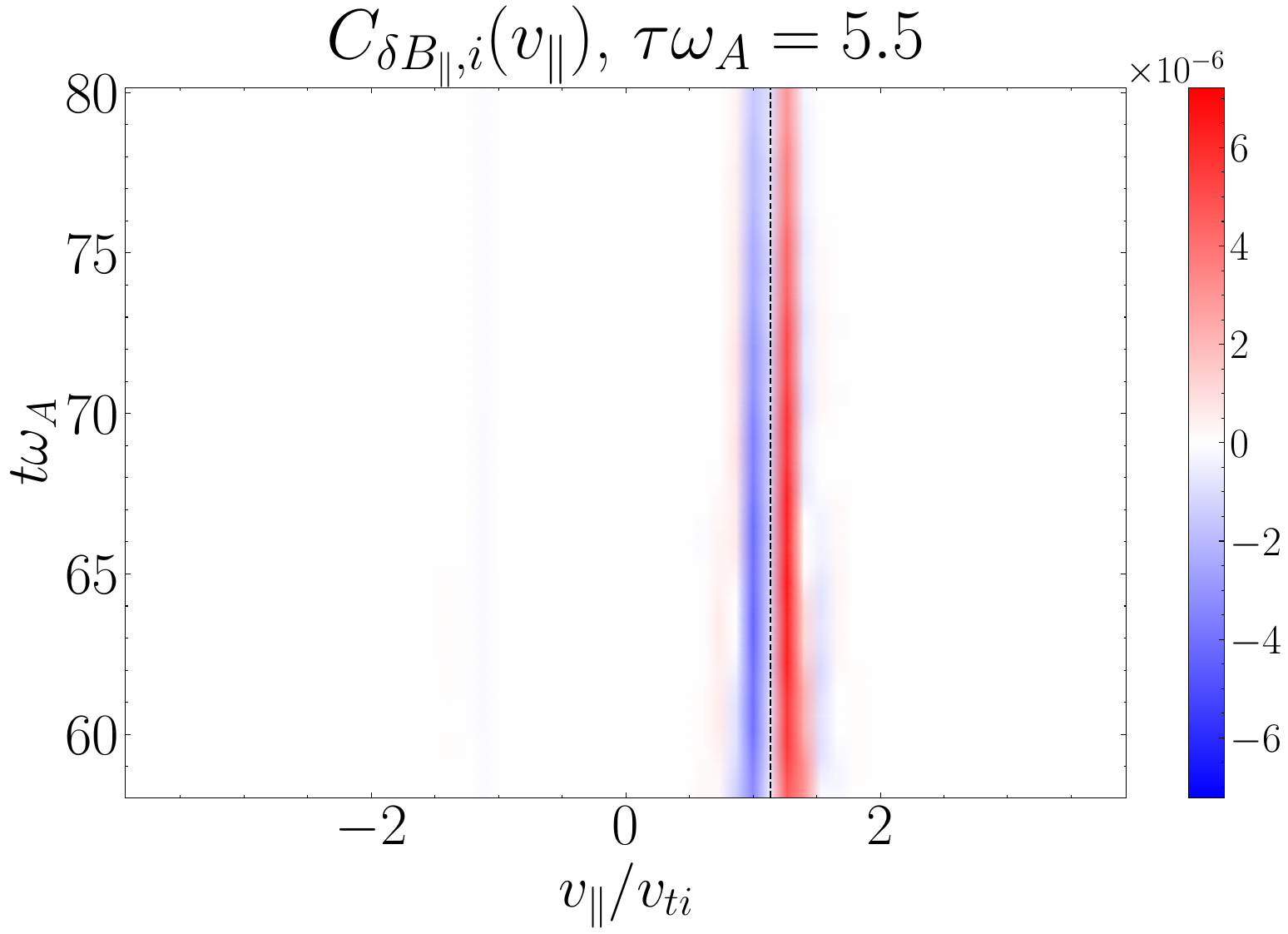}
    \end{center}
    \vskip -1.8in
\hspace*{0.15in} (c)\hspace*{2.3in} (d)
\vskip +1.8in
   \caption{Analysis of correlation interval selection for $\beta_i = 1$ \T{AstroGK} single KAW simulation. Top row: time evolution of (a) the rate of change of ion kinetic energy density due to TTD, denoted as $\partial W_i /\partial t$, and (b) the reduced correlation $C_{\delta B_{\parallel, i}} (v_\parallel, t)$ at $v_\parallel = 1.1 v_{ti}$. Both quantities are presented over a range of $\tau \omega_A$ values from 0 to 10. The selected $\tau \omega_A$ value of 5.5 is marked with a black line. Bottom row: timestack plots of the reduced correlation $C_{\delta B_{\parallel, i}} (v_\parallel, t)$ for (c) $\tau \omega_A = 0$ and (d) $\tau \omega_A = 5.5$, with the vertical dashed line at $v_\parallel/v_\text{ti} = 1.137$ labelling the normalized parallel phase velocity $\omega/(k_\parallel v_\text{ti})$.}
   \label{fig:choice_of_nc_single}
\end{figure}

We select the $\beta_i = 1$ single KAW simulation as a fiducial case to determine the velocity-space signature of TTD with \eqref{eq:corrttd_final} and compare it to the known velocity-space signature of LD using only the parallel contribution to the dot product in \eqref{eq:corrld} \citep{Klein:2016a, Howes:2017a, Howes:2018a, Klein:2017b, Chen:2019, Klein:2020, Horvath:2020, Afshari:2021}.  The linear Vlasov-Maxwell dispersion relation yields a parallel phase velocity normalized to the \Alfven velocity of $\overline{\omega} \equiv \omega/(k_\parallel v_A) = 1.137$ for a KAW with 
$k_\perp \rho_i =1$, $\beta_i=1$, and $T_i / T_e = 1$, corresponding to a normalized wave period of  $T \omega_A = 5.526$.

A key step in the field-particle correlation analysis is to choose an appropriate \emph{correlation interval} $\tau$ over which to time-average the rate of energization to eliminate a possibly larger amplitude signal of oscillatory energy transfer in order to reveal the often smaller secular rate of energy transfer that corresponds to the collisionless damping of the wave \citep{Klein:2016a, Howes:2017a, Klein:2017b}. In \figref{fig:choice_of_nc_single}, we present a test of different correlation intervals over the range  $0 \le \tau \omega_A \le 10$ for the $\beta_i=1$ single KAW simulation.  In panel (a), we plot the velocity-space integrated rate of ion energization due to TTD, $(\partial W_i/\partial t)_{\mbox{TTD}} = \int C_{\delta B_\parallel, i}(v_\parallel, v_\perp) d^3 \V{v}$, vs.~time. The unaveraged correlation ($\tau=0$, dark blue) exhibits pronounced oscillations of the net energy transfer vs.~time.  Setting the correlation interval to one linear wave period $\tau \omega_A = T \omega_A \simeq  5.5$ (black) minimizes the oscillations, providing an optimal choice for $\tau$ for a single wave with a well-defined period, as might be expected on theoretical grounds. We show in panel (b) the evolution of the reduced parallel correlation $C_{\delta B_\parallel, i}(v_\parallel,t)$ at a parallel velocity $v_\parallel = 1.1 v_\text{ti}$ slightly below the resonant velocity. Here again, setting $\tau \omega_A = 5.5$ (black) effectively minimizes the oscillations, revealing clearly the rate of secular energy transfer at that parallel velocity.  We illustrate the impact of choosing an appropriate correlation interval $\tau$ on the timestack plot of the correlation $C_{\delta B_\parallel, i}(v_\parallel,t)$ by comparing (c) the instantaneous ($\tau=0$) field-particle correlation to (d) the correlation using $\tau \omega_A = 5.5$,  showing a clear bipolar signature about the normalized parallel phase velocity $\omega/(k_\parallel v_{ti})=1.137$ that persists over the course of the simulation.
 \begin{figure}
(a)\\ \vskip -0.1in
\hbox{\hfill
            \includegraphics[width=0.9 \textwidth]{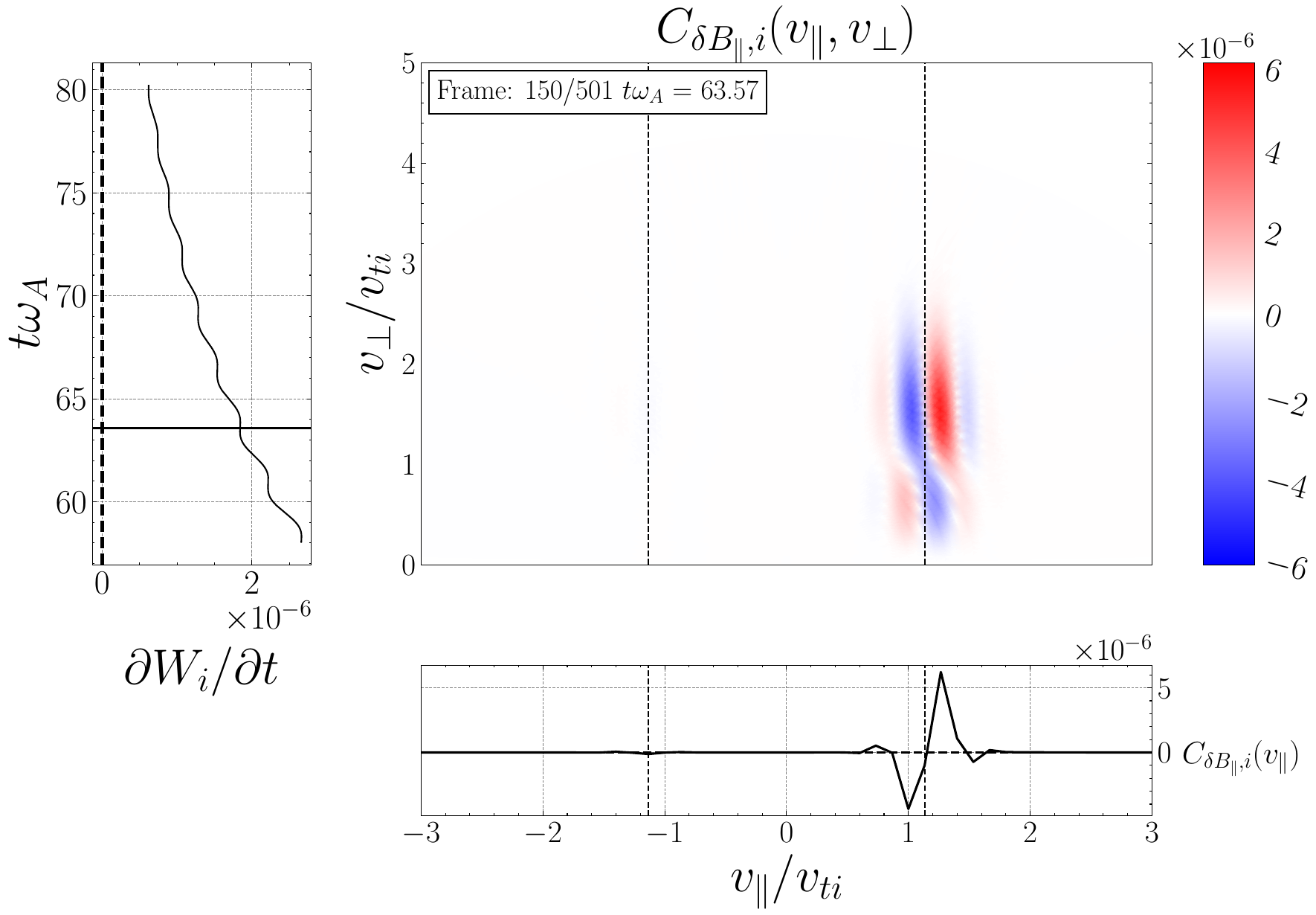} \hfill}
(b)\\ \vskip -0.1in
\hbox{\hfill
        \includegraphics[width=0.9 \textwidth]{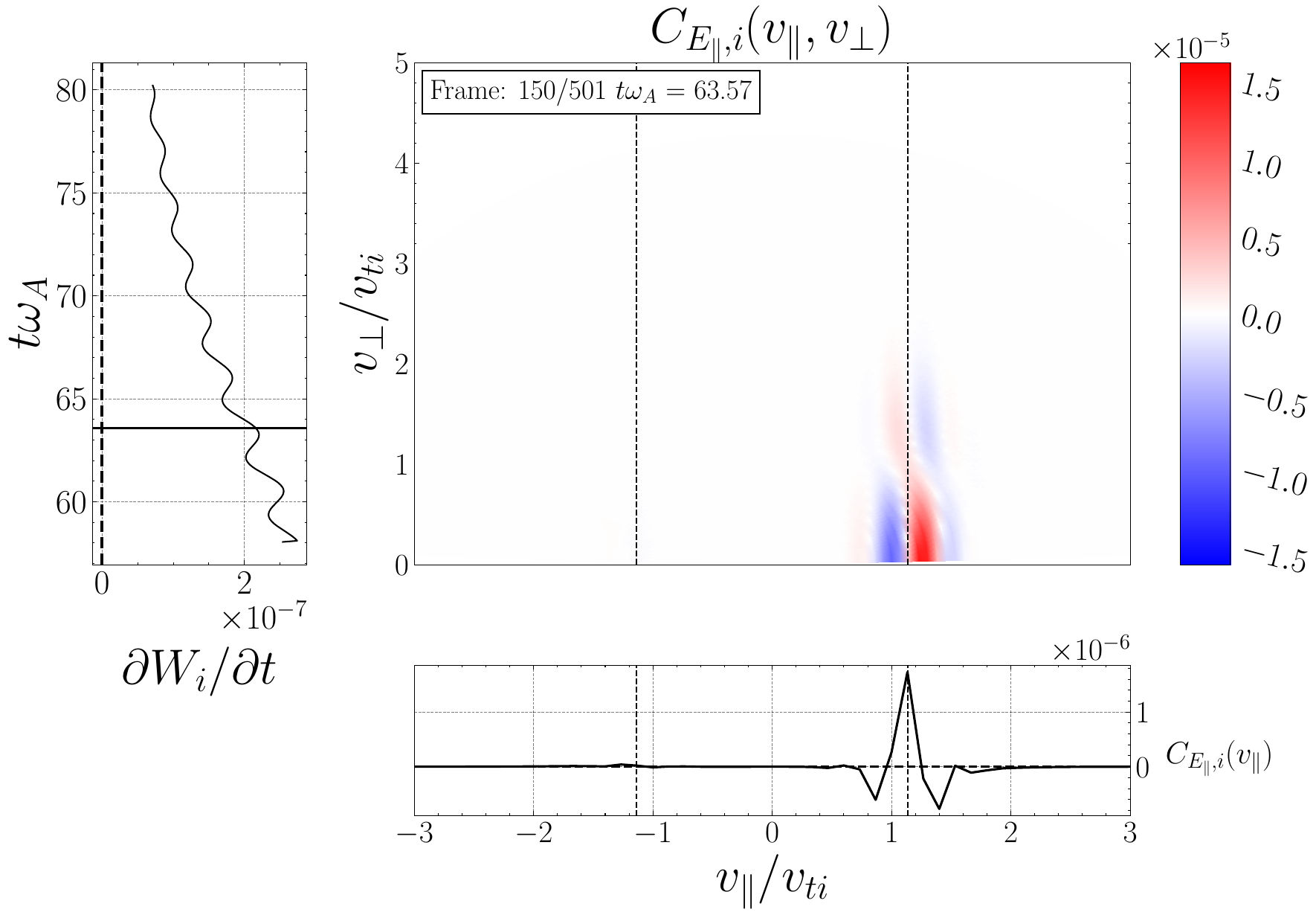} \hfill}
   \caption{Velocity-space signatures of (a) transit-time damping (TTD) and (b) Landau damping (LD), from the \T{AstroGK} simulation of a single kinetic \Alfven wave with $k_\perp \rho_i =1$, $\beta_i=1$, and $T_i / T_e = 1$, each showing the gyrotropic signatures in the main panel, the time-integrated reduced parallel signatures in the lower panel, and the net rate of ion energization vs.~time for each mechanism in the left panel. The correlation interval is chosen as $\tau \omega_A = 5.5$. The normalized parallel phase velocity is labeled by the two vertical dashed lines at $v_\parallel/v_\text{ti} = \pm 1.137$.
    \label{fig:fiducial}}
\end{figure}

Specifying the correlation interval to be approximately one wave period $\tau \omega_A = 5.5$,\footnote{Note that the velocity-space signature for a single KAW is independent of the probe position if the correlation interval is taken to be an integral multiple of the wave period.} we now present in \figref{fig:fiducial} the velocity-space signatures of (a) TTD and (b) LD  from the $\beta_i = 1$ single KAW simulation.  Each panel presents three sub-plots, explained here for the TTD case in panel (a): (i) the main plot presents the gyrotropic velocity-space signature  $C_{\delta B_\parallel, i}(v_\parallel,v_\perp)$ at time $t \omega_A = 63.57$, with the parallel phase velocity indicated (vertical dotted line); (ii) the lower plot shows the time-integrated parallel velocity-space signature $C_{\delta B_\parallel, i}(v_\parallel)$ to highlight the variation of the net energy transferred as a function of $v_\parallel$, showing a clear bipolar signature at the parallel phase velocity; and (iii) the left plot shows the velocity-space integrated net energy density transfer rate due to TTD $(\partial W_i(t)/\partial t)_{\mbox{TTD}}$ vs.~time, with the centered time of the correlation interval shown in the gyrotropic signature indicated (horizontal solid line). Panel (b) presents the corresponding plots for the LD case.

A key result of this paper is the \emph{velocity-space signature of transit time damping} plotted on gyrotropic velocity space $C_{\delta B_\parallel, i}(v_\parallel,v_\perp)$ in \figref{fig:fiducial}(a).  From the same simulation, the gyrotropic velocity-space signature of LD   $C_{E_\parallel, i}(v_\parallel,v_\perp)$, given by the parallel contribution to the dot-product in \eqref{eq:corrld}, is presented in (b) for comparison. The TTD signature agrees qualitatively with our prediction presented in \figref{fig:ttd_cartoon}(c), showing the key features: (i) the bipolar signature of the rate of loss (blue) and gain (red) of phase-space energy density is centered about the parallel wave phase velocity $v_\parallel \sim \omega/k_\parallel$ (vertical dotted black line); (ii) the gyrokinetic velocity-space signature does not extend down to $v_\perp=0$, due to a combination of the $v_\perp^2$ weighting arising from the magnetic moment $\mu=m v^2_\perp/(2B)$ in the mirror force and from the loss cone angle of the mirror force, as explained in \secref{sec:predict_ttd_sig}.  The LD signature in (b) likewise yields a bipolar signature near the parallel wave phase velocity. Besides the fact that $E_\parallel$ governs energization through LD and  $E_\perp$ governs energization through TTD \citep{Howes:2024}, a key way to distinguish these two mechanisms is that in gyrotropic velocity space the TTD signature does not extend down to $v_\perp=0$, whereas the LD signature extends right down to $v_\perp=0$.

Careful inspection of these single KAW velocity-space signatures reveals a ``twist'' seen in both panels of \figref{fig:fiducial}, where the bipolar pattern of energy loss (blue) to energy gain (red) across the parallel phase velocity reverses sign as $v_\perp$ changes.  This feature arises from the linear combination of three terms contributing to the linear response for perturbed distribution function $f_1$, which involves the components of the electric field perturbation and the zeroth-order Bessel function and its derivative, as seen in 
eq.~(4.180) from \citet{Swanson:2003}. Although a detailed decomposition of these contributions to understand this twist for a single KAW could be performed, for the case of transit-time or Landau damping in a turbulent plasma, this twist is obscured when damping a broadband spectrum of wave modes, as shown in \secref{sec:turb}, so we do not pursue this line of investigation further.


 \begin{figure}
    \begin{center}
      \includegraphics[width=0.48\textwidth]{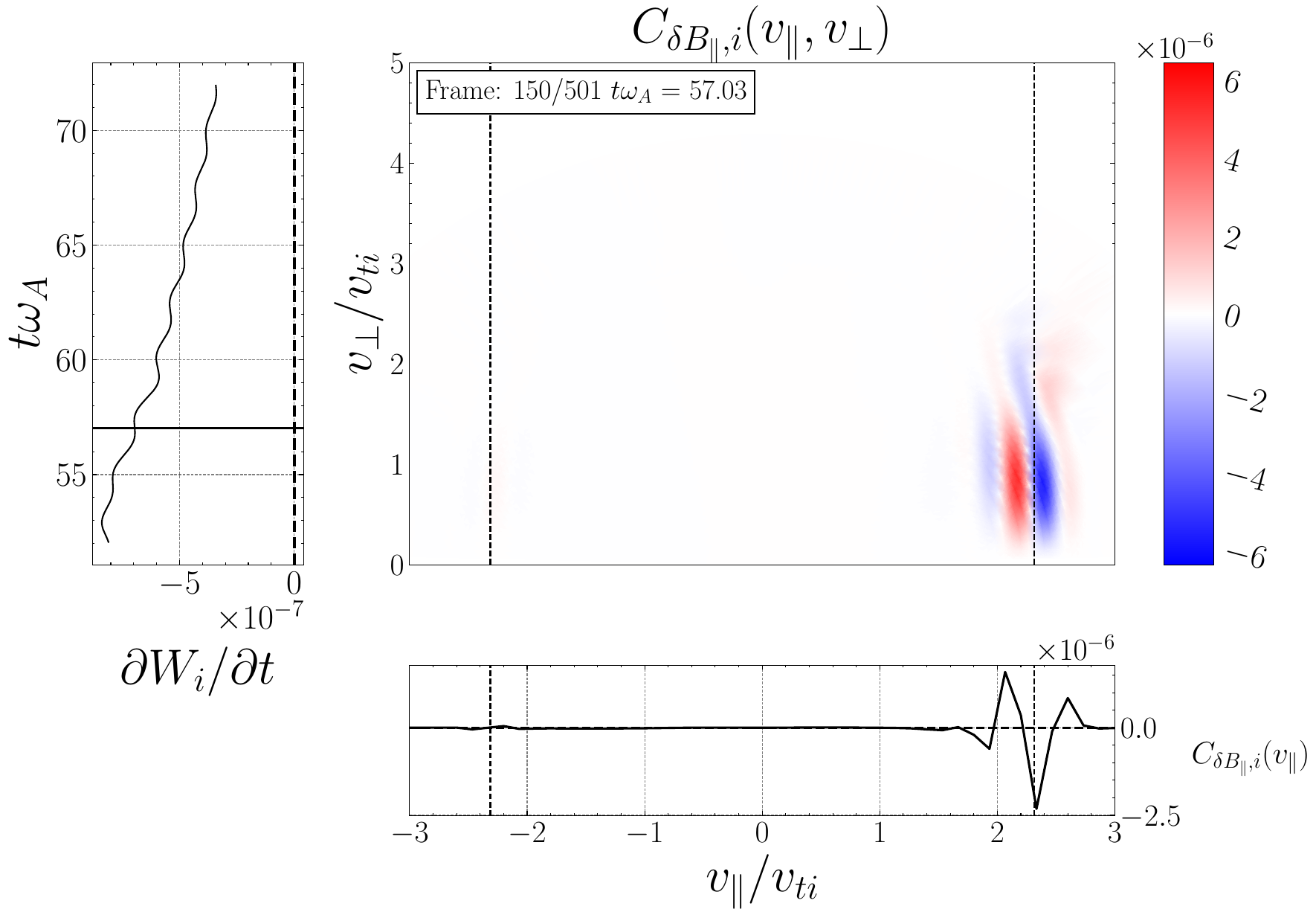}
      \includegraphics[width=0.48\textwidth]{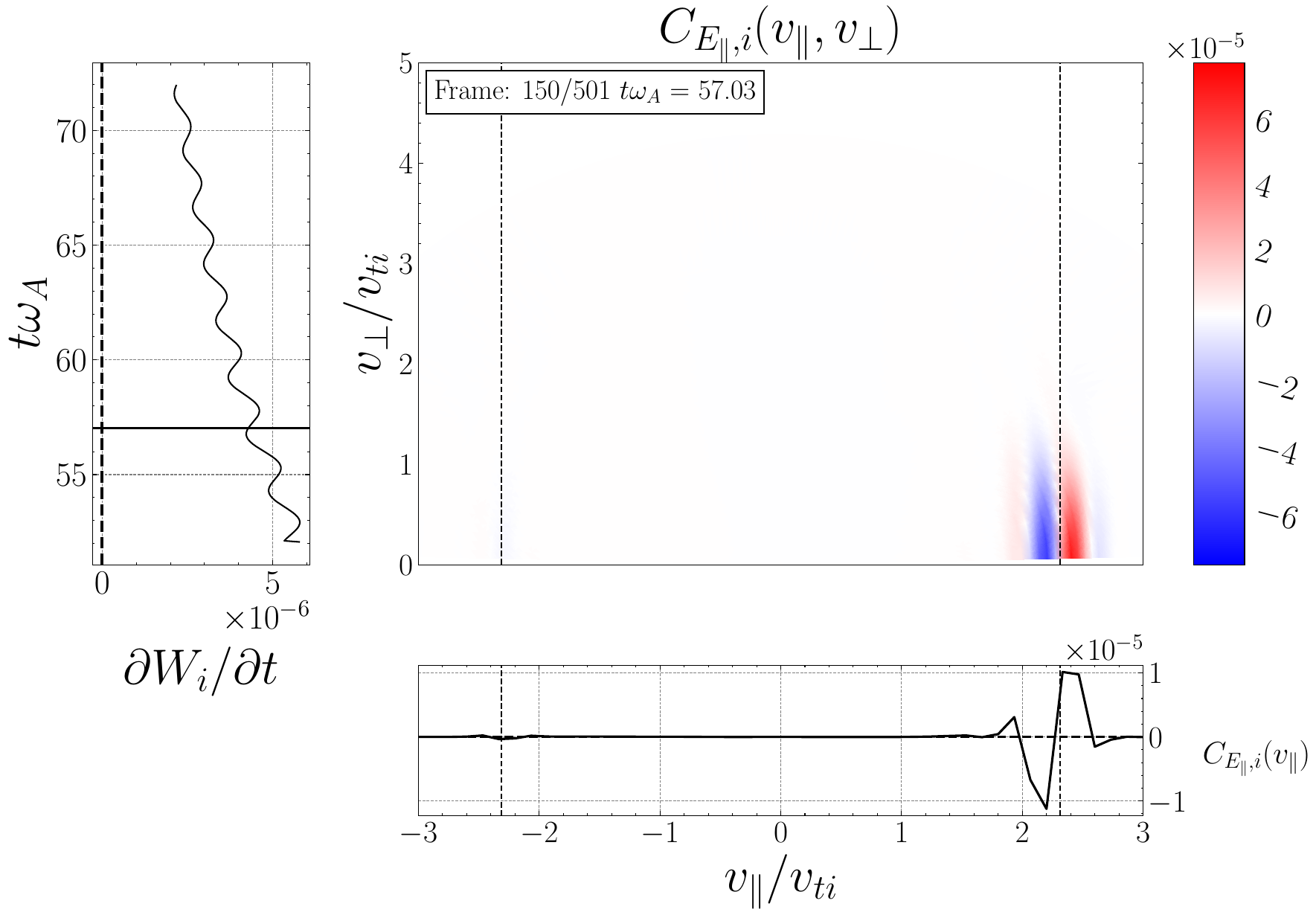}
   \end{center}
 \vskip -1.8in
\hspace*{0.05in} (a)\hspace*{2.3in} (b)
\vskip +1.8in
\begin{center}
        \includegraphics[width=0.48 \textwidth]{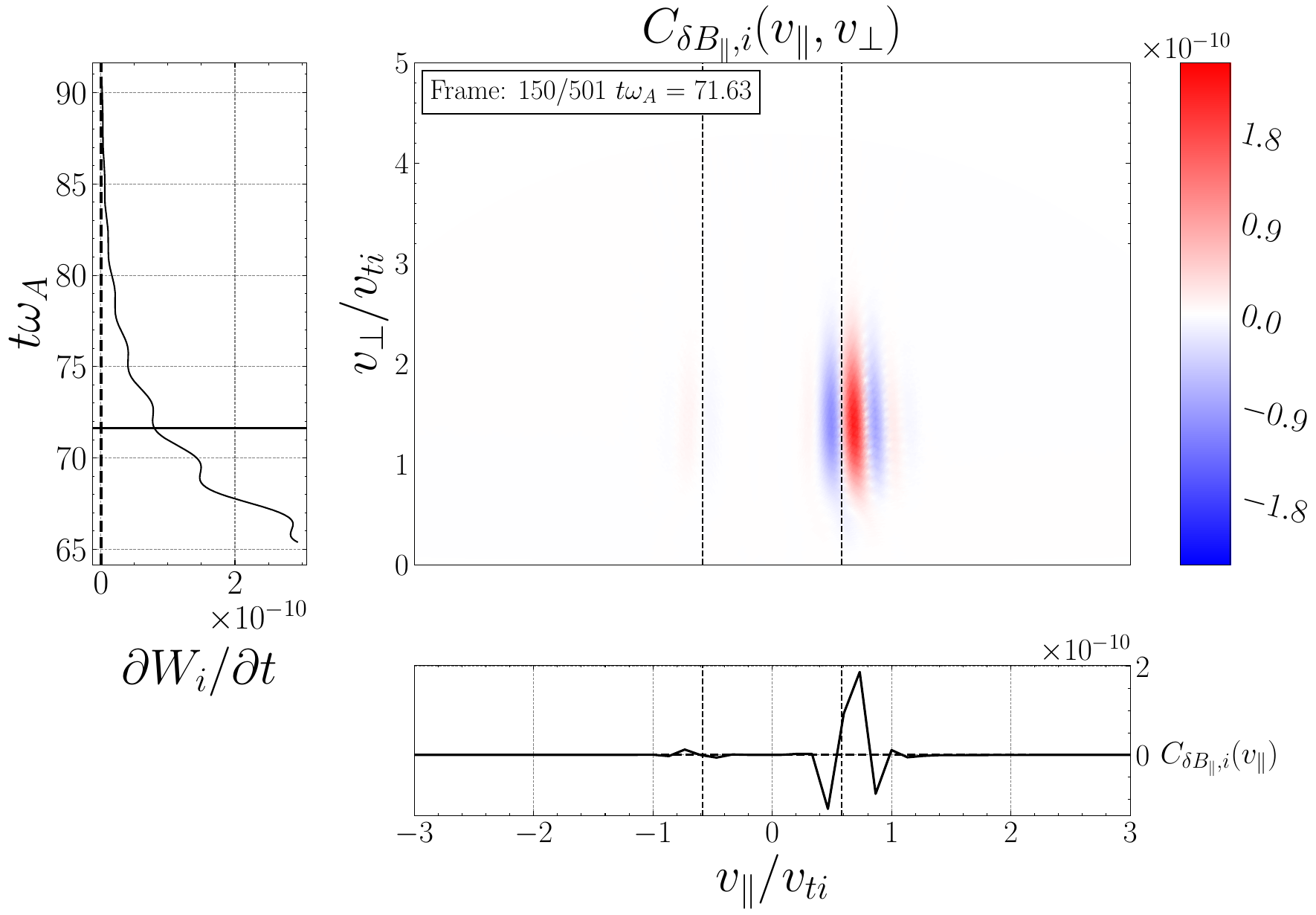}
        \includegraphics[width=0.48 \textwidth]{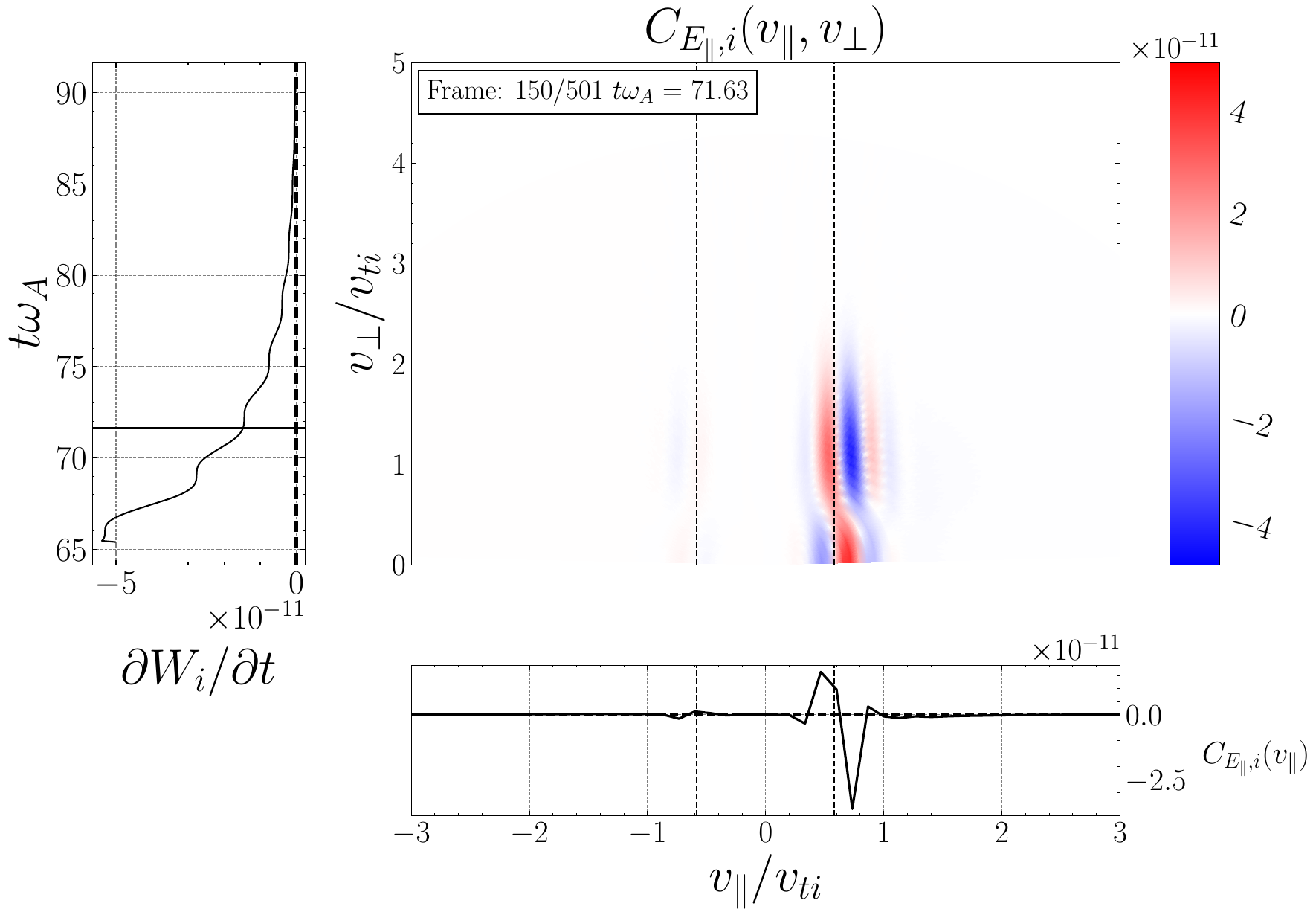}
    \end{center}
    \vskip -1.8in
\hspace*{0.05in} (c)\hspace*{2.3in} (d)
\vskip +1.8in
   \caption{Velocity-space signatures of transit-time damping (TTD, left column) and Landau damping (LD, right column) in \T{AstroGK} single KAW simulations with $k_\perp \rho_i =1$, $T_i / T_e = 1$, and  $\beta_i = 0.3$ (top row) and $\beta_i = 3$ (bottom row). The correlation intervals are set to the corresponding linear wave periods, with $\tau \omega_A = 5.0$ for $\beta_i = 0.3$ case and $\tau \omega_A = 6.2$ for $\beta_i = 3$ case. The normalized parallel phase velocity is labeled by the two vertical dashed lines at $v_\parallel/v_\text{ti} = \pm 2.313$ for $\beta_i = 0.3$ and $v_\parallel/v_\text{ti} = \pm 0.583$ for $\beta_i = 3$. Each panel follows the layout format of \figref{fig:fiducial}.}
    \label{fig:single_wave_varying_beta}
\end{figure}

\subsection{Variation of Signature with Ion Plasma Beta $\beta_i$ \label{sec:singleKAW_ttd_sig_varing_beta}}

Resonant damping of electromagnetic fluctuations through TTD and LD depends strongly on the plasma beta, typically with LD dominant at $\beta_i \ll 1$ and TTD dominant at $\beta_i \gg 1$ \citep{Quataert:1998}, so we vary the value of $\beta_i$ here to determine its impact on the characteristics of the velocity-space signature of the Landau-resonant damping mechanisms.  Using the same parameters $T_i/T_e=1$, $m_i/m_e=1836$, and $k_\perp \rho_i =1$, we perform additional single KAW simulations with $\beta_i = 0.3$ and $\beta_i = 3$.  For $\beta_i = 0.3$, the normalized parallel phase velocity is given by $\overline{\omega} \equiv \omega/(k_\parallel v_A) = 1.267$ yielding a normalized wave period of $T \omega_A = 4.959$; for $\beta_i = 3$, we obtain  $\overline{\omega}= 1.009$ and $T \omega_A = 6.227$.  

In \figref{fig:single_wave_varying_beta}, we plot the velocity-space signatures for (a) TTD and (b) LD for the  $\beta_i = 0.3$ case and for (c) TTD and (d) LD for the  $\beta_i = 3$ case, where each panel has three subplots in the same format as in \figref{fig:fiducial}. These bipolar velocity-space signatures look qualitatively similar to the $\beta_i = 1$ case in \figref{fig:fiducial}, with two important quantitative differences.

First, the position of the bipolar signature in $v_\parallel/v_{ti}$ changes as $\beta_i$ is varied, consistent with the variation of the parallel phase velocity normalized to the ion thermal velocity as $\beta_i$ varies, given by $\omega/(k_\parallel v_{ti})= \overline{\omega}/ \beta_i^{1/2}$.  For $\beta_i = 0.3$, we obtain $\omega/(k_\parallel v_{ti})= 2.313$, indicated by the vertical black dashed line in \figref{fig:single_wave_varying_beta}(a) and (b); for  $\beta_i = 3$, we obtain $\omega/(k_\parallel v_{ti})= 0.583$, as shown in (c) and (d).  In both cases, the bipolar signature remains closely associated in  $v_\parallel$ with the parallel phase velocity $\omega/k_\parallel$, as expected for a Landau-resonant energy transfer between the fields and the ions. 

Second, looking at the vertical subplots on the left for each panel, which shows the net rate of change of ion energy density $W_i$ mediated by each mechanism averaged over a correlation interval equal to one wave period $\tau=T$, we find the surprising result for $\beta_i = 0.3$ that, although LD leads to a net gain of energy by the ions (as expected for collisionless damping of a wave), TTD leads to a net \emph{loss} of energy from the ions.  This finding suggests that the ions are losing energy through the magnetic mirror force while gaining energy through acceleration by the parallel electric field.  The rate of ion energy gain by LD is about ten times larger than the rate of loss by TTD, so the summed effect of these two mechanisms is energization of the ions by collisionless damping of the wave, as expected.
For the $\beta_i = 3$ case, we find the equally surprising result that although TTD leads to a net gain of energy by the ions, LD is leading to a net loss of energy from the ions; the rate of ion energy gain by TTD is larger than the rate of ion energy loss by LD, so the summed contributions yield a net gain of ion energy as expected for collisionless damping.

 \begin{figure}
 \begin{center}
\includegraphics[width=0.65\textwidth]{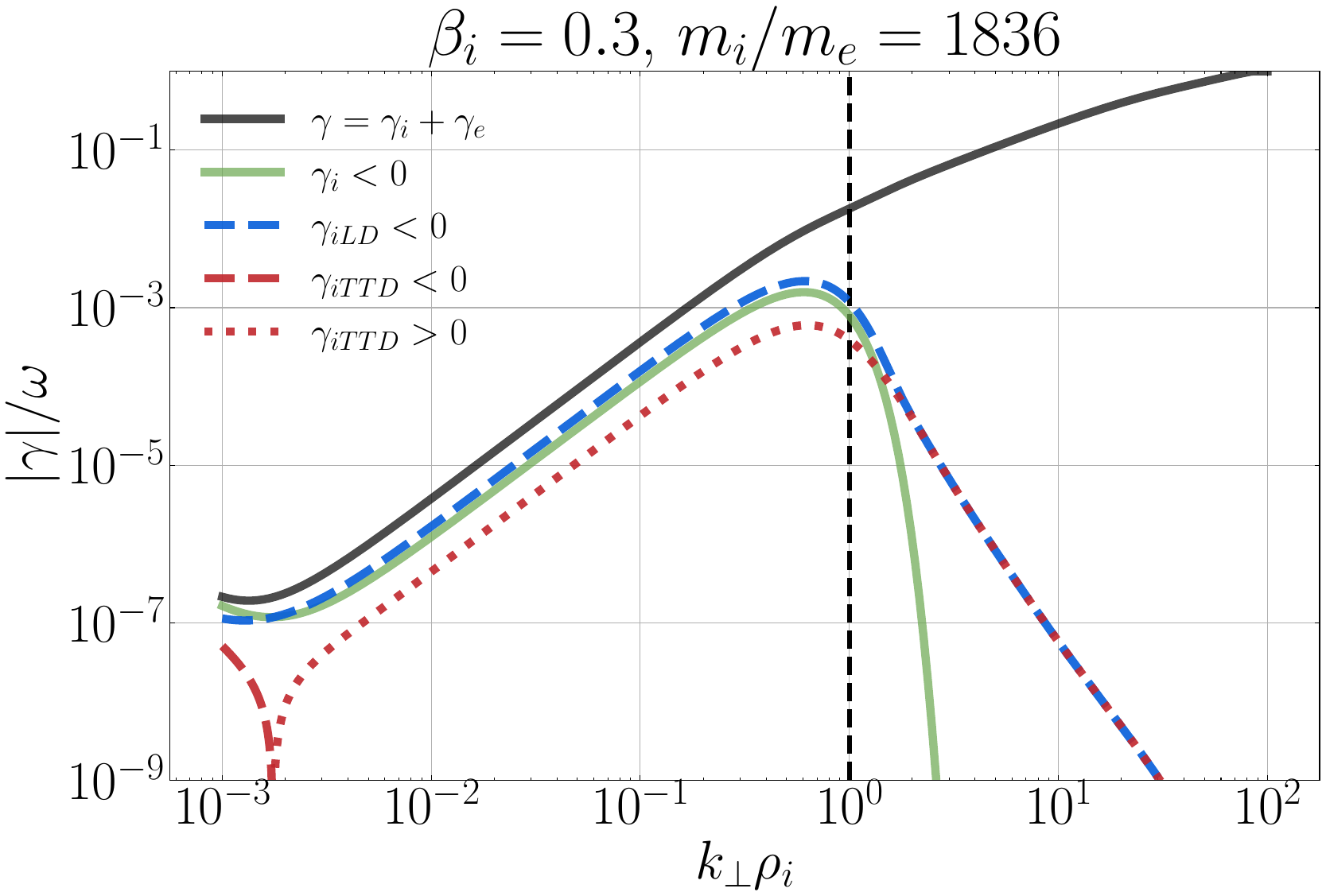} \\
\includegraphics[width=0.65 \textwidth]{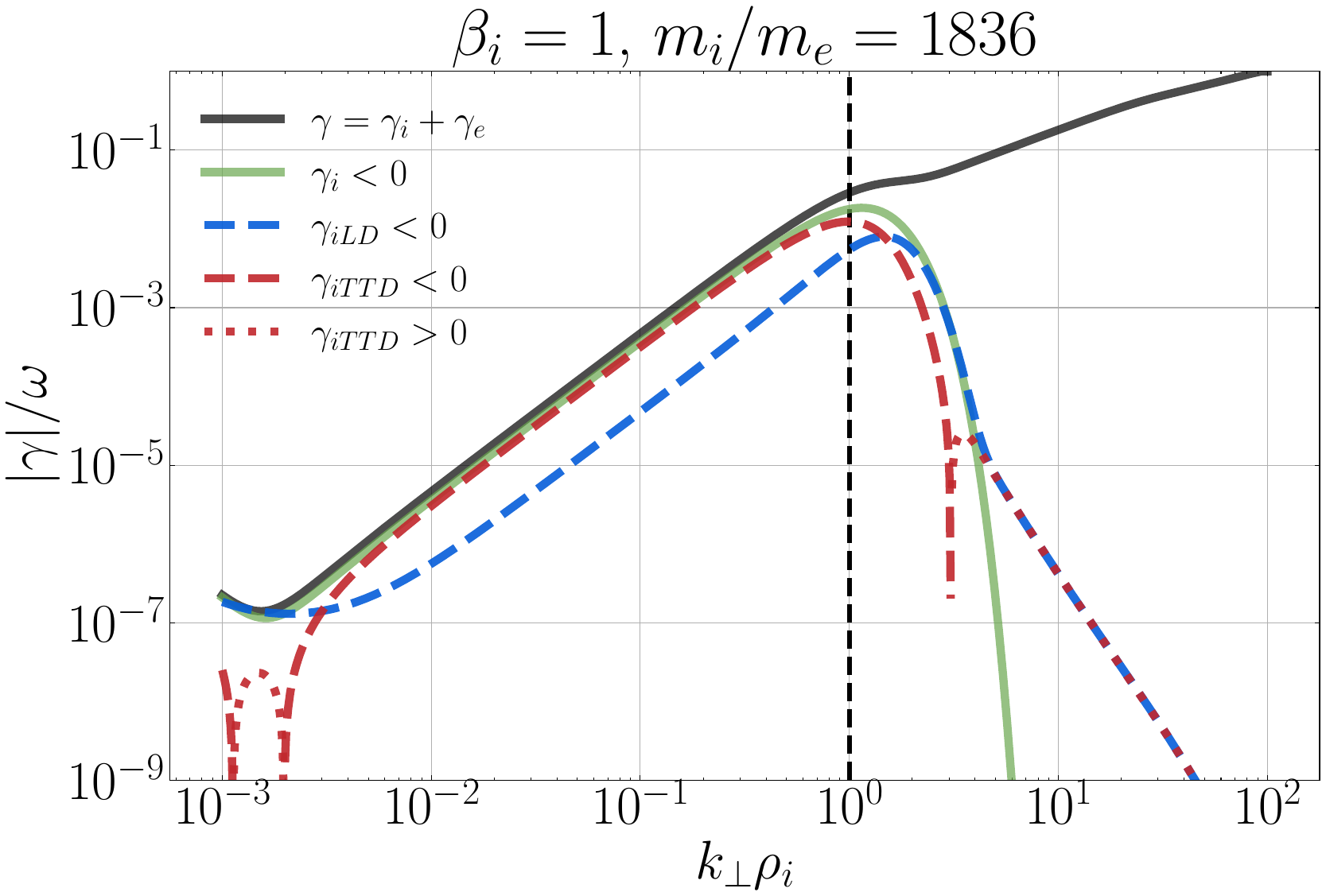}\\
\includegraphics[width=0.65 \textwidth]{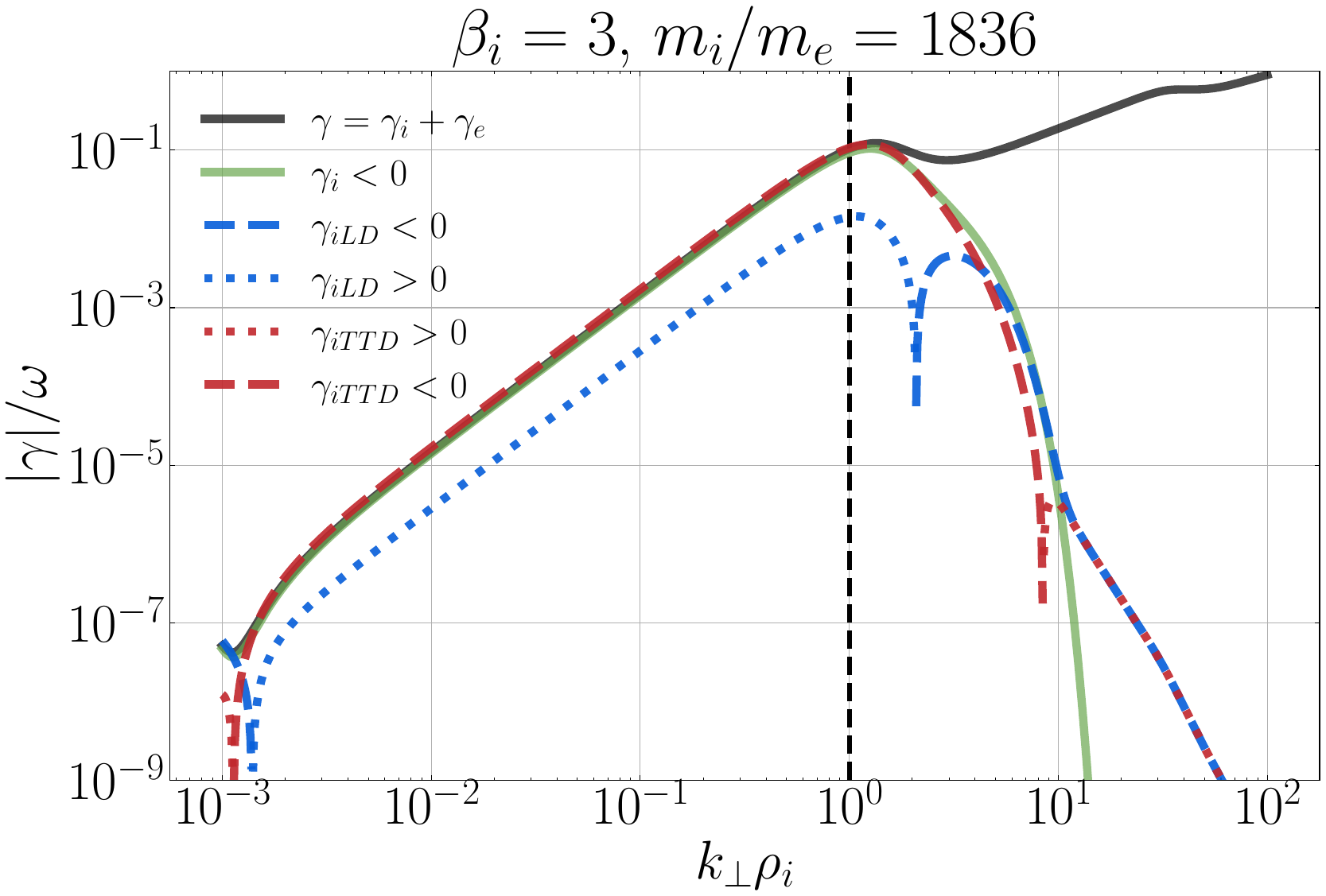}\\
\end{center}
\vskip -7.0in \hspace*{0.8in}(a)\\
\vskip 2.1in  \hspace*{0.8in}(b)\\
\vskip 2.1in  \hspace*{0.8in}(c) \\
\vskip 2.in 
   \caption{Linear dispersion relations for KAWs from PLUME calculations with the realistic mass ratio $m_i/m_e = 1836$, showing the absolute value of the normalized wave growth rate $|\gamma|/\omega$ as a function of the dimensionless perpendicular wave vector $k_\perp \rho_i$ for (a) $\beta_i = 0.3$, (b) $\beta_i = 1$, and (c) $\beta_i = 3$. The vertical black dashed line at $k_\perp \rho_i = 1$ indicates the values used in the single KAW \T{AstroGK} simulations. We plot $\gamma$ (total damping rate, black), $\gamma_i$ (total ion damping rate, green), $\gamma_{iTTD}$ (ion growth or damping rate via the magnetic mirror force, red), and $\gamma_{iLD}$ (ion growth or damping rate via the electrostatic force, blue). Line styles---solid, dashed, and dotted---represent the total damping rates, damping rates separated by mechanism, and growth rates separated by mechanism, respectively.}
    \label{fig:ldr_mr1836}
\end{figure}

How do we reconcile these surprising results with the general expectation of Landau-resonant collisionless damping of waves for a Maxwellian equilibrium velocity distribution?   As it turns out, this behavior is exactly what is predicted by the linear Vlasov-Maxwell dispersion relation.  To be specific, we take the complex eigenfrequency from the linear dispersion relation to be given by $\omega + i \gamma$, so that time evolution of a plane-wave mode is given by $\exp(-i\omega t) \exp (\gamma t)$: positive imaginary components $\gamma>0$ correspond to growth of the wave, and negative imaginary components $\gamma<0$ correspond to damping of the wave.  In \figref{fig:ldr_mr1836}, we plot the normalized absolute value of the imaginary component of the wave frequency $|\gamma|/\omega$ vs. $k_\perp \rho_i$ for KAWs using the \T{PLUME} solver \citep{klein:2015} for a fully ionized proton-electron plasma with isotropic Maxwellian velocity distributions with $T_i/T_e=1$, $m_i/m_e = 1836$, $v_{ti}/c=10^{-4}$, $k_\parallel \rho_i=10^{-3}$  over the range $10^{-3} \le k_\perp \rho_i \le  10^2$ for the ion plasma beta values (a) $\beta_i=0.3$, (b) $\beta_i=1$, and (c) $\beta_i=3$.  We plot separately the total collisionless damping rate (black solid) due to both ions and electrons, the total ion damping rate (green solid), and the separate contributions to the ion damping rate from TTD (red) and LD (blue).  
For the separated TTD and LD contributions, we plot negative imaginary components (which correspond to collisionless damping of the wave) using dashed lines, and positive imaginary components (which correspond to collisionless growth of the wave) using dotted lines.

Although all of the KAW dispersion relations plotted in \figref{fig:ldr_mr1836} yield a net effect of collisionless damping by the ions, over some ranges in  $k_\perp \rho_i$ (dotted lines) either TTD or LD individually may lead to a net transfer of energy \emph{from} the ions \emph{to} the waves over the course of a single wave period.  For example, for the $\beta_i=3$ case in \figref{fig:ldr_mr1836}(c), the imaginary component due to LD is positive over $1.4 \times 10^{-3} \lesssim k_\perp \rho_i \lesssim 2.1$, corresponding to a transfer of energy from the ions to the wave, and is negative outside of that range, corresponding to a transfer of energy from the wave to the ions. A transfer from ions to the wave would lead to a growth of the wave, but the sum of both of the LD and TTD contributions is always negative for these cases with a Maxwellian velocity distribution, leading to a net collisionless damping of the wave.  

The perpendicular wave number $k_\perp \rho_i=1$ of our simulated single waves is indicated in \figref{fig:ldr_mr1836} by the vertical black dashed line, and with these plots we can understand the results presented in \figref{fig:ldr_mr1836}.  For the $\beta_i=0.3$ case, at $k_\perp \rho_i=1$ we have $\gamma_{iLD} <0 $ (blue dashed) and $\gamma_{iTTD} >0 $ (red dotted), suggesting that ions gain energy due to $E_\parallel$ but lose energy due to the mirror force when averaged over the full wave period.  This finding agrees with the net gain of ion energy by LD in   \figref{fig:single_wave_varying_beta}(b) and the net loss of ion energy by TTD in (a).    Similarly, for the  $\beta_i=3$ case, at $k_\perp \rho_i=1$ we have $\gamma_{iTTD} <0 $ (red dashed) and $\gamma_{iLD} >0 $ (blue dotted), suggesting that ions gain energy due to the mirror force but lose energy due to $E_\parallel$ when averaged over the full wave period.  Again, this finding from the linear Vlasov-Maxwell dispersion relation agrees with the net gain of ion energy by TTD in   \figref{fig:single_wave_varying_beta}(c) and the net loss of ion energy by LD in (d).  

To further understand the physical meaning of $\gamma_{iTTD} >0$ in the $\beta_i=0.3$ case, we first point out that the only mechanism that can change the particle energy is work done by the electric field, and the rate of change of the ion energy density $W_i$ is given by $\V{j}_i \cdot \V{E}$.  For TTD, this energization arises from the component of the electric field that is perpendicular to the magnetic field \citep{Howes:2024} (whereas LD energizes particles through the parallel component of the electric field), so here we consider the perpendicular contribution to the energization ${j}_{\perp i} {E}_\perp$. For a single plane wave, the net transfer of energy to or from the ions by ${j}_{\perp i} {E}_\perp$ over a single wave period depends on the phase of the perpendicular electric field fluctuation ${E}_\perp$ relative to the phase of the self-consistent perpendicular component of the ion current density associated with the wave, ${j}_{\perp i}$.  If the phase difference $\delta  \phi$ is such that there is an in-phase component $ 0 < \delta  \phi < \pi /2$, there will be a net energization of the ions; if there is an out-of-phase component $ \pi/2 < \delta  \phi < \pi$, the ions will lose energy.
The eigenfunctions arising from solutions of the linear Vlasov-Maxwell dispersion relation dictate the phases of the components of the electric field and current density. For the $\beta_i=0.3$ case, this eigenfunction dictates that 
${j}_{\parallel i}$ and ${E}_\parallel$ are in-phase, leading to ion energization and wave damping by ${E}_\parallel$ (yielding LD), but ${j}_{\perp i}$ and $ {E}_\perp$ are out-of-phase, so the magnetic mirror force partly counteracts the damping of the wave.

\section{Turbulence Simulations \label{sec:turb}}
Now that we have determined the gyrotropic velocity-space signature of TTD for single KAWs, with the fiducial example for $\beta_i=1$ shown in \figref{fig:fiducial}(a), we will seek similar signatures of TTD in simulations of strong plasma turbulence.

\subsection{Turbulence Simulation Set Up \label{sec:turb_sim}}

We perform kinetic simulations of strong plasma turbulence using the Astrophysical Gyrokinetics Code \T{AstroGK} \citep{numata:2010} for three values of the ion plasma beta $\beta_i=0.3,1,3$.   Each simulation has numerical resolution $(n_x, n_y, n_z, n_{\lambda}, n_E, n_s) = (96, 96, 32, 64, 32, 2)$ within a simulation domain $L_\perp^2 \times L_\parallel = (8 \pi \rho_i)^2 \times (2\pi a_0)$, where the elongation of the domain along the equilibrium magnetic field $\V{B}_0=B_0 \zhat$ is characterized by the arbitrary 
gyrokinetic expansion parameter $\epsilon \sim \rho_i/a_0 \ll 1$.  The proton-to-electron temperature ratio of the Maxwellian equilibrium is $T_i/T_e=1$, and we choose a reduced mass ratio  $m_i/m_e = 36$ to ensure that we fully resolve the kinetic damping mechanisms needed to achieve a steady-state turbulent cascade in a driven simulation, as discussed in \citet{Howes:2018a}. These parameters lead to a fully resolved range of perpendicular wavenumbers  $0.25 \le k_\perp \rho_i \le 7.75$, or $0.042\le k_\perp \rho_e \le 1.29$. For the $\beta_i = 0.3$ and $\beta_i = 1$ simulations, the proton and electron collisionalities are set to $\nu_s/(k_{\parallel 0} v_{ti})=0.1$, and for the $\beta_i = 3$ simulation, $\nu_s/(k_{\parallel 0} v_{ti}) = 0.05$. These collisionalities ensure weakly collisional plasma conditions, yet prevent the small-scale variations that develop in velocity space from becoming unresolved on the velocity grid.

Turbulence in the simulations is driven from zero initial conditions using an oscillating Langevin antenna \citep{TenBarge:2014a} with characteristic frequency $\omega_0/(k_{\parallel 0} v_A) = 0.9$ and decorrelation rate  $\sigma_0/(k_{\parallel 0} v_A) = -0.3$ to drive four \Alfven wave modes at the domain scale with wave vectors $(k_x \rho_i, k_y\rho_i, k_z a_0) = (0.25, 0, \pm 1) $ and $(0, 0.25, \pm 1) $, generating four perpendicularly polarized \Alfven waves propagating in both directions along the equilibrium magnetic field.  The amplitude of the driving is chosen to satisfy critical balance with a nonlinearity parameter $\chi \equiv k_\perp \delta B_\perp/(k_\parallel B_0) \simeq 1$  \citep{goldreich:1995,Howes:2008b} at the driving scale $k_\perp \rho_i=0.25$. This driving has been shown to generate effectively a strong plasma turbulent cascade to small scales in previous kinetic simulations through nonlinear interactions between the counterpropagating \Alfven waves \citep{Howes:2008a, Howes:2011a, Howes:2013a, TenBarge:2013a, TenBarge:2013b, Howes:2018a, Verniero:2018a, Verniero:2018b, Horvath:2020, Conley:2023}.
To provide the data needed to apply the field-particle correlation analysis, the electromagnetic fluctuations and proton velocity distributions are sampled at a high cadence at twenty-four probe points that are distributed throughout the domain, sixteen in the $xy$-plane at $z=0$, and the remaining eight along the $z$-axis, as illustrated in Fig.~2 of \citet{Horvath:2020}.  

\begin{figure}
 \begin{center}
\includegraphics[width=0.65\textwidth]{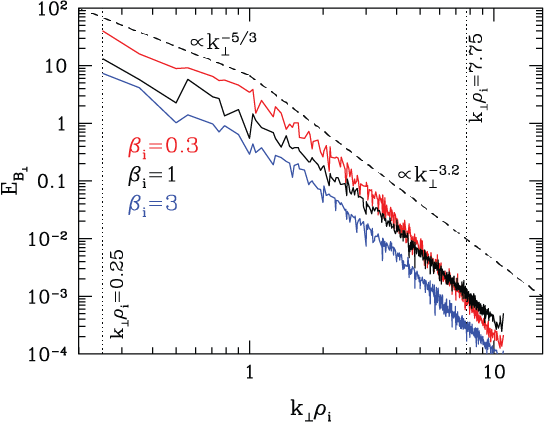} \\
\end{center}
   \caption{Perpendicular magnetic energy spectra at the end of each of the turbulence simulations, showing $\beta_i=0.3$ (red),  $\beta_i=1$ (black), and $\beta_i=3$ (blue). Vertical dotted lines indicate the limit of fully resolved perpendicular wavenumbers in the simulation,  $0.25 \le k_\perp \rho_i \le 7.75$, or $0.042\le k_\perp \rho_e \le 1.29$.}
    \label{fig:kspec}
\end{figure}

The timescale associated with outer scale of the turbulent cascade in each simulation is the wave period of the domain-scale \Alfven wave $T$, and the simulations are run for $6.78 T$ for the $\beta_i=0.3$ simulation,  $5.51 T$ for the $\beta_i=1$ simulation, and $3.61 T$ for the $\beta_i=3$ simulation.  The perpendicular magnetic energy spectrum $E_{B_\perp}(k_\perp)$ at the end of each of the simulations is shown in 
\figref{fig:kspec}, showing the spectrum for  $\beta_i=0.3$ (red),  $\beta_i=1$ (black), and $\beta_i=3$ (blue).  These spectra demonstrate that each simulation yields a broadband turbulent spectrum, with the spectral slope for each simulation consistent with the expectation of $-5/3$ for strong plasma turbulence \citep{goldreich:1995} in the inertial range at $k_\perp \rho_i<1$, and steepening of the spectrum at the transition to the dissipation range at $k_\perp \rho_i\sim 1$.  In the dissipation range at  $k_\perp \rho_i> 1$, the spectral slopes begin around $-3.2$, steepening as  $k_\perp \rho_e \rightarrow 1$ due to the resolved kinetic dissipation mechanisms that remove energy from the turbulent cascade.  Note that these dissipation range slopes are slightly steeper than the values ranging from $-2.7$ to $-3.1$ typically observed in the solar wind \citep{Sahraoui:2013b}, but this is to be expected due to the unphysical mass ratio of $m_i/m_e = 36$, which effectively enhances the damping rate due to electrons relative to the realistic mass ratio case \citep{TenBarge:2013b}, leading to slightly steeper dissipation range spectra for stronger damping \citep{Howes:2011b}.

\begin{figure}
 \begin{center}
   \includegraphics[width=0.65\textwidth]{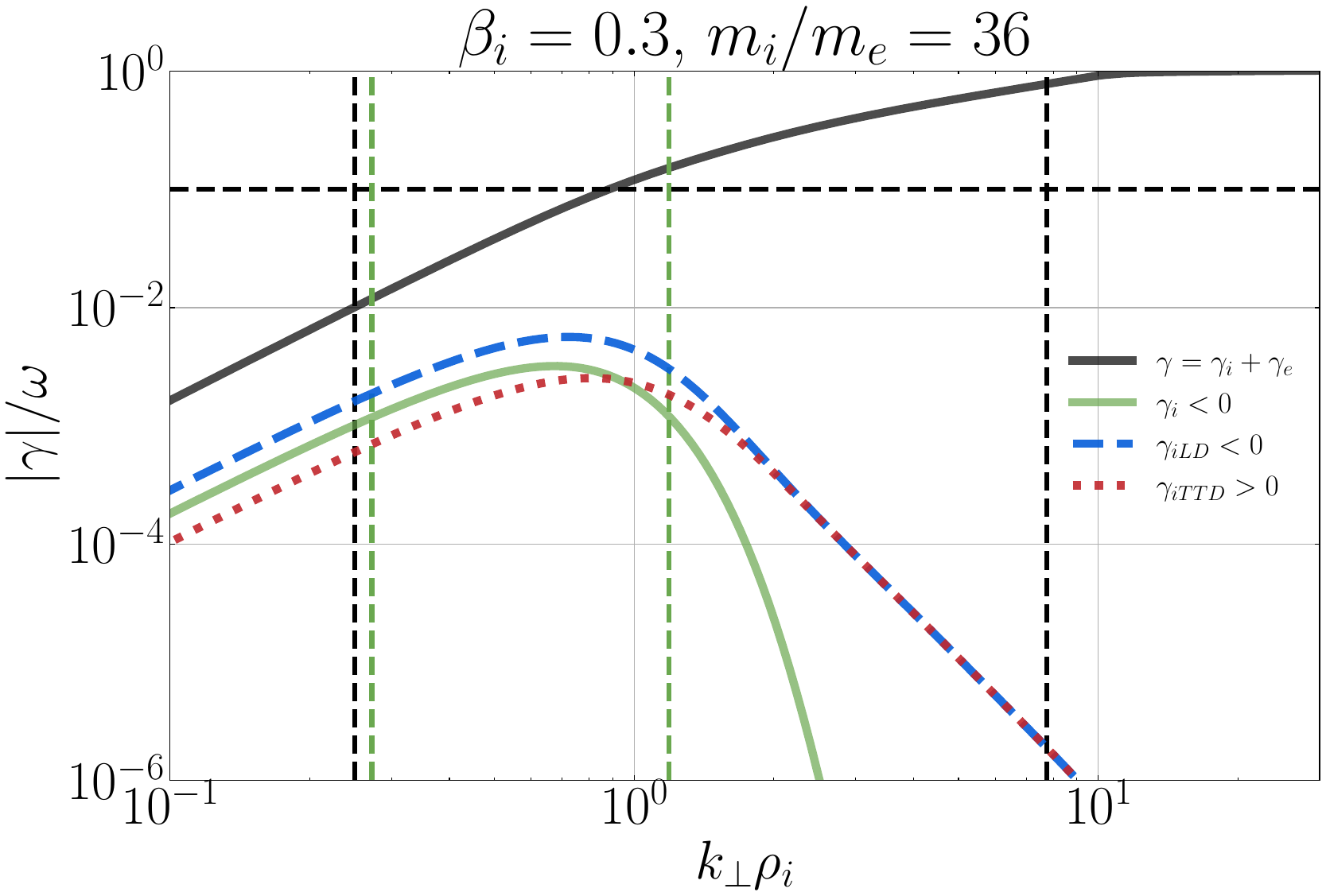}\\
      \includegraphics[width=0.65 \textwidth]{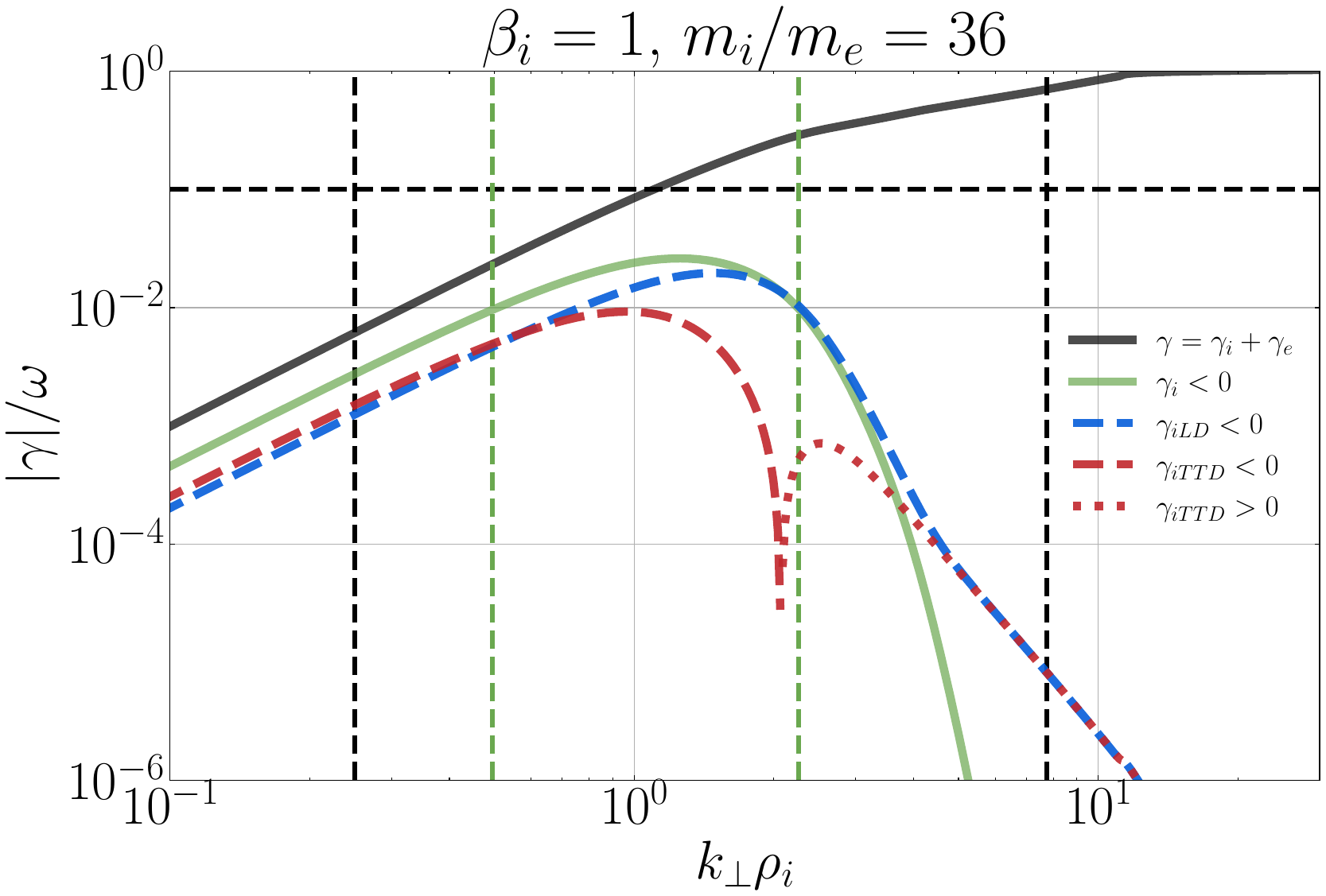}\\
      \includegraphics[width=0.65 \textwidth]{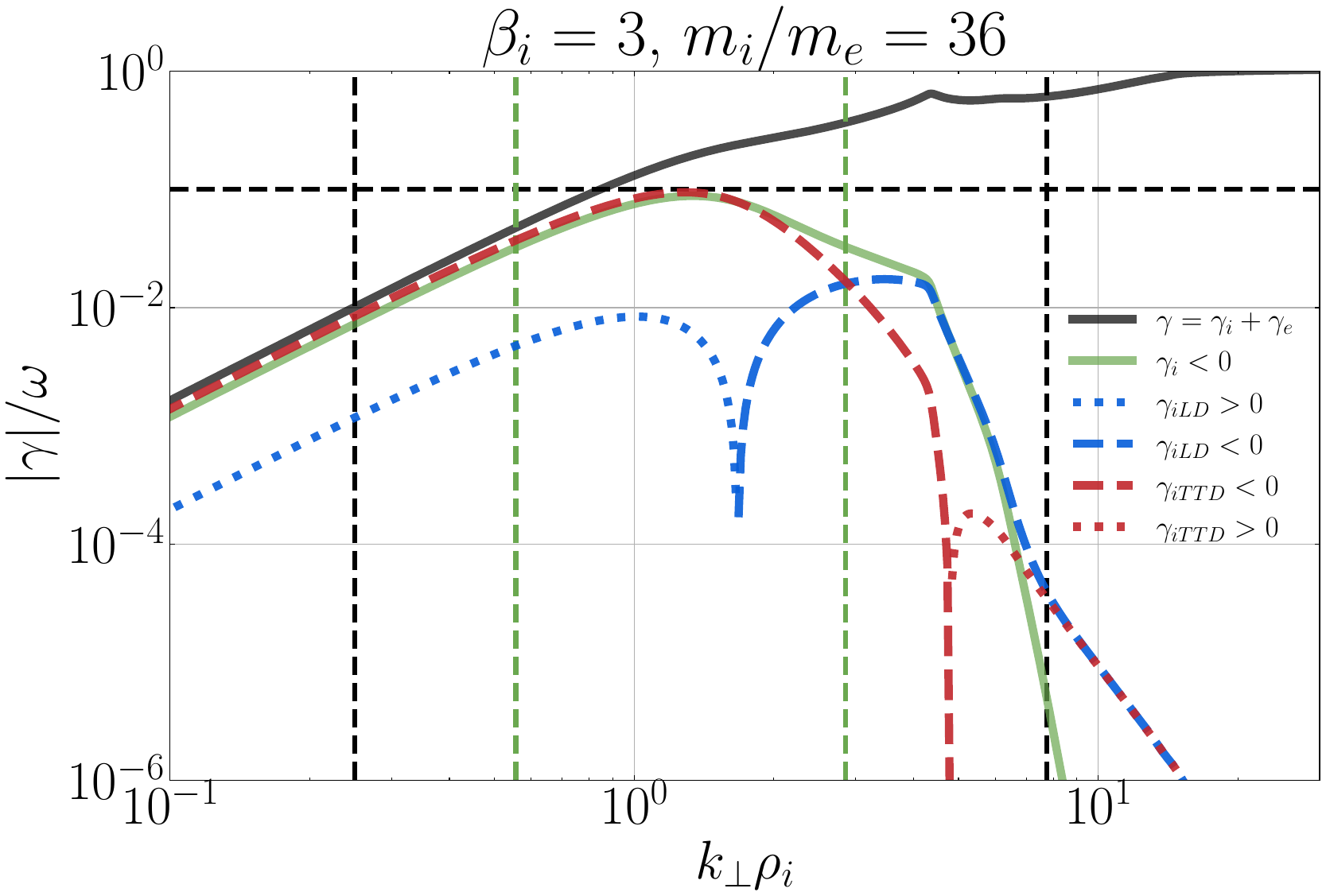}\\
\end{center}
\vskip -7.0in
\hspace*{1.0in}(a)\\
\vskip 2.0in 
\hspace*{1.0in}(b)\\
\vskip 2.0in 
\hspace*{1.0in}(c)\\
\vskip 2.1in 
   \caption{Linear dispersion relations for KAWs from PLUME calculations with the reduced mass ratio $m_i/m_e = 36$, showing the absolute value of the normalized wave damping or growth rate $|\gamma|/\omega$ as a function of the dimensionless perpendicular wavenumber $k_\perp \rho_i$ for (a) $\beta_i = 0.3$, (b) $\beta_i = 1$, and (c) $\beta_i = 3$. The two vertical black dashed lines at $k_\perp \rho_i = 0.25$ and $7.75$ label the range consistent with the \T{AstroGK} turbulence simulations, and the two vertical green dashed lines mark the range of $1/e$ of the peak value of $\gamma_i$. The horizontal dashed black line at $|\gamma|/\omega = 10^{-1}$ indicates the threshold above which significant damping or growth occurs. Each panel follows the layout format of \figref{fig:ldr_mr1836}.}
    \label{fig:ldr_mr36}
\end{figure}

In \figref{fig:ldr_mr36}, we plot the normalized damping rates $|\gamma|/\omega$ from the linear dispersion relation for the simulation parameters with the reduced mass ratio $m_i/m_e=36$, presenting the results for (a) $\beta_i=0.3$, (b) $\beta_i=1$, and (c) $\beta_i=3$, in the same format as presented in \figref{fig:ldr_mr1836}.  Here we plot vertical black dashed lines at the perpendicular wavenumber limits of the simulation at $k_\perp \rho_i=0.25$ and $k_\perp \rho_i=7.75$.  Note the salient features that TTD yields a loss of ion energy for  $\beta_i=0.3$ in (a), and LD yields a loss of ion energy for  $\beta_i=3$ at $k_\perp \rho_i \lesssim 1.5$.  These calculations of the linear wave properties and the effective ion energization rates by TTD and LD provide an important theoretical framework for the interpretation of our field-particle analysis results.

\subsection{Choosing the Correlation Interval $\tau$ \label{sec:turb_tau}}

\begin{figure}
    \begin{center}
     \includegraphics[width=0.45\textwidth]{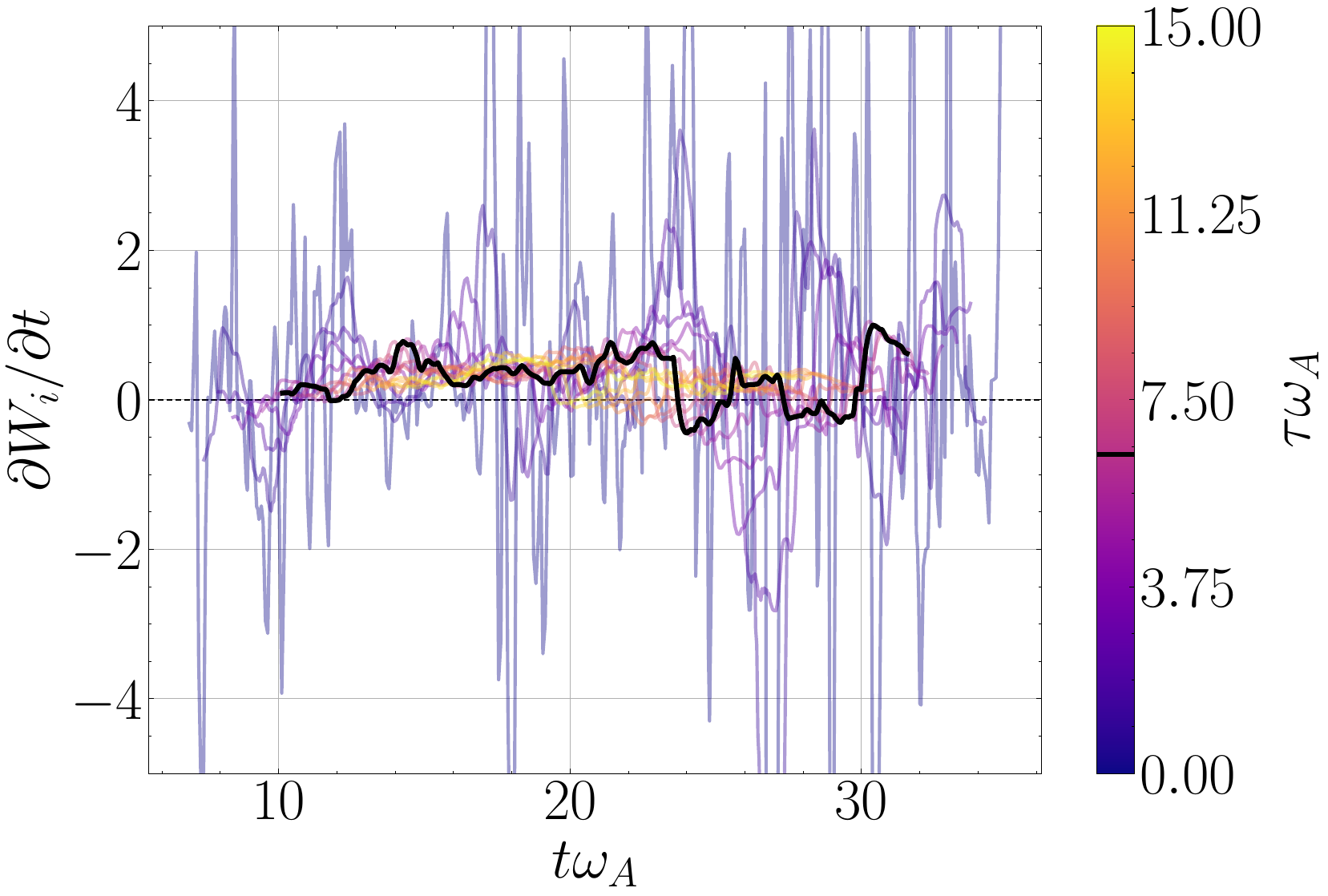}
      \includegraphics[width=0.45\textwidth]{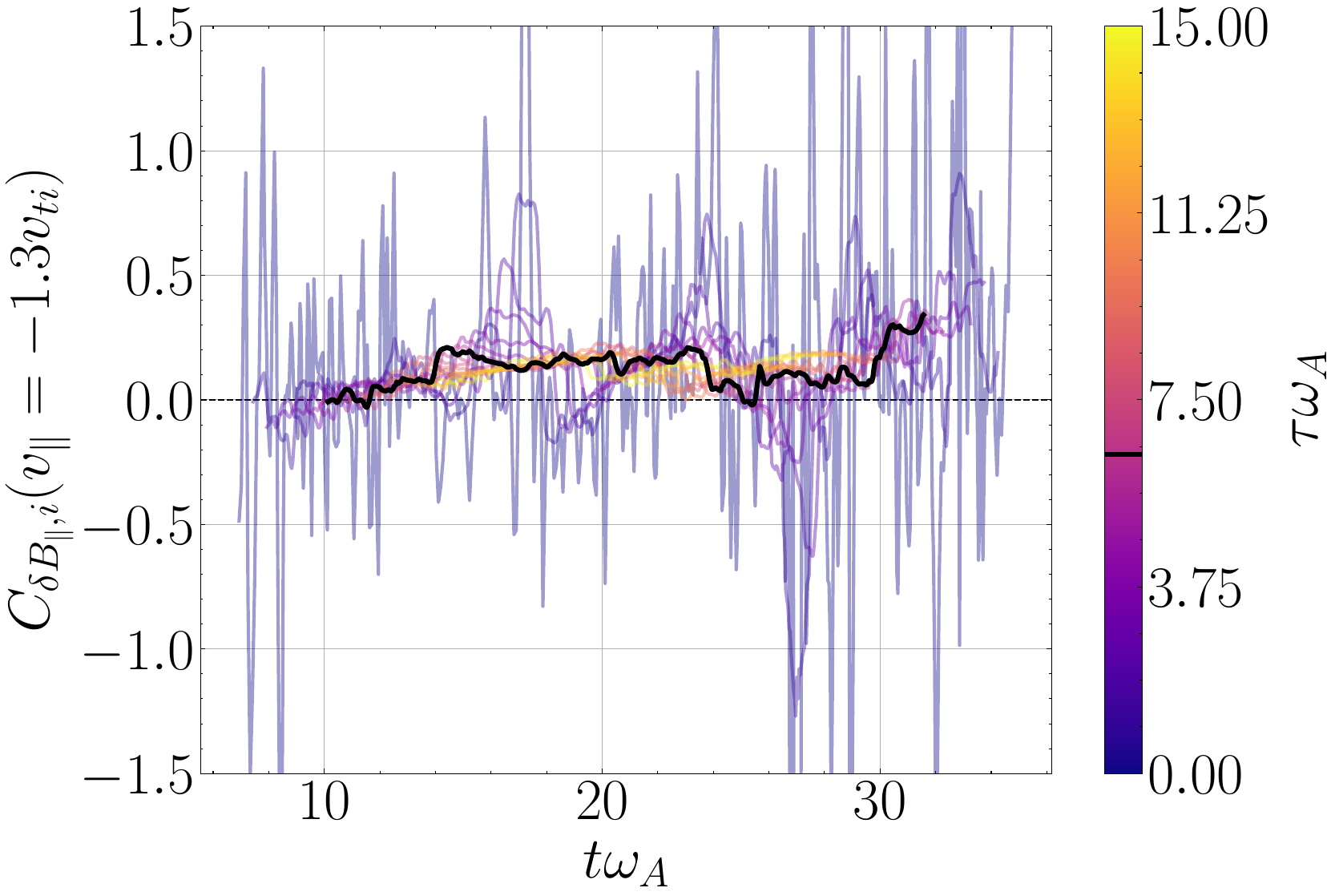}
   \end{center}
 \vskip -1.7in
\hspace*{0.15in} (a)\hspace*{2.3in} (b)
\vskip +1.8in
\begin{center}
        \includegraphics[width=0.45 \textwidth]{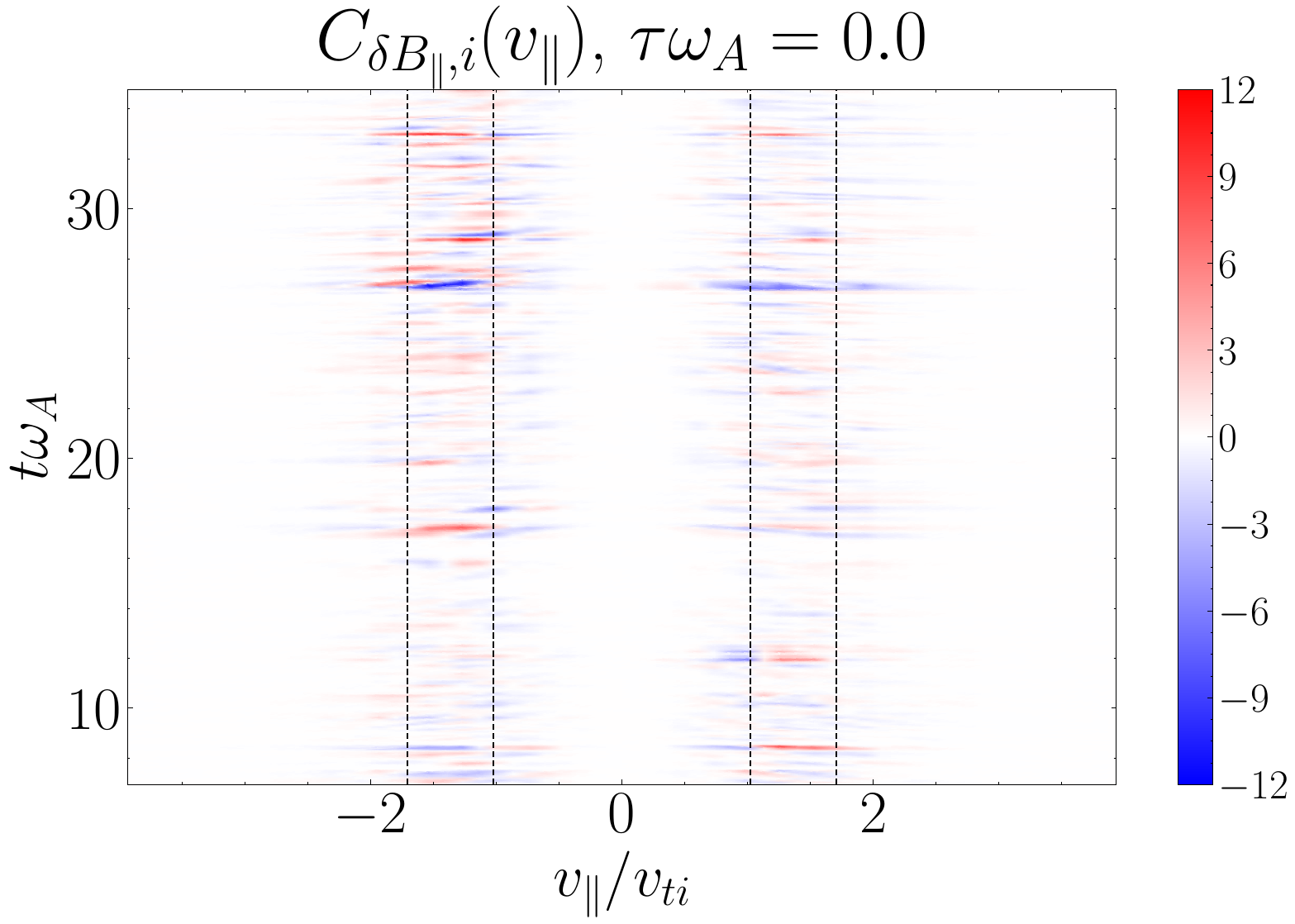}
        \includegraphics[width=0.45 \textwidth]{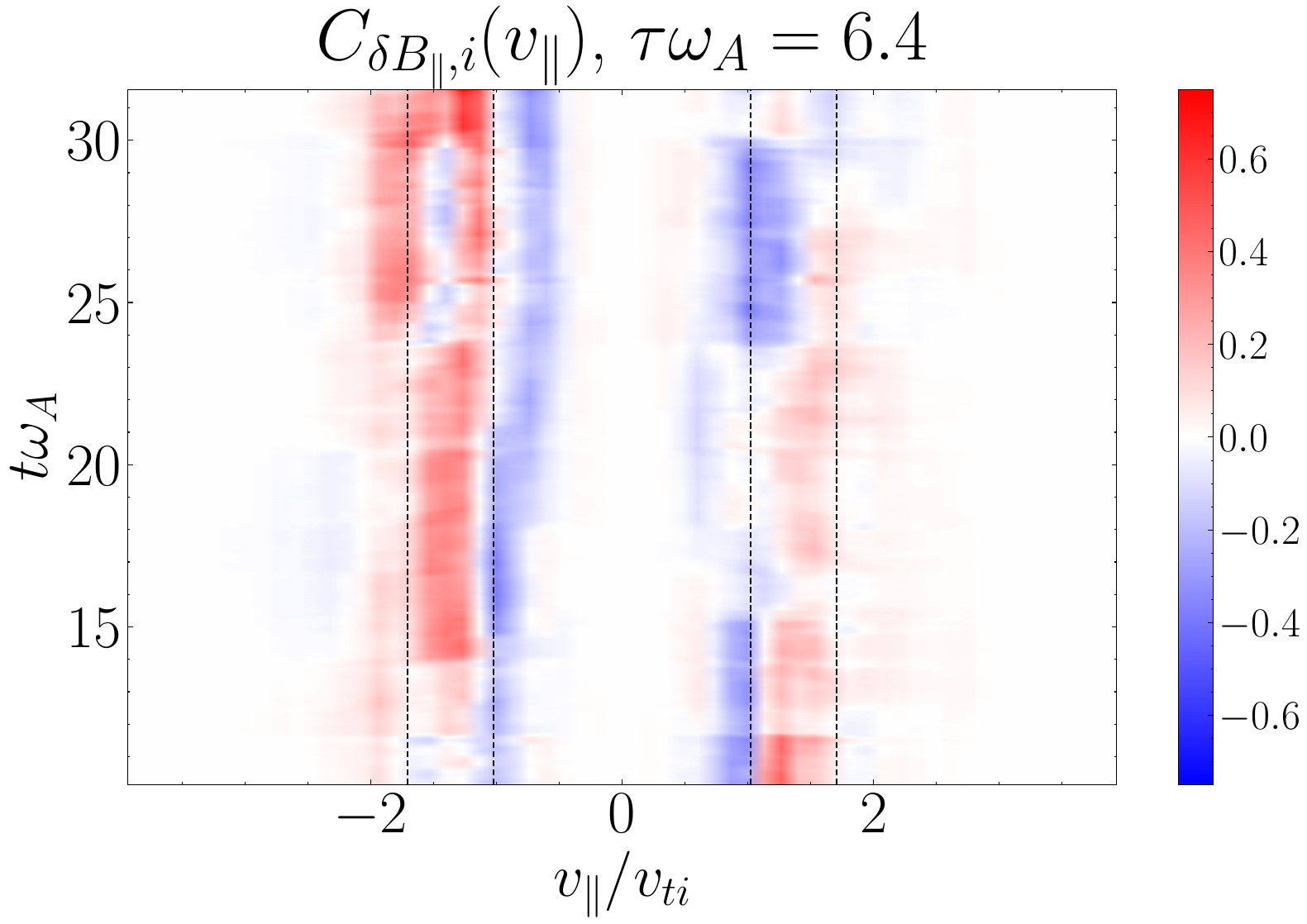}
    \end{center}
    \vskip -1.8in
\hspace*{0.15in} (c)\hspace*{2.3in} (d)
\vskip +1.8in
   \caption{Analysis of correlation interval selection for the $\beta_i = 1$ \T{AstroGK} turbulence simulation at Probe 13 $(\pi \rho_i, 7 \pi \rho_i, 0)$. Top row: time evolution of (a) the rate of change of ion kinetic energy density due to TTD, denoted as $\partial W_i /\partial t$, and (b) the reduced correlation $C_{\delta B_{\parallel, i}} (v_\parallel, t)$ at $v_\parallel = -1.3 v_{ti}$. Both quantities are presented over a range of $\tau \omega_A$ values from 0 to 15. The selected $\tau \omega_A$ value of 6.4 is marked with a black line. Bottom row: timestack plots of the reduced correlation $C_{\delta B_{\parallel, i}} (v_\parallel, t)$ for (c) $\tau \omega_A = 0$ and (d) $\tau \omega_A = 6.4$, where the range of parallel phase velocities of KAWs that experience significant damping by ions is indicated by vertical dashed lines at $v_\parallel / v_\text{ti} = \pm 1.020$ and $v_\parallel / v_\text{ti} = \pm 1.704$.}
   \label{fig:choice_of_tau_turb}
\end{figure}

Unlike in the single-KAW simulations, where the single wave period $T$ is the obvious choice for the correlation interval $\tau$ to eliminate the oscillatory contribution to the transfer of energy from fields to particles, choosing $\tau$ for a plasma supporting broadband turbulent fluctuations is less straightforward.  The longest wave period for an \Alfven wave at the domain scale in our $\beta_i=1$ simulation is $\tau \omega_A \simeq 6.28$.  In \figref{fig:choice_of_tau_turb}, for a single probe position in the $\beta_i=1$ simulation,  we present (a) the total energy transfer rate to ions due to TTD $(\partial W_i/\partial t)_{\mbox{TTD}}$ and (b) the energy transfer rate at $v_\parallel/v_{ti}=-1.3$, for a correlation interval spanning $0 \le \tau \omega_A \le 15$. The instantaneous values of the local energy transfer rates (with $\tau=0$, blue) exhibit large fluctuations with both positive and negative signs, but for $\tau \omega_A = 6.4$ (black) and longer correlation intervals, those large fluctuations are averaged out, leading to a time-averaged energy transfer rate that is about an order-of-magnitude smaller in amplitude than the peaks of the instantaneous value.  In \figref{fig:choice_of_tau_turb}, we also show timestack plots $C_{\delta B_\parallel, i}(v_\parallel, t; \tau)$ for (c) $\tau=0$ and (d) $\tau\omega_A =6.4$, showing that a relatively persistent bipolar signature of TTD is revealed at  $v_\parallel/v_{ti}=-1.1$ in the $\tau\omega_A =6.4$ case with peak amplitudes about an order-of-magnitude smaller than the instantaneous case with  $\tau=0$. Thus, we choose a correlation interval $\tau \omega_A = 6.4$ to perform the field-particle correlation analysis of our  $\beta_i=1$ turbulence simulation.

\subsection{Results for $\beta_i=1$ Simulation \label{sec:turb_ttd_sig}}

 \begin{figure}
    \begin{center}
      \includegraphics[width=0.5\textwidth]{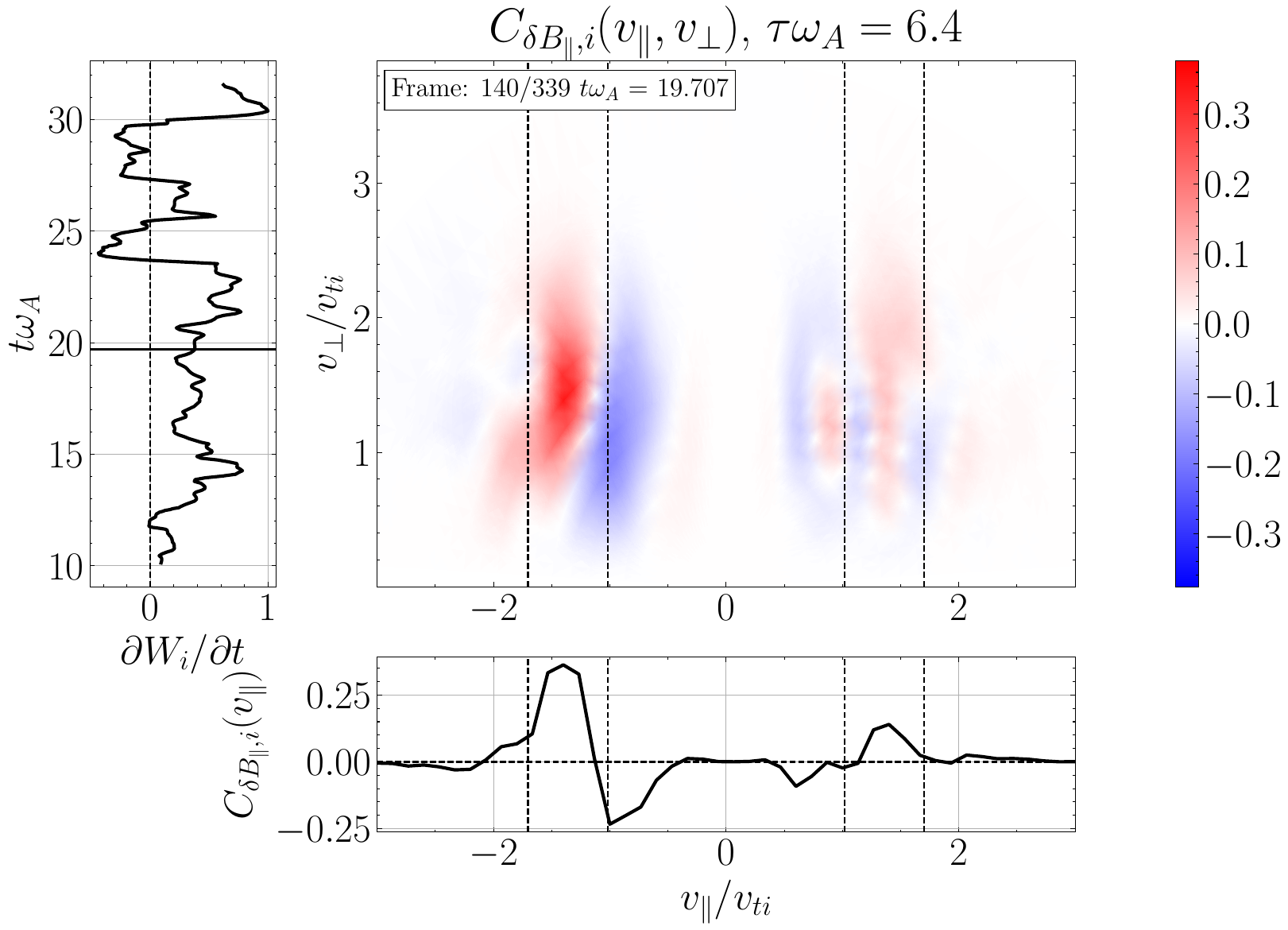}
      \includegraphics[width=0.37\textwidth]{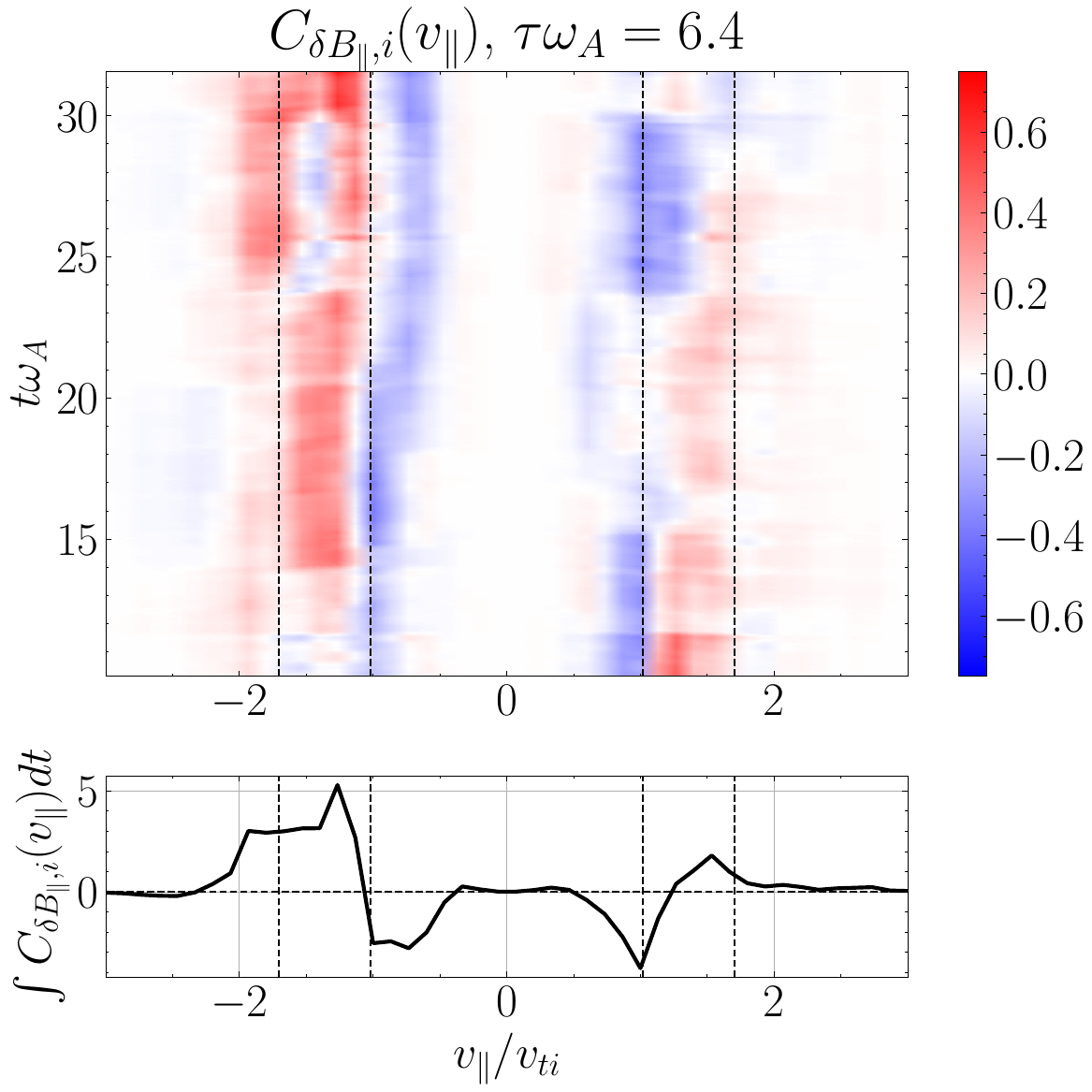}
   \end{center}
 \vskip -2.0in
\hspace*{0.05in} (a)\hspace*{2.7in} (b)
\vskip +2.2in
    \begin{center}
        \includegraphics[width=0.5 \textwidth]{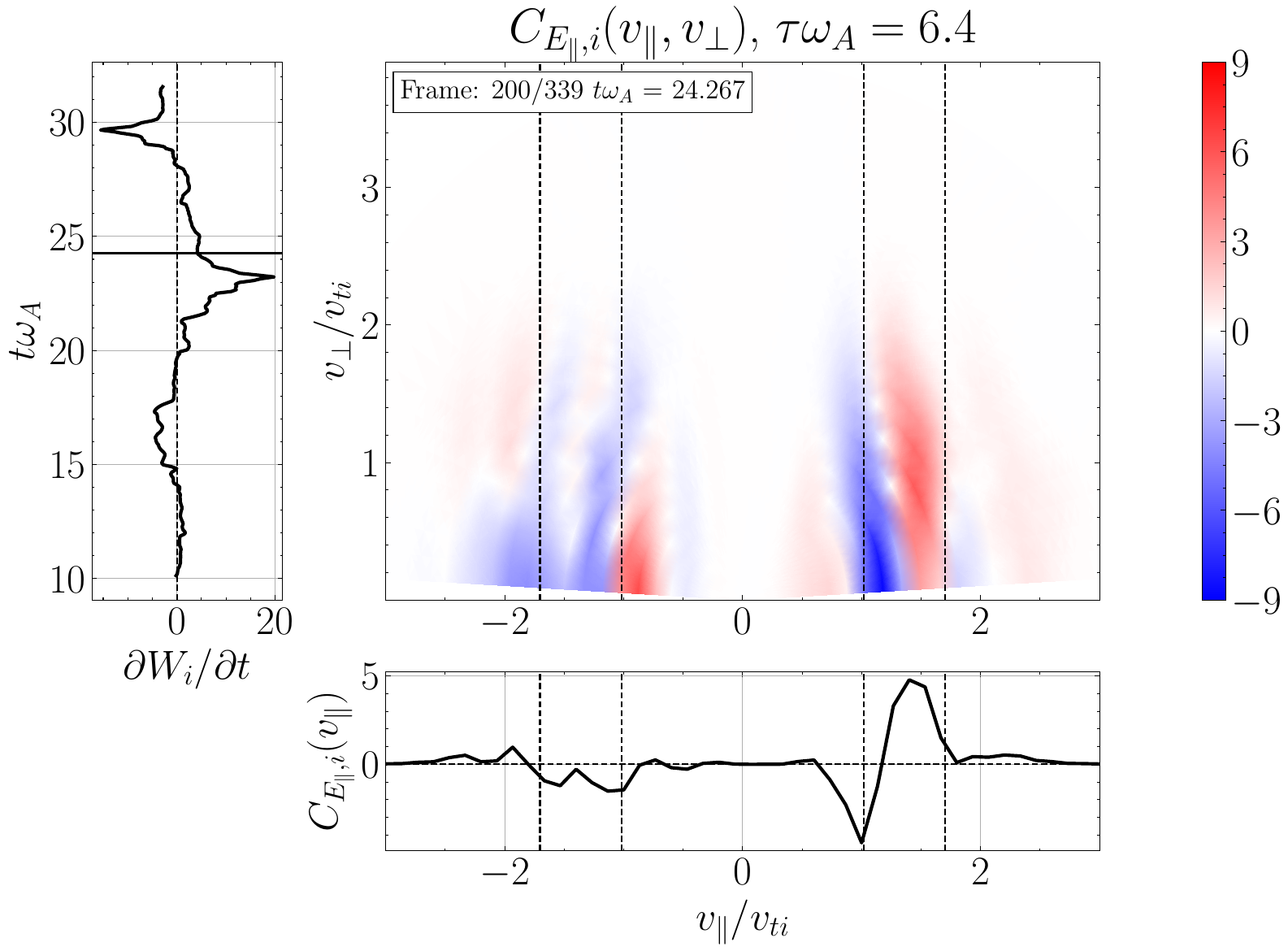}
        \includegraphics[width=0.37 \textwidth]{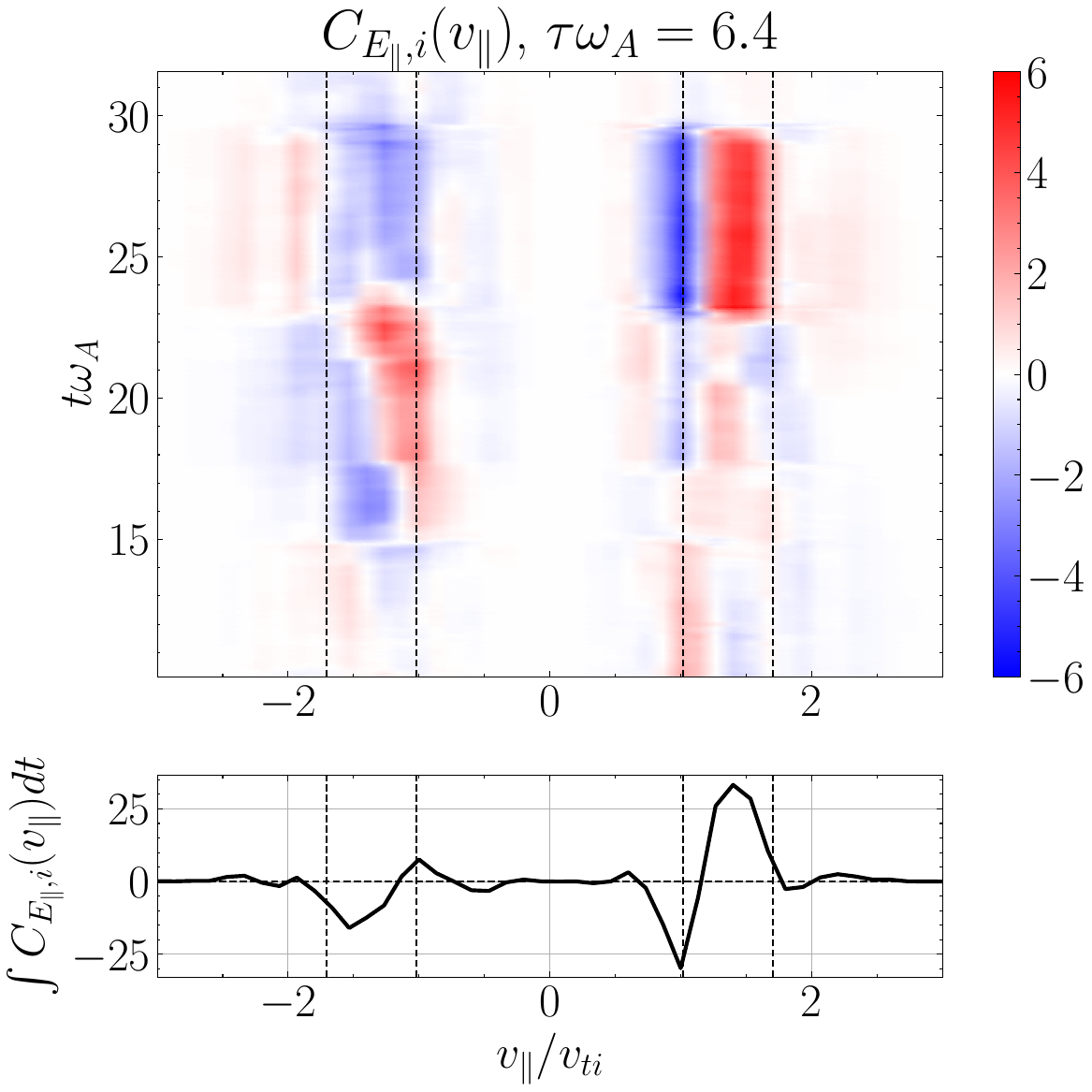}
    \end{center}
    \vskip -2.0in
\hspace*{0.05in} (c)\hspace*{2.7in} (d)
\vskip +2.2in
   \caption{Velocity-space signatures of transit-time damping (TTD, top row) at Probe 13 $(\pi \rho_i, 7 \pi \rho_i, 0)$ and Landau damping (LD, bottom row)  at Probe 24 $(4 \pi \rho_i, 4 \pi \rho_i, 0.875 \pi a_0)$ in \T{AstroGK} turbulence simulation with $0.25 \leq k_\perp \rho_i \leq 7.75$, $T_i / T_e = 1$, and $\beta_i = 1$. The correlation interval is set as $\tau \omega_A = 6.4$. The left column presents gyrotropic plane $(v_\parallel, v_\perp)$ signatures, following the layout format of panels in \figref{fig:fiducial}. The right column features the timestack plots of the $v_\perp-$integrated reduced correlation; the main panel here shows the reduced correlation on $(v_\parallel, t)$ grids, and the lower panel shows the time-integrated reduced correlation. Four vertical dashed lines at $v_\parallel / v_\text{ti} = \pm 1.020$ and $v_\parallel / v_\text{ti} = \pm 1.704$ indicate the resonant parallel phase velocity ranges where significant ion damping occurs.}
    \label{fig:turb_sig_beta_1}
\end{figure}

Performing the field-particle correlation analysis with $\tau \omega_A = 6.4$ for TTD and LD at all 24 probe positions in our $\beta_i = 1$ turbulence simulation, we seek the gyrotropic velocity-space signatures of TTD and LD shown in \figref{fig:fiducial}.  In \figref{fig:turb_sig_beta_1}(a), we plot the gyrotropic velocity-space signature $C_{\delta B_\parallel, i}(v_\parallel,v_\perp;\tau)$ at one of the 24 probes, centered in time at $t\omega_A = 19.71$, showing a clear bipolar signature comparable to that shown in \figref{fig:fiducial}(a).  The range of resonant parallel phase velocities for the kinetic \Alfven wave mode over which significant ion collisionless damping is expected is indicated by the vertical dashed lines; specifically, these lines mark the resonant velocities of the kinetic \Alfven wave mode at $k_\perp \rho_i \simeq 0.5$ and $k_\perp \rho_i \simeq 2.3$, the points at which the total ion collisionless damping rate drops to a factor $1/e$ of its peak value at $k_\perp \rho_i \simeq 1.3$, as illustrated by the vertical green dashed lines on the linear dispersion relation plot for $m_i/m_e=36$ in \figref{fig:ldr_mr36}(b).  In the lower panel is shown the reduced parallel velocity-space signature $C_{\delta B_\parallel, i}(v_\parallel;\tau)$ integrated over $v_\perp$, yielding a clear bipolar signature with a zero crossing at $v_\parallel/v_{ti} \simeq -1.1$.  Note that the zero crossing of this velocity-space signature falls within the expected range of resonant velocities (vertical dashed lines), as expected theoretically for a resonant collisionless damping mechanism.  The left panel shows the net rate of ion energization due to TTD  $(\partial W_i/\partial t)_{\mbox{TTD}}$ as a function of time, showing net positive ion energization due to TTD at this position over almost the full duration of the simulation. Thus, \figref{fig:turb_sig_beta_1}(a) demonstrates that TTD indeed plays a role in the damping of turbulent fluctuations and consequent energization of the ions, a second key result of this paper.

In \figref{fig:turb_sig_beta_1}(b), we show a timestack plot $C_{\delta B_\parallel, i}(v_\parallel,t;\tau)$ of the ion energization by TTD at the same probe position as in (a), showing the persistence in time of the reduced parallel velocity-space signature with the zero crossing at $v_\parallel/v_{ti} \simeq -1.1$.  Note that the zero crossing at $v_\parallel/v_{ti} \simeq -1.1$ in this timestack plot shifts to slightly lower phase velocities near the end of the simulation at $t\omega_A > 25$, likely due to damping associated with KAWs that have slightly lower perpendicular wavenumbers $k_\perp \rho_i$, as can occur in broadband turbulence.  There is also a relatively short-lived bipolar signature observed at $v_\parallel/v_{ti} \simeq 1$ at times $t\omega_A < 15$, indicating that TTD is also acting on KAWs propagating the other direction along the magnetic field \citep{Afshari:2021}.

We perform the analogous field-particle correlation for LD, showing  $C_{E_\parallel, i}(v_\parallel,v_\perp;\tau)$ in \figref{fig:turb_sig_beta_1}(c), yielding a bipolar gyrotropic velocity-space signature at $v_\parallel/v_{ti} \simeq 1.2$ at time $t\omega_A=24.3$, also within the expected range of resonant parallel phase velocities.  This finding of the velocity-space signature of LD confirms previous field-particle correlation analyses showing that ion LD plays a role in the dissipation of plasma turbulence \citep{Klein:2017b, Howes:2018a, Klein:2020, Cerri:2021}.  
The timestack plot  $C_{E_\parallel, i}(v_\parallel,t;\tau)$ in \figref{fig:turb_sig_beta_1}(d) shows this strong bipolar signature of LD at $v_\parallel/v_{ti} \simeq 1.2$ persists over $22 \lesssim t\omega_A \lesssim 30$.  In closing, it is worthwhile noting that, to distinguish the velocity-space signature of TTD from that of LD, it is necessary to examine the signatures in gyrotropic velocity space  $(v_\parallel,v_\perp)$, showing that ion energization is limited to ions with $v_\perp/v_{ti} \gtrsim 1$ for TTD, but that ion energization extends down to $v_\perp \rightarrow  0$ for LD, as expected by the physical arguments outlined in \secref{sec:iTTD}.

\subsection{Variations with Ion Plasma Beta $\beta_i$ \label{sec:turb_ttd_sig_varing_beta}}

Next, we explore how the velocity-space signatures of TTD and LD in plasma turbulence vary with changing the ion plasma beta $\beta_i = v_{ti}^2/v_A^2$. Because $\beta_i$ is a function of the ratio of the ion thermal velocity to the \Alfven velocity, it directly characterizes where the parallel wave phase velocity falls with the ion velocity distribution, making it the most important parameter controlling resonant wave-particle interactions in a weakly collisional plasma.  Specifically, $\omega/(k_\parallel v_{ti})  = \overline {\omega} \beta_i^{-1/2}$, where the parallel phase velocity normalized to the \Alfven velocity $\overline {\omega} \equiv\omega/(k_\parallel v_{A})$ typically has a value  $\overline {\omega} \sim 1$ for the perpendicular wavelengths $k_\perp \rho_i \sim 1$ at which the ions strongly interact with the waves.  Note that, at the perpendicular scale of the domain $k_{\perp 0} \rho_i=0.25$, the \Alfven wave has the $\overline {\omega} \simeq 1$ for all values of $\beta_i$, so we simply choose a correlation interval $\tau \omega_A = 6.4$ for all of the turbulence simulation analysis below.

 \begin{figure}
     \begin{center}
      \includegraphics[width=0.42\textwidth]{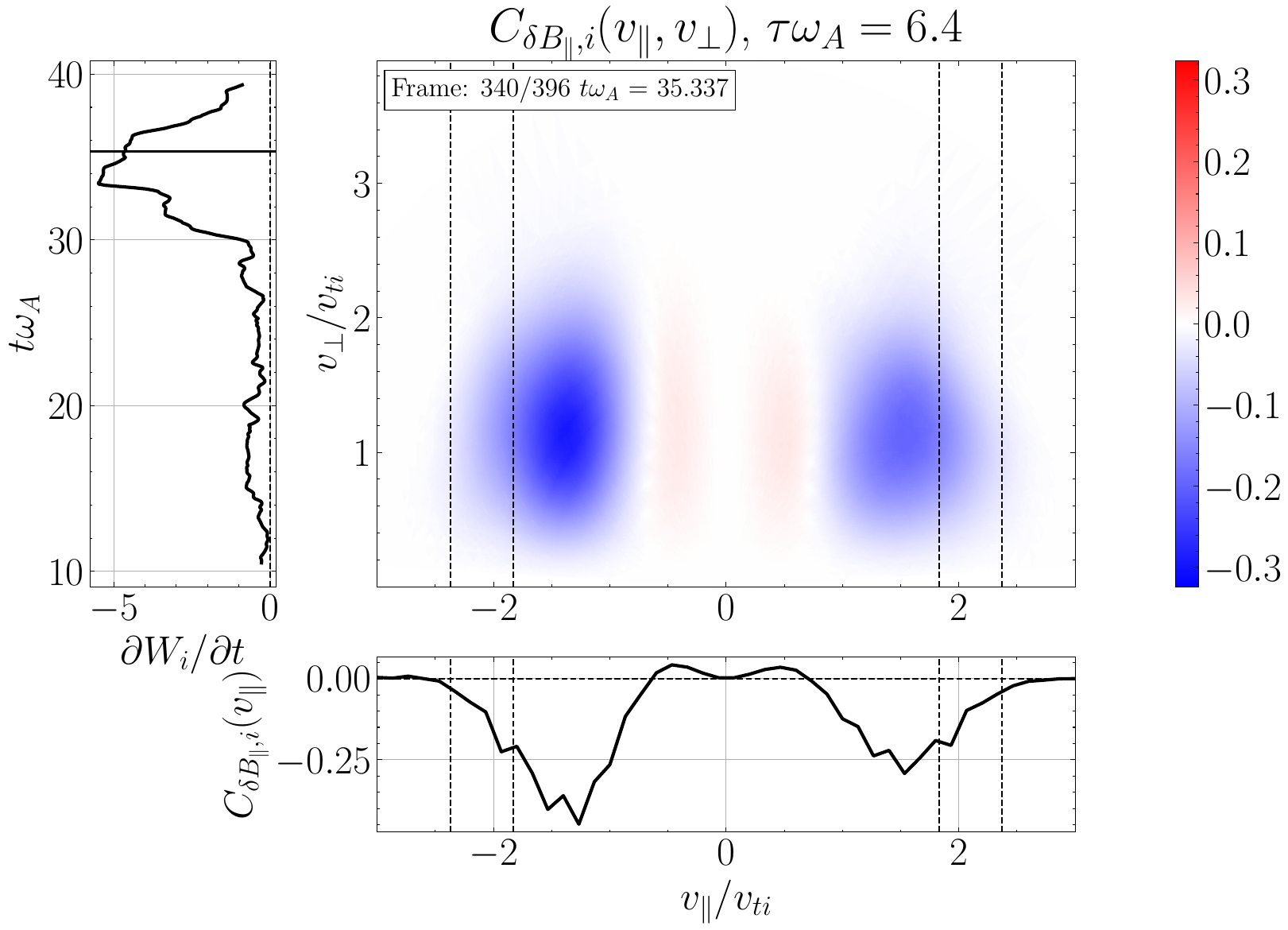}
      \includegraphics[width=0.31\textwidth]{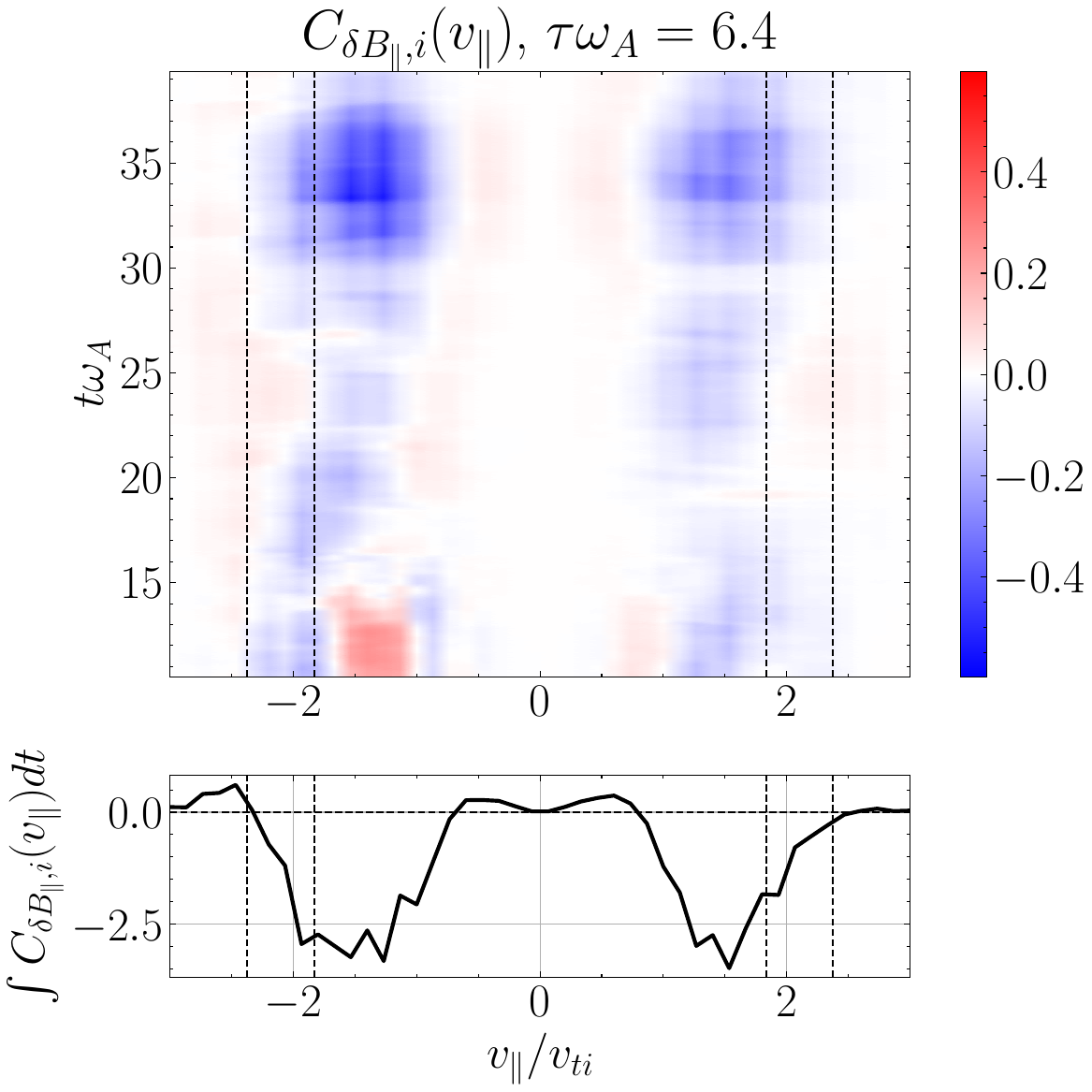}
   \end{center}
 \vskip -1.6in
 \hspace*{0.4in} (a)\hspace*{2.3in} (b)
 \vskip +1.6in
    \begin{center}
      \includegraphics[width=0.42\textwidth]{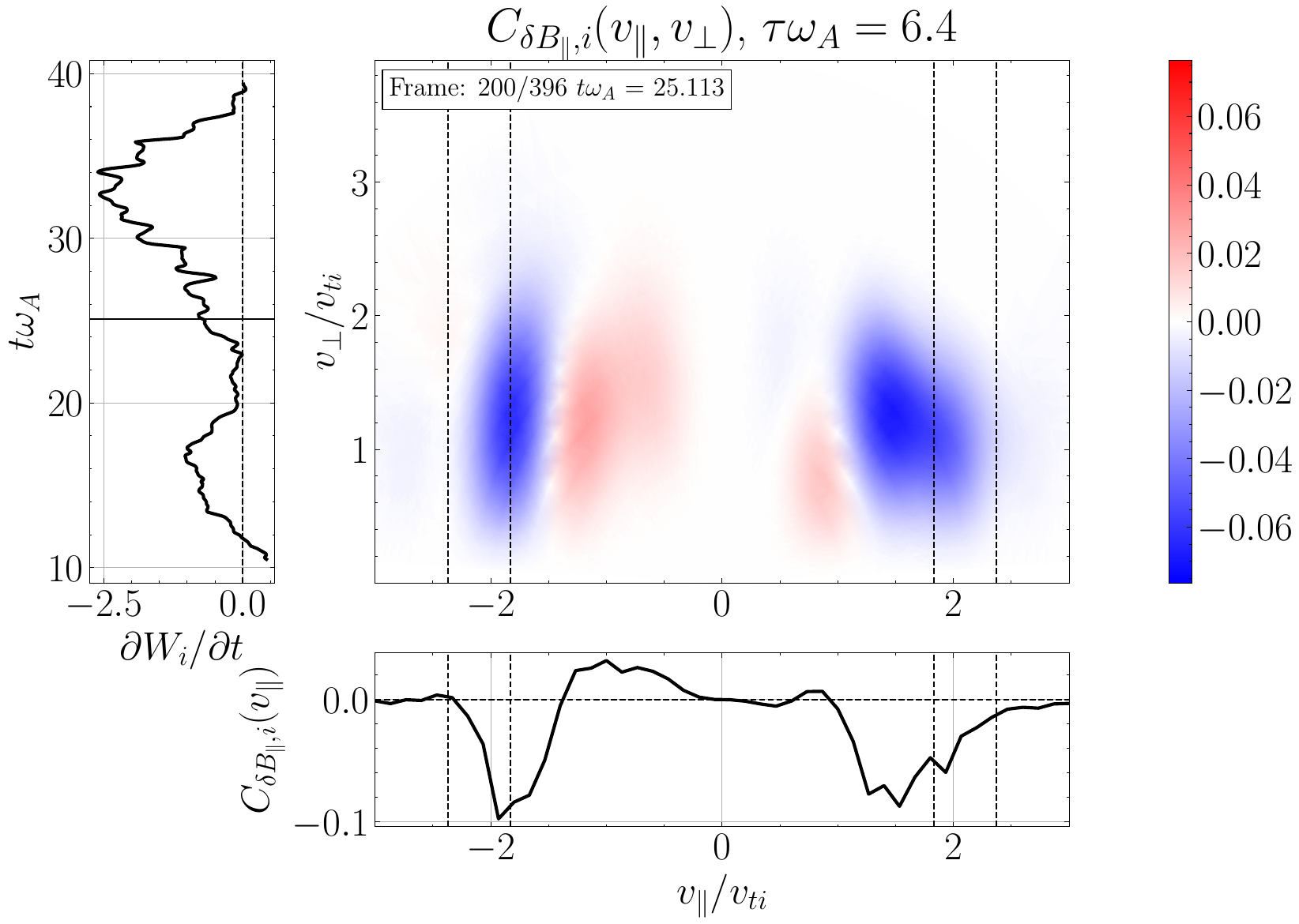}
      \includegraphics[width=0.31\textwidth]{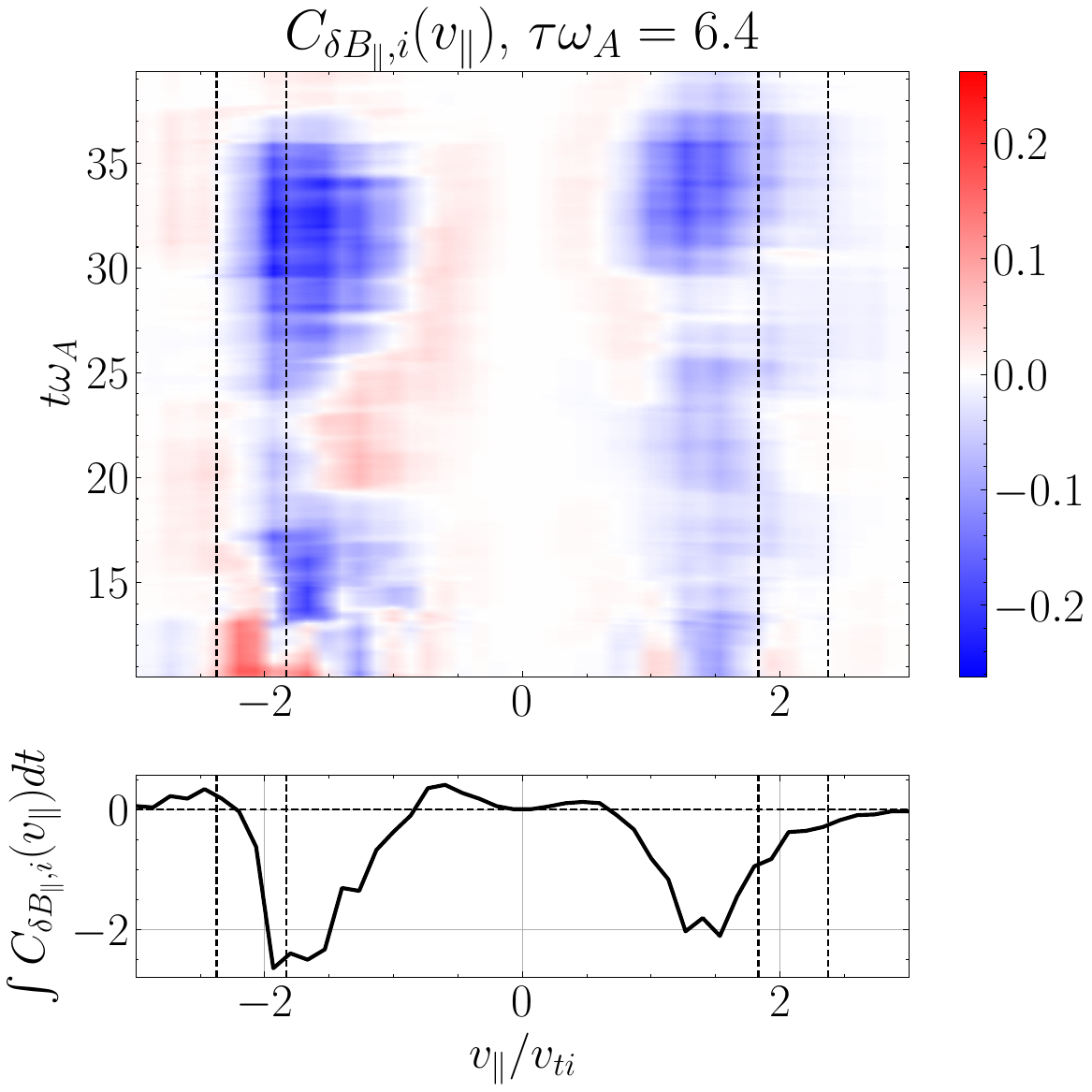}
   \end{center}
 \vskip -1.6in
 \hspace*{0.4in} (c)\hspace*{2.3in} (d)
 \vskip +1.6in
     \begin{center}
      \includegraphics[width=0.42\textwidth]{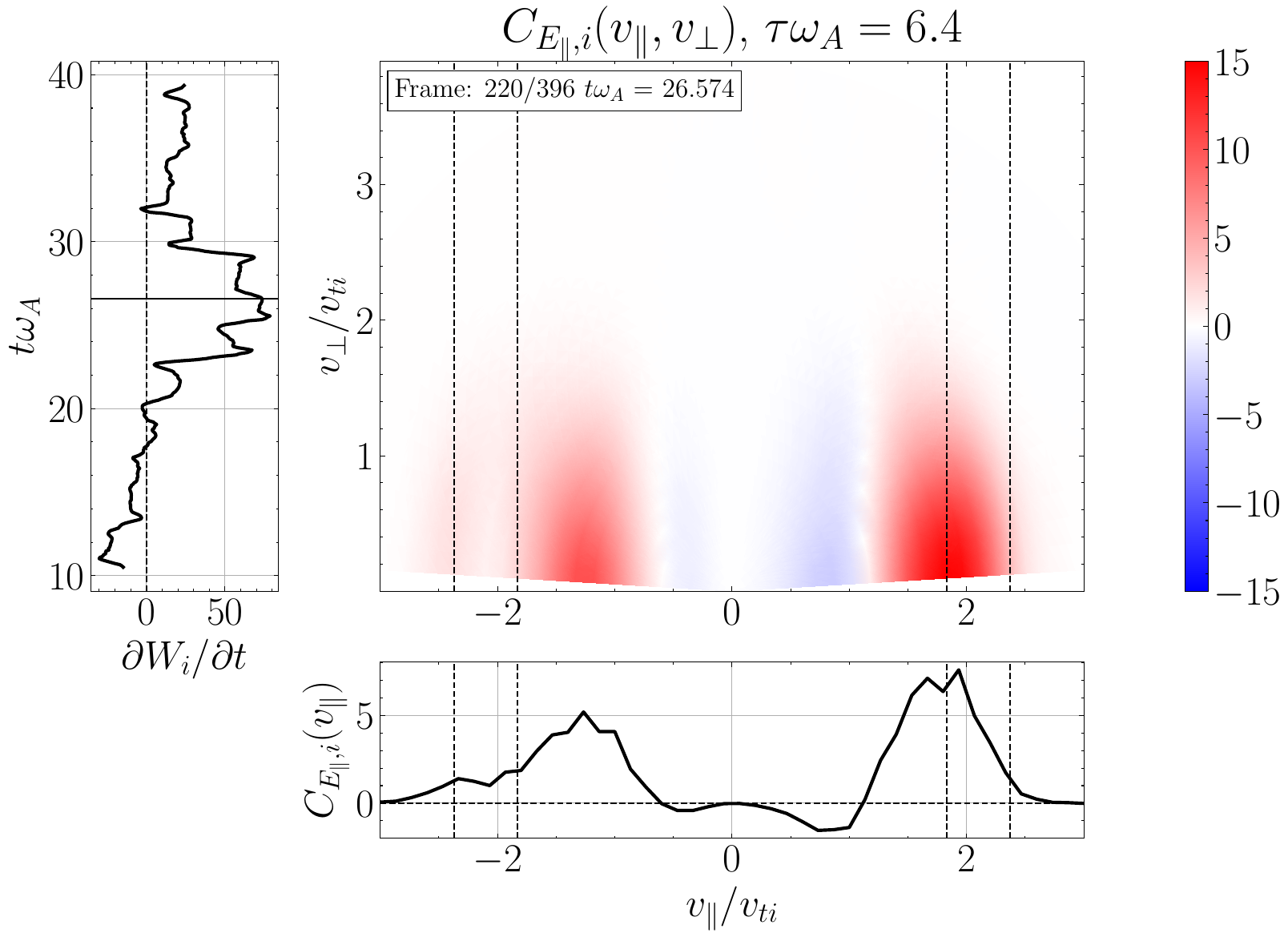}
      \includegraphics[width=0.31\textwidth]{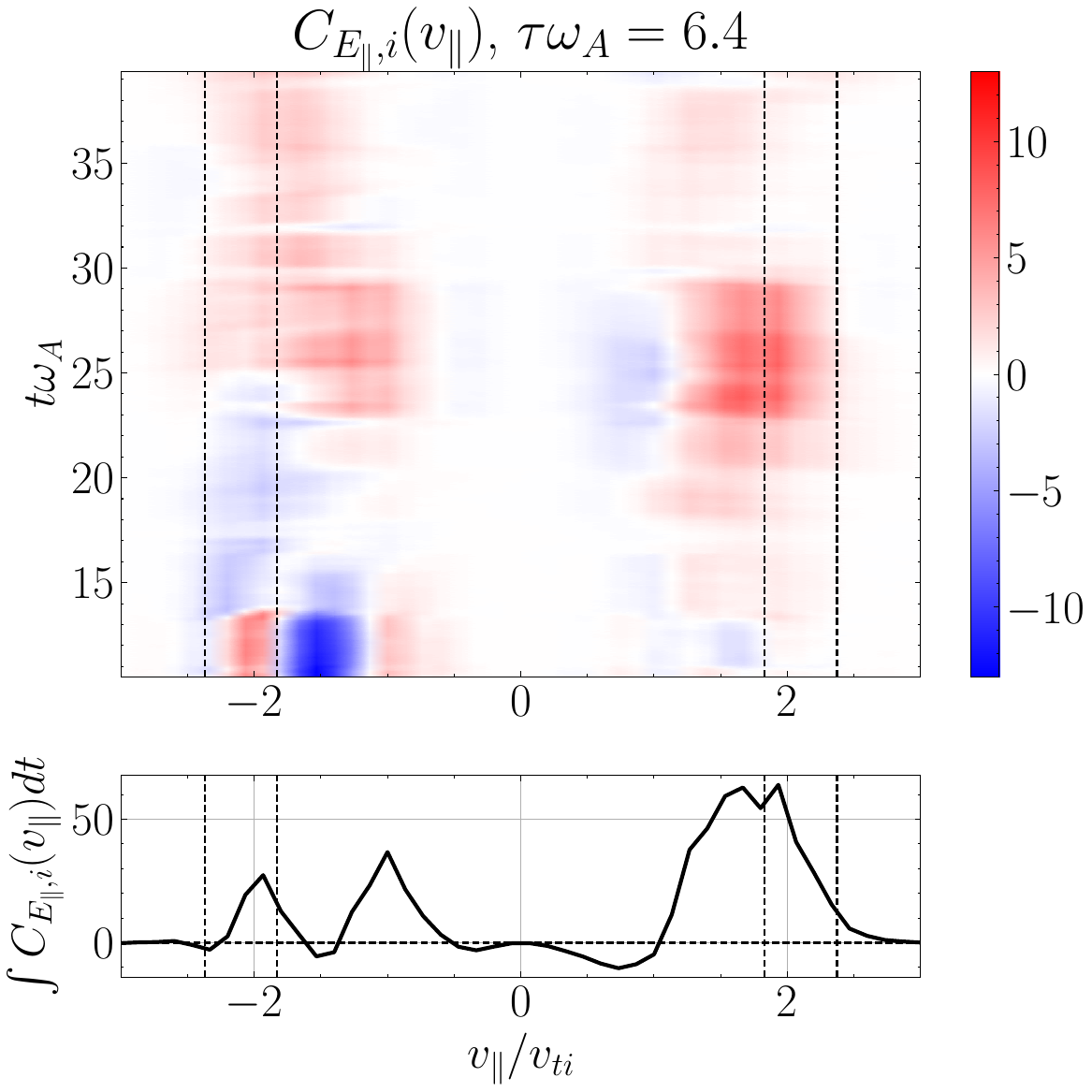}
   \end{center}
 \vskip -1.6in
 \hspace*{0.4in} (e)\hspace*{2.3in} (f)
 \vskip +1.6in
    \begin{center}
        \includegraphics[width=0.42\textwidth]{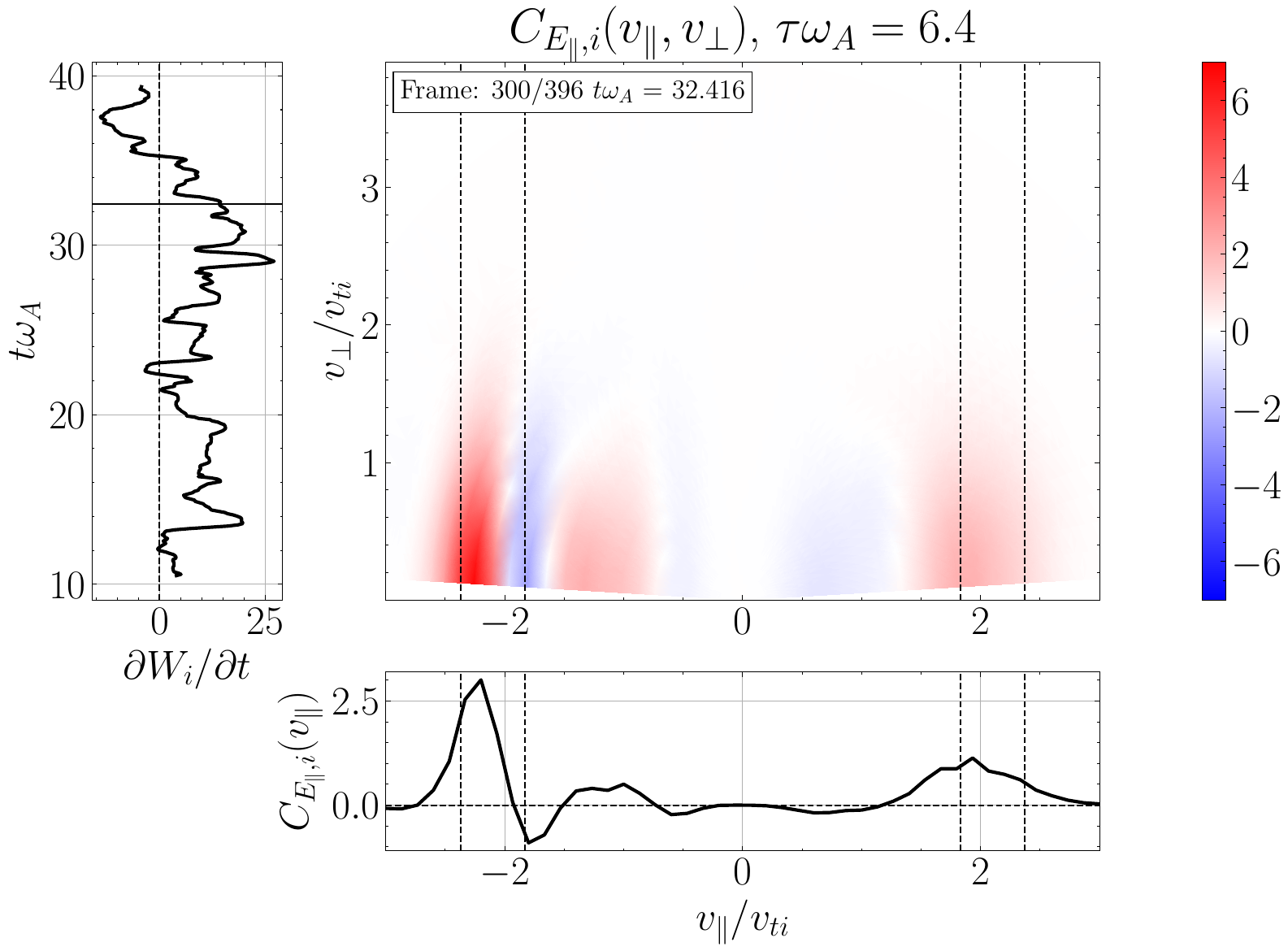}
        \includegraphics[width=0.31\textwidth]{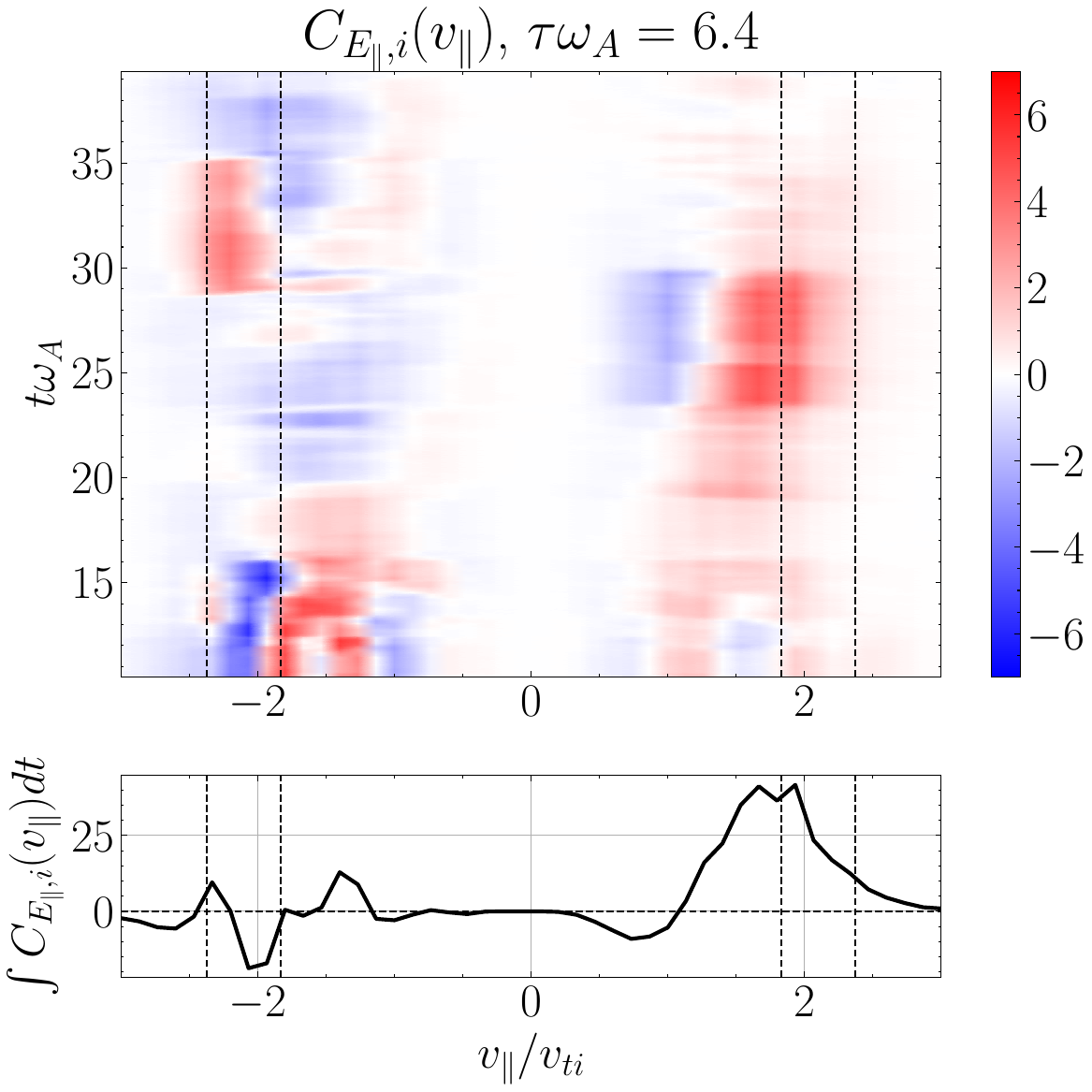}
    \end{center}
 \vskip -1.6in
 \hspace*{0.4in} (g)\hspace*{2.3in} (h)
 \vskip +1.4in
   \caption{Velocity-space signatures of transit-time damping (TTD, top two rows) and Landau damping (LD, bottom two rows) sampled from \T{AstroGK} turbulence simulation with $0.25 \leq k_\perp \rho_i \leq 7.75$, $T_i / T_e = 1$, and $\beta_i = 0.3$. The correlation interval is set as $\tau \omega_A = 6.4$. The left column presents gyrotropic plane $(v_\parallel, v_\perp)$ signatures, and the right column features the timestack plots of the $v_\perp-$integrated reduced correlation; both following the layout format of \figref{fig:turb_sig_beta_1}. The resonant parallel phase velocity ranges are marked by the four vertical dashed lines at $v_\parallel/v_\text{ti} =\pm 1.832$ and $v_\parallel/v_\text{ti} =\pm 2.373$. From top to bottom, data are taken from Probe 7 $(5 \pi \rho_i, 3 \pi \rho_i, 0)$, Probe 15 $(5 \pi \rho_i, 7 \pi \rho_i, 0)$, Probe 20 $(4 \pi \rho_i, 4 \pi \rho_i, -0.125 \pi a_0)$, and Probe 5 $(\pi \rho_i, 3 \pi \rho_i, 0)$, respectively.}
    \label{fig:turb_sig_beta_03}
\end{figure}

We plot some typical velocity-space signatures for TTD and LD for the  $\beta_i = 0.3$ simulation in \figref{fig:turb_sig_beta_03}.  Note that, for this value of $\beta_i = 0.3$, the contribution of TTD is always to remove energy from the ions, as shown in  \figref{fig:ldr_mr36}(a), so we expect only negative TTD signatures, similar to that shown in \figref{fig:single_wave_varying_beta}(a).  Consequently, we expect LD to dominate the removal of energy from the turbulence. Performing the TTD analysis to determine 
$C_{\delta B_\parallel, i}(v_\parallel,v_\perp;\tau)$, we display a typical gyrotropic velocity-space signature in  \figref{fig:turb_sig_beta_03}(a) with the associated timestack plot at the same probe in (b), showing two bipolar signatures with a net negative energy transfer rate and zero crossings at $v_\parallel/v_{ti} \simeq \pm 0.7$.  These look like typical reversed TTD signatures, but the phase velocity is \emph{not} within the expected range  $1.8 \lesssim |v_\parallel/v_{ti}| \lesssim 2.4 $ for a KAW with $\beta_i = 0.3$. Only 3 of the 24 probes showed reversed TTD signatures with phase velocities closer to the expected range, as shown in \figref{fig:turb_sig_beta_03}(c) and (d) with a bipolar zero crossing around $v_\parallel/v_{ti} \simeq -1.5$.  

Performing the LD analysis to determine $C_{E_\parallel, i}(v_\parallel,v_\perp;\tau)$ on the same  $\beta_i = 0.3$ turbulence simulation, we find a similar intriguing result that we commonly find bipolar signatures associated with positive energy transfer to ions, but with zero crossings well below the expected range of $1.8 \lesssim |v_\parallel/v_{ti}| \lesssim 2.4 $.  In \figref{fig:turb_sig_beta_03}(e), we see two bipolar signatures at $v_\parallel/v_{ti} \simeq -0.7$  and $v_\parallel/v_{ti} \simeq 1.2$. Only 2 of the 24 probes recover LD velocity-space signatures in the expected range, such as that shown in  \figref{fig:turb_sig_beta_03}(g) and (h).  

 \begin{figure}
    \begin{center}
      \includegraphics[width=0.5\textwidth]{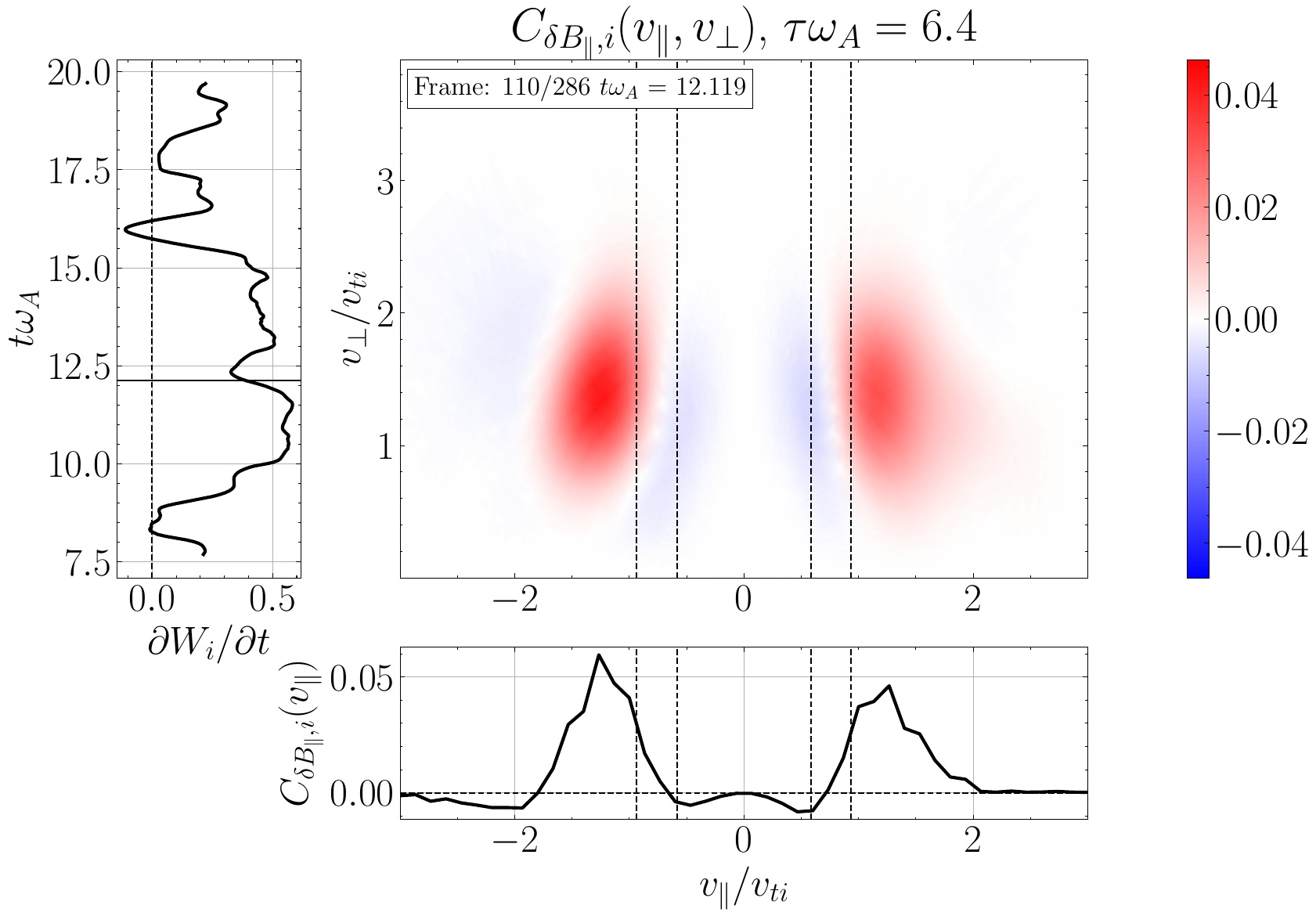}
      \includegraphics[width=0.37\textwidth]{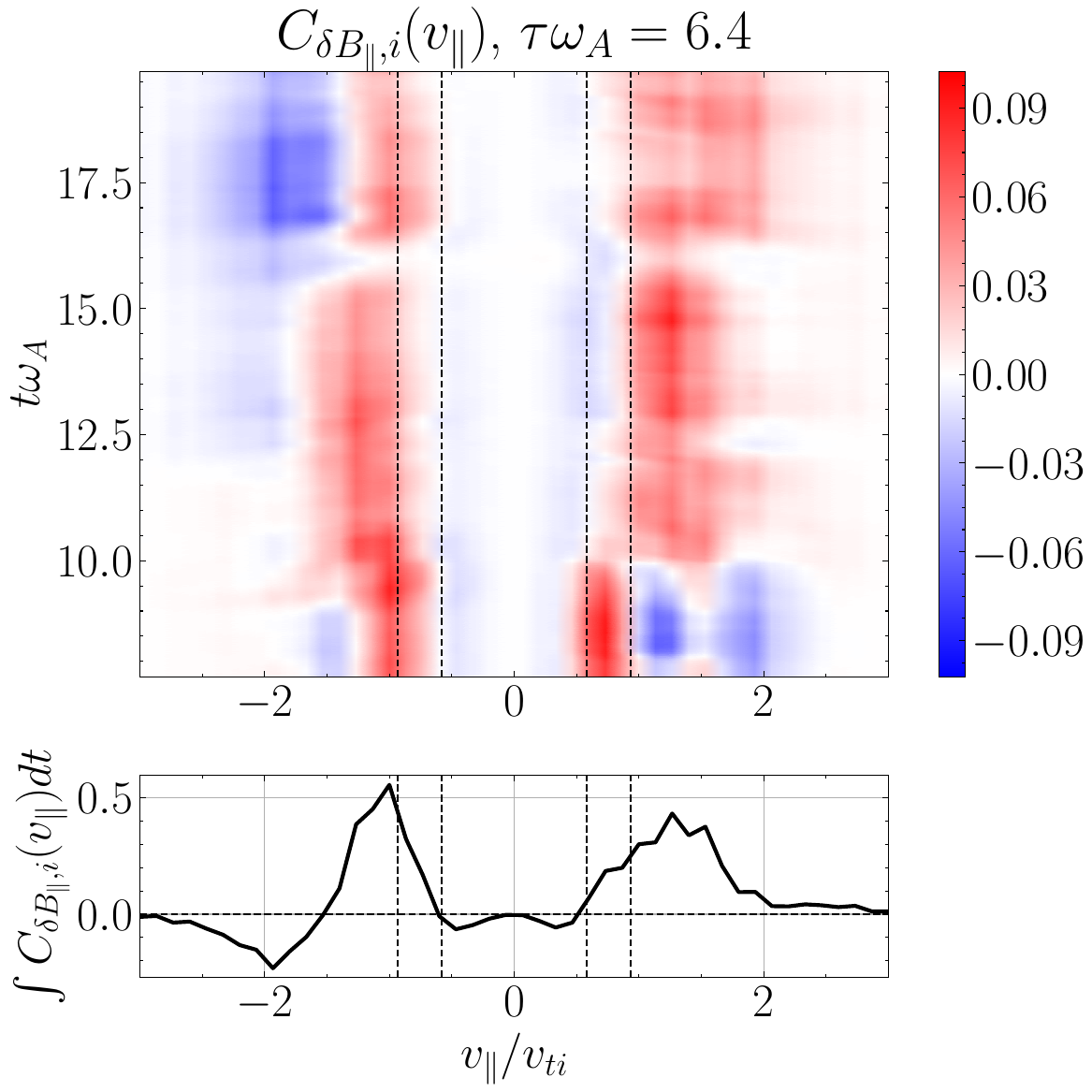}
   \end{center}
 \vskip -2.0in
\hspace*{0.05in} (a)\hspace*{2.7in} (b)
\vskip +2.2in
    \begin{center}
        \includegraphics[width=0.5 \textwidth]{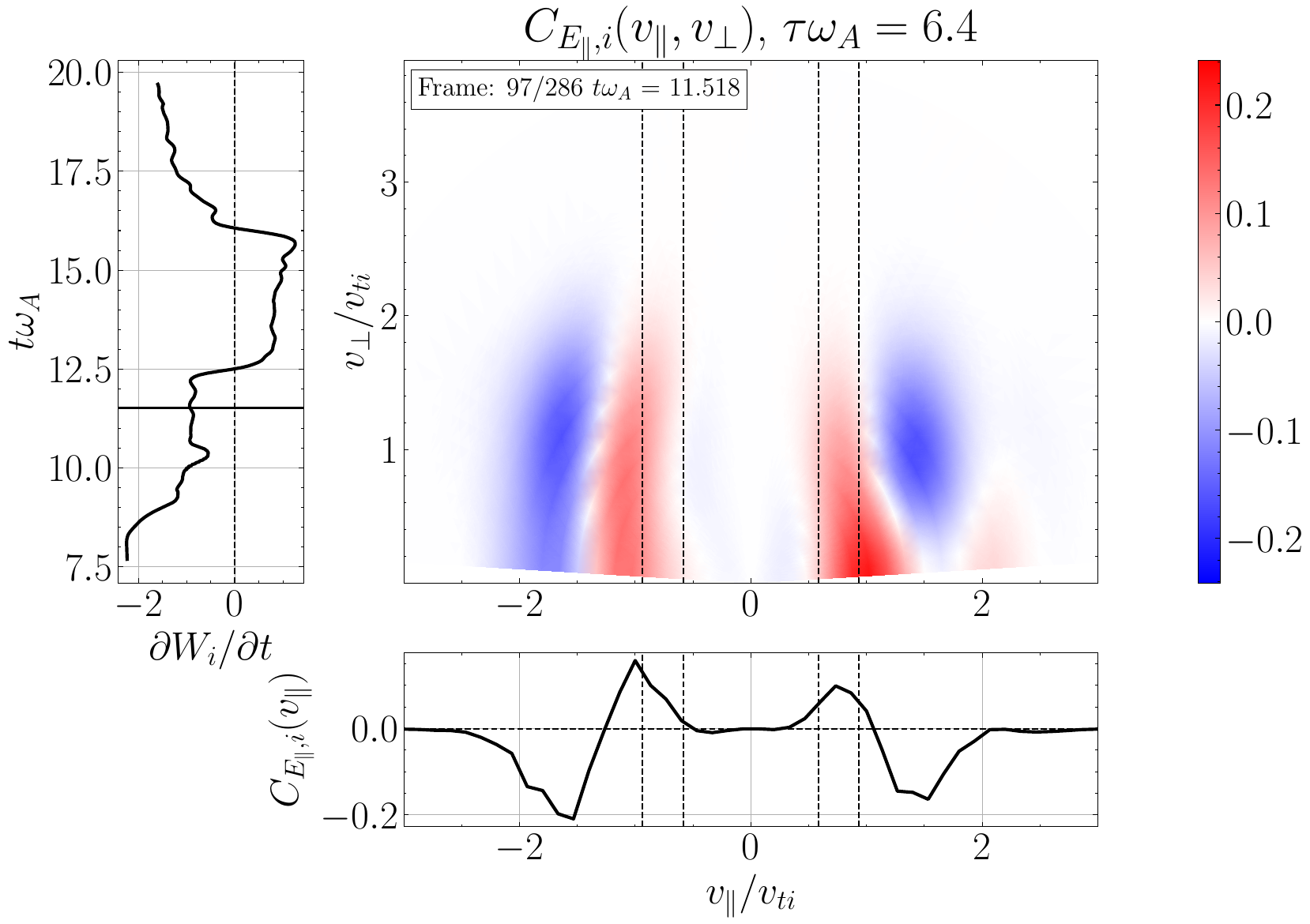}
        \includegraphics[width=0.37 \textwidth]{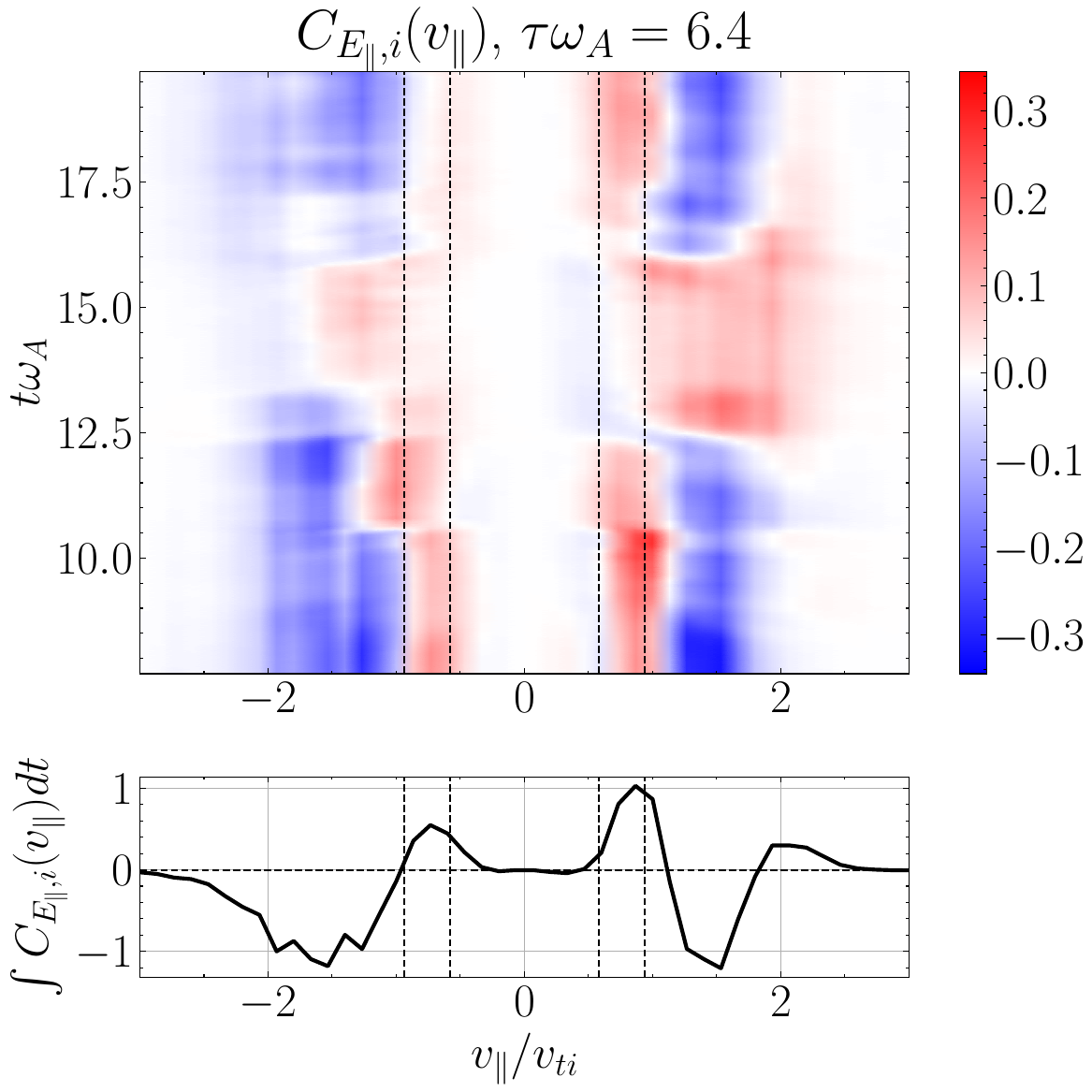}
    \end{center}
    \vskip -2.0in
\hspace*{0.05in} (c)\hspace*{2.7in} (d)
\vskip +2.2in
   \caption{Velocity-space signatures of transit-time damping (TTD, top row) at Probe 22 $(4 \pi \rho_i, 4 \pi \rho_i, 0.375 \pi a_0)$ and Landau damping (LD, bottom row) at Probe 5 $(\pi \rho_i, 3 \pi \rho_i, 0)$ sampled from \T{AstroGK} turbulence simulation with $0.25 \leq k_\perp \rho_i \leq 7.75$, $T_i / T_e = 1$, and $\beta_i = 3$. The correlation interval is set as $\tau \omega_A = 6.4$. The left column presents gyrotropic plane $(v_\parallel, v_\perp)$ signatures, and the right column features the timestack plots of the $v_\perp-$integrated reduced correlation; both following the layout format of \figref{fig:turb_sig_beta_1}. The resonant parallel phase velocity ranges are marked by the four vertical dashed lines at $v_\parallel/v_\text{ti} =\pm 0.583$ and $v_\parallel/v_\text{ti} =\pm 0.936$.}
    \label{fig:turb_sig_beta_3}
\end{figure}

Turning next to the field-particle correlation analysis of the  $\beta_i = 3$ turbulence simulation, the linear dispersion relation plot for KAWs in  \figref{fig:ldr_mr36}(c) shows that LD removes energy from ions for waves with $k_\perp \rho_i \lesssim 1.7$, but TTD positively energizes ions over perpendicular wavenumbers with $k_\perp \rho_i \lesssim 4$.  At all probes in the simulation, we find bipolar velocity-space signatures of positive energy transfer by TTD to ions at both positive and negative parallel velocities near the expected range  $0.58 \lesssim |v_\parallel/v_{ti}| \lesssim 0.94$, with a typical case illustrated in \figref{fig:turb_sig_beta_3}, showing (a) the gyrotropic velocity-space signature $C_{\delta B_\parallel, i}(v_\parallel,v_\perp;\tau)$ and (b) the timestack plot $C_{\delta B_\parallel, i}(v_\parallel,t;\tau)$ indicating that these signatures are persistent in time.  On the other hand, analyzing the signatures of LD using (c) the gyrotropic velocity-space signature $C_{E_\parallel, i}(v_\parallel,v_\perp;\tau)$ and (d) the timestack plot $C_{E_\parallel, i}(v_\parallel,t;\tau)$, the pattern of energy transfer to ions is widely variable, with ions losing energy more than gaining energy, in agreement with the expectations from the linear dispersion relation.  During the majority of the time when clear reversed bipolar patterns are visible for LD, they appear to have zero crossings that are close to the expected range of parallel resonant velocities.

Since the field-particle correlation analysis of our single-wave simulations in \secref{sec:singleKAW_ttd_sig_varing_beta} shows that the velocity-space signatures of TTD and LD appear near the resonant parallel velocities of the waves, as illustrated in \figref{fig:single_wave_varying_beta}, it raises the question of why the velocity-space signatures in the $\beta_i = 0.3$ simulation in \figref{fig:turb_sig_beta_03} often do not appear in the expected range of resonant velocities.   There exist at least two possible explanations for this finding: (i) nonlinear transfer from \Alfven waves to other wave modes, and (ii) nonlinear transit-time damping.  Before going into  the details of each of these possibilities below, it is worthwhile emphasizing that the field-particle correlation technique determines the total energy transfer to particles over the full velocity distribution, meaning that the net energy transfer within a region of velocity space is effectively weighted by the number of particles within that region of the velocity distribution.  To illustrate this point, let us  focus on the specific example of \figref{fig:turb_sig_beta_03}(a), where the range of resonant parallel phase velocities for the \Alfven mode (vertical black lines) occurs in the range  $1.8 \lesssim v_\parallel /v_{ti} \lesssim 2.4$. Even if there exists resonant energy transfer to the particles at these suprathermal parallel velocities, a potentially competing mechanism (such as the two possibilities named above) could dominate the net particle energization if it transfers energy to particles lying within the bulk of the velocity distribution at  $v_\parallel /v_{ti} \sim 1$, where many more particles can participate in the energy transfer.  Thus, the velocity-space signatures arising from the field-particle correlation analysis are generally weighted by the underlying velocity distribution to emphasize energy transfer mechanisms interacting with the main part of the velocity distribution at $v_\parallel /v_{ti} \sim 1$.

First, let us analyze how nonlinear energy to other wave modes may explain the velocity-space signatures observed in \figref{fig:turb_sig_beta_03}. Due to the predominantly \Alfvenic nature of fluctuations in space plasma turbulence \citep{Tu:1995,schekochihin:2009, Bruno:2013}, we focus on the collisional damping rates of  \Alfvenic fluctuations by TTD and LD in this study.  However, the plasma turbulence in our gyrokinetic simulations is broadband and may contain slow magnetosonic fluctuations that arise from nonlinear couplings among the turbulent fluctuations.  Although slow magnetosonic waves are not generated by nonlinear couplings among the dominantly \Alfvenic fluctuations in the MHD limit at $k_\perp\rho_i \ll 1$ \citep{schekochihin:2009}, at the ion kinetic scales $k_\perp\rho_i \sim 1$ it is possible that energy can be nonlinearly transferred into slow magnetosonic fluctuations.  These kinetic slow wave fluctuations\footnote{The kinetic slow wave fluctuations specified here are the weakly collisional equivalent of the slow magnetosonic wave in the MHD limit; these are compressible waves distinguished from the kinetic fast magnetosonic wave by the having thermal and magnetic pressure fluctuations acting in opposition as restoring forces for the wave.} may have a different parallel phase velocity from the \Alfvenic fluctuations and thereby mediate damping of the turbulent energy through either TTD or LD at phase velocities that fall within the bulk of the velocity distribution at $v_\parallel /v_{ti} \sim 1$. The second possibility is that \emph{nonlinear transit-time damping} may occur, whereby the beat mode fluctuations (which are not natural wave modes of the system)---arising from nonlinear interactions between \Alfvenic fluctuations with different frequencies and wave vectors---can have an effective phase velocity that falls in the core of the velocity distribution, leading to efficient net energy transfer to the ions via collisionless wave-particle interactions.  

In closing this discussion about the velocity-space signatures observed in \figref{fig:turb_sig_beta_03}, a final point to emphasize is that the clear bipolar signatures of negative energy transfer from ions by TTD in \figref{fig:turb_sig_beta_03}(a) and of positive energy transfer to ions by LD in \figref{fig:turb_sig_beta_03}(e) for $\beta_i=0.3$ suggest that some resonant energy transfer mechanism is governing the energization in these simulations; the details of this turbulent damping mechanism, which may include nonlinear couplings to other linear wave modes or nonlinear beat modes as suggested above, are clearly a ripe avenue for exploration in future work.

\begin{figure}
    \begin{center}
      \includegraphics[width=0.65\textwidth]{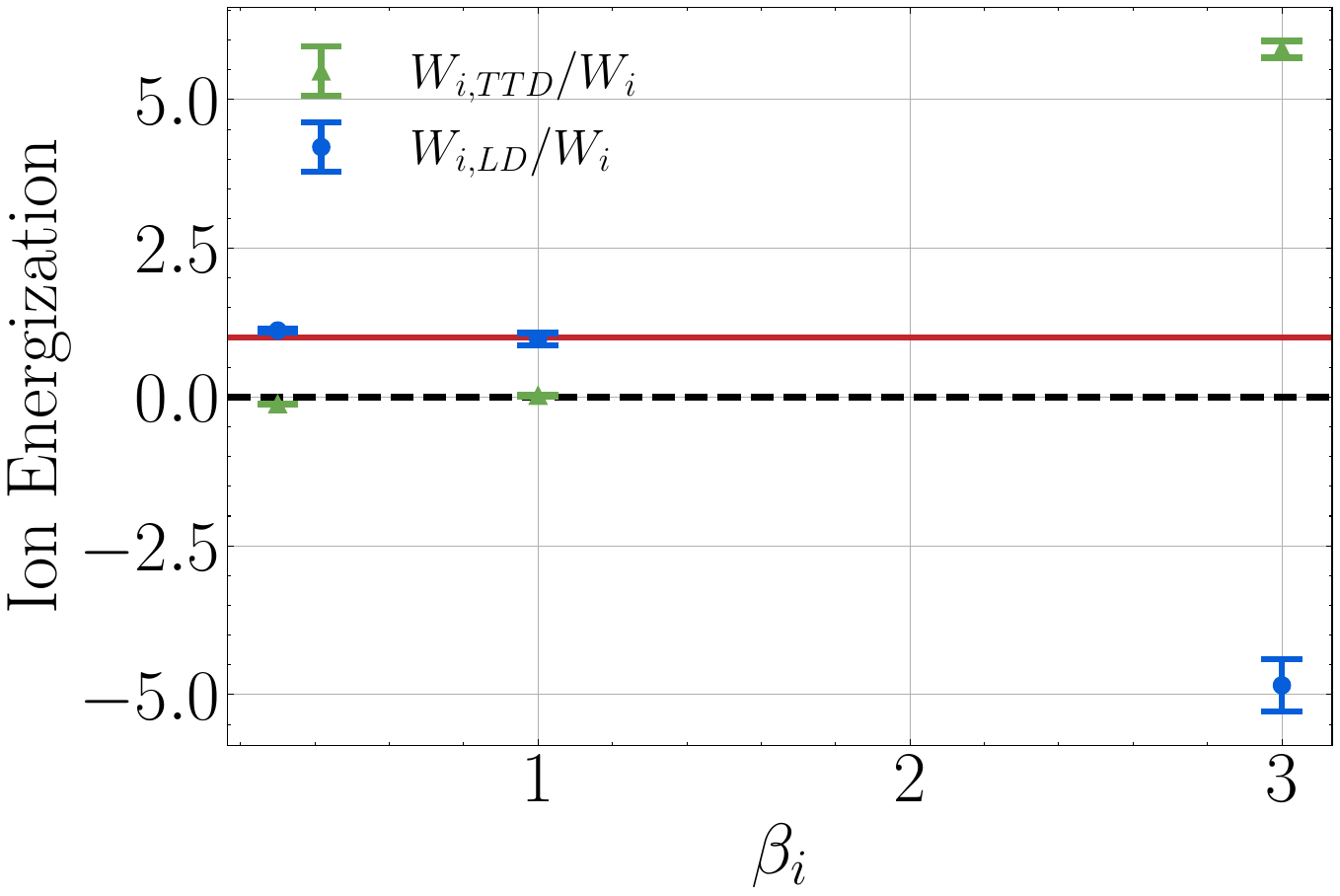}
   \end{center}
 \vskip -1.7in
\vskip +1.6in
    \caption{Ratio of the change of the ion kinetic energy density due to TTD and LD to the total change of the ion kinetic energy density during the analysis time, both averaged over all 24 probes, plotted against $\beta_i$. The error bars represent the standard deviations calculated across all probes.}
   \label{fig:WiTTD_WiLD_beta}
\end{figure}
\subsection{Net Turbulent Damping as a Function of Ion Plasma Beta $\beta_i$ \label{sec:turb_damp_beta}}

To estimate the net energy density transfer to or from ions mediated by TTD or LD averaged over the volume of the simulation, we integrate the energy density transfer rate due to each mechanism over the duration of each turbulence simulation (after the energy spectra have reached a statistically steady state) and average over all 24 probes for each simulation.
This procedure provides a statistical estimate of the total change in ion energy density by each mechanism,   denoted by $W_{i, TTD}$ and $W_{i, LD}$.  We compute the ratio of these ion energy density changes by each mechanism to the total ion energy density change $W_i = W_{i, TTD} + W_{i, LD}$, and plot $W_{i, TTD}/W_i$ (green) and $W_{i, LD}/W_i$ (blue) versus the ion plasma beta $\beta_i$ in \figref{fig:WiTTD_WiLD_beta}.  Variability in the energy density changes due to TTD and LD is indicated by error bars on each point, computed using the standard deviation of the time-integrated energy density changes at all 24 probes. The red solid line at $1$ presents the sum of $W_{i, TTD}/W_i$ and $W_{i, LD}/W_i$, and the black dashed line at zero highlights whether a mechanism yields a positive net energy density transfer to ions or a negative net energy density transfer from ions.

For $\beta_i=0.3$, the ion energization is dominated by LD, with TTD yielding a small and slightly negative energy transfer. The ion LD and TTD contributions to the damping rate from the linear dispersion relation in \figref{fig:ldr_mr36}(a) provide intuition about the relative role of these two collisionless damping mechanisms in a turbulent plasma.  Note that the broadband turbulence in the simulation consists of wave modes spanning the range $0.25 \le k_\perp \rho_i \le 7.75$, denoted by the two vertical dashed black lines in \figref{fig:ldr_mr36}(a), with a monotonically decreasing magnetic energy spectrum, as shown in \figref{fig:kspec}.  The net ion energization by each mechanism expected in a given simulation can be intuitively estimated by an integration over $k_\perp \rho_i$ of the product of the damping rate for that mechanism with the electromagnetic energy at each scale.  For $\beta_i=0.3$, \figref{fig:ldr_mr36}(a) shows that $\gamma_{iTTD}>0$ and $\gamma_{iLD}<0$ over the full range of perpendicular wavenumbers in the simulation, so the numerical finding of net energy transfer rates in \figref{fig:WiTTD_WiLD_beta} appears to be consistent with the expectation from the linear dispersion relation.

For $\beta_i=1$, both LD and TTD yield net positive energization of the ions, but LD dominates over TTD once more, as seen in \figref{fig:WiTTD_WiLD_beta}.  This result is broadly consistent with the linear damping rates from the linear dispersion relation plotted in \figref{fig:ldr_mr36}(b), where the LD rate integrated over $k_\perp \rho_i$ would be expected to dominate over the TTD rate, but both mechanisms would be expected to yield net positive ion energization rates.

At $\beta_i=3$, on the other hand, \figref{fig:WiTTD_WiLD_beta} shows that TTD mediates a net positive energy transfer to the ions, while LD governs a net negative energy transfer with a slightly smaller magnitude than the positive TTD energy transfer.  The linear dispersion relation results in \figref{fig:ldr_mr36}(c) predict a more complicated behavior than the $\beta_i=0.3$ and $\beta_i=1$ cases: TTD yields ion energization for all $k_\perp \rho_i \lesssim 5$, but LD leads to ion-to-wave energy transfer for $k_\perp \rho_i \lesssim 1.5$ and wave-to-ion energy transfer for $k_\perp \rho_i \gtrsim 1.5$.   The velocity-space integrated rates of energy transfer by each mechanism at each probe---for example, the vertical left panels of \figref{fig:turb_sig_beta_3}(a) and (c)---show variations in time that appear to be consistent with these expectations from linear theory. The TTD rates of ion energy density change are generally positive with moderate amplitudes. On the other hand, the LD rates of ion energy density change are somewhat larger in amplitude but vary with both positive and negative rates, such that the net LD rate averaged over all probes is negative and slightly smaller in amplitude than the net TTD rate.  This broader variation is evident in the larger standard deviation of the $\beta_i=3$ values in \figref{fig:WiTTD_WiLD_beta} relative to the $\beta_i=0.3$ and $\beta_i=1$ cases.

In conclusion, it is worthwhile emphasizing that the expectations from linear theory of the relative contributions to the energy density transfer rates from each mechanism, including their overall sign, is only expected to hold in statistical sense, averaged over a sufficiently large volume. This is particularly true for plasma parameter values where the expected linear damping rate for a given mechanism changes sign within the range of $k_\perp \rho_i$ where significant damping is expected, such as the case for ion LD in the $\beta_i=3$ simulation.  Based on kinetic numerical simulations of turbulence, the rate of change of energy density for species $s$ given by $\V{j}_s \cdot \V{E}$ is highly variable spatially, with variations that can be much larger than the spatially or temporally averaged net value.  An example from a simulation of a strongly nonlinear \Alfven wave collision---the nonlinear interaction between counterpropagating \Alfven waves that is argued to be the fundamental building block of astrophysical plasma turbulence \citep{Howes:2012b,Howes:2013a,Nielson:2013a,Howes:2013b,Drake:2013,Howes:2016b,Verniero:2018a,Verniero:2018b}---
is seen in Fig.~8 of \citet{Howes:2018a}, where the wave-period averaged energy density transfer rates are found to be an order of magnitude smaller than the instantaneous rates.  Furthermore, even for an overall damping rate that energizes the plasma particles at the expense of the turbulent electromagnetic energy, the net energy transfer rate averaged over time at single position may be negative, even when the spatially averaged energy transfer rate is positive, as seen in  Fig.~8(d) of \citet{Howes:2018a}.

That the turbulent energy damping rate is highly oscillatory, both in space and time, with a much smaller amplitude net energy damping rate that is generally positive (indicating a damping of turbulence and consequent energization of particles), should not be especially surprising. The nonlinear energy transfer from large to small scales by the turbulence is likewise found to be highly oscillatory, with third-moment calculations of the instantaneous energy transfer rates \citep{Kolmogorov:1941b,Politano:1998a} yielding large values of both signs, with the net averaged value having  a much smaller amplitude of positive energy transfer to small scales \citep{Coburn:2014,Coburn:2015}.  Thus, field-particle correlation calculations of the energy density transfer rate at a single point, even averaged over a long duration, may still lead to a result in a turbulent plasma that appears to conflict with the expectation from linear theory. Only with a sufficient large statistical average, here estimated using the average over all 24 probe positions, should we expect the results to be consistent with the general expectations of linear theory---the results in \figref{fig:WiTTD_WiLD_beta} indeed appear to be consistent with the general expectations for the relative contributions of LD and TTD  from linear theory.


\section{Conclusion \label{sec:conclusion}}

Transit-time damping is a well-known mechanism for the resonant collisionless damping of electromagnetic waves exhibiting variations of the magnetic field magnitude along the mean magnetic field direction, mediated by the magnetic mirror force.  This mechanism has been proposed as a possible means for removing energy from the fluctuations in weakly collisional plasma turbulence, but to date there exists little direct evidence clearly showing the damping of turbulence via transit-time damping.  Here we employ the recently developed field-particle correlation technique to use measurements of the gradient of the magnetic field magnitude and the ion velocity distribution at a single point to determine a velocity-space signature that can be used to identify definitively that transit-time damping plays a role in the damping of plasma turbulence.  

We first derive the particular mathematical form of the field-particle correlation for the rate of energy transfer due to transit-time damping in \secref{sec:ttd_fpc_formula}, and then we predict the velocity-space signature of the rate of change of phase-space energy density due to transit-time damping using a simple model in \secref{sec:predict_ttd_sig}.  Next, we perform gyrokinetic simulations of single kinetic \Alfven waves to determine the resulting velocity-space signature of transit-time damping numerically, confirming the qualitative features of our prediction, and presenting the first key result of this study: the gyrotropic velocity-space signature of transit-time damping in \figref{fig:fiducial}(a).  

We contrast the velocity-space signature of transit-time damping with the known bipolar velocity-space signature of Landau damping, showing the same bipolar pattern of phase-space energy density loss below and gain above the resonant parallel phase velocity, but the transit-time damping signature does not extend down to $v_\perp \rightarrow 0$ because it is mediated via the magnetic moment of the charged particle $\mu=m v_\perp^2/(2B)$; thus, signatures of transit-time damping and Landau damping can be distinguished in gyrotropic velocity space by examining the behavior at the resonant parallel phase velocity as  $v_\perp \rightarrow 0$.  Furthermore, we find the unexpected result that transit-time damping can lead to a net \emph{loss} of ion energy over the period of the wave for $\beta_i < 1$ and Landau damping can lead to a net \emph{loss} of ion energy over the period of the wave for $\beta_i > 1$. This surprising result is explained, however, by examining the separate contributions of transit-time damping and Landau damping to ion damping from the linear Vlasov-Maxwell dispersion relation: for a single linear kinetic \Alfven wave, the net effect of transit-time damping and Landau damping combined for a plasma with a Maxwellian equilibrium ion velocity distribution always leads to a net damping of the wave and net gain of energy by the ions.

Next, we perform three gyrokinetic simulations of weakly collisional plasma turbulence with three values of $\beta_i=0.3, 1, 3$ to seek the velocity-space signature of transit-time damping in the damping of the strong turbulent fluctuations.  In the $\beta_i=1$ turbulence simulation, we indeed find a velocity-space signature of transit-time damping as shown in \figref{fig:turb_sig_beta_1}(a), indicating that this mechanism does indeed play a role in the dissipation of kinetic plasma turbulence, along with confirming previously demonstrated signatures of Landau damping with ions in \figref{fig:turb_sig_beta_1}(c). This second key result of this paper shows clearly the transit-time damping does serve to damp the fluctuations in weakly collisional plasma turbulence.

The relative strength of transit-time damping and Landau damping is predicted to be a strong function of $\beta_i$ \citep{Quataert:1998}, so we analyze our  $\beta_i=0.3$ and  $\beta_i=3$ simulations to confirm this prediction.  For  $\beta_i=3$, we indeed find signatures of transit-time damping in the predicted range of resonant parallel phase velocities in \figref{fig:turb_sig_beta_3}(a), but Landau damping signatures vary widely, with both positive and negative energy transfer rates to the ions, and a negative overall average consistent with expectations from the linear dispersion relation in \figref{fig:ldr_mr36}(c).  For $\beta_i=0.3$, however, we discover puzzling bipolar velocity-space signatures of negative energy transfer but with a zero-crossing well below the parallel phase velocity of kinetic \Alfven waves.  This may indicate that energy transfer via transit-time damping is occurring through alternative wave modes, such as kinetic slow magnetosonic fluctuations, or through nonlinear transit-time damping via beat wave modes that are generated by nonlinear interactions among the turbulent fluctuations.  These possibilities will be explored in future work.  

Determining the time-integrated change of ion kinetic energy density due to transit-time damping and Landau damping as a function of $\beta_i$ from the three simulations, averaged over all of the probes to obtain a reasonable statistical estimate,  we find results in \figref{fig:WiTTD_WiLD_beta} that are generally consistent with the expectations from the linear dispersion relation: (i) at $\beta_i=0.3$, transit-time damping is small and slightly negative, while Landau damping is about an order-of-magnitude larger and positive; (ii) at $\beta_i=1$, both transit-time damping and Landau damping are positive, but again Landau damping is about an order-of-magnitude larger than transit-time damping; and (iii) at $\beta_i=3$, transit-time damping is large and positive while Landau damping is somewhat smaller and negative.  Note that despite one of the mechanisms possibly leading to a net negative transfer of energy from ions to waves, for a sufficiently large statistical sample it is always the subdominant mechanism that is negative, so the net effect of the sum of both of these $n=0$ Landau resonant collisionless wave-particle interactions (transit-time damping and Landau damping) averaged over all of the probe positions is a damping of the turbulence for equilibrium Maxwellian velocity distributions.

\section*{Acknowledgements}
Numerical simulations were performed using the Extreme Science and Engineering Discovery Environment (XSEDE), which is supported by National Science Foundation grant number ACI-1548562, through allocation TG-PHY090084.

\section*{Funding}
 Supported by NASA grants 80NSSC18K0643, 80NSSC18K1217, and 80NSSC18K1371 
and NSF grant AGS-1842561.

\section*{Declaration of interests}
The authors report no conflict of interests.

\appendix

\section{Explicit Form of Landau Damping and Transit-Time Damping Terms in Nonlinear Gyrokinetics}
\label{sec:gk_theory}
The nonlinear, collisionless gyrokinetic equation  \citep{howes:2006} can be manipulated into a form in which the terms governing Landau damping (LD) and transit-time damping (TTD) are readily apparent.  We begin with the nonlinear, collisionless gyrokinetic equation in cgs units,  Eq.~(25) in \citet{howes:2006},
\begin{equation}
  \frac{\partial h_s}{\partial t} + v_\parallel  \frac{\partial h_s}{\partial z}
+ \frac{c}{B_0} \left[\langle \chi \rangle_{\V{R}_s},h_s\right] =
   \frac{q_s F_{0s}}{T_{0s}}  \frac{\partial \langle \chi \rangle_{\V{R}_s}}{\partial t}
   \label{eq:dhdt}
\end{equation}
where the nonlinear term is expressed in the Poisson bracket, defined by
\begin{equation}
 \left[ U,V\right] = \zhat \cdot \left[ \frac{\partial U}{\partial \V{R}_s} \times \frac{\partial V}{\partial \V{R}_s} \right] =  \frac{\partial U}{\partial X}  \frac{\partial V}{\partial Y}  -  \frac{\partial U}{\partial Y}  \frac{\partial V}{\partial X} 
\end{equation}
where the guiding center coordinates are given by $\V{R}_s=(X,Y,z)$. The gyroaverage of a given quantity at the guiding center position for a particle of species $s$ is denoted by $\langle \ldots \rangle_{\V{R}_s}$.  The gyrokinetic potential is defined by $\chi(\V{r},t)= \phi - \V{v} \cdot \V{A}/c$, 
where $\phi(\V{r},t)$  is the scalar electrostatic potential and $\V{A}(\V{r},t)$ is the vector potential. 
In this formulation, the total velocity distribution function for species $s$ is separated into 
$f_s(\V{r},\V{v},t) = F_{0s}(v) + (-q_s \phi(\V{r},t)/ T_{s}) F_{0s}(v) + h_s(\V{R}_s,v_\parallel,v_\perp,t) + O(\epsilon^2)$, where $F_{0s}(v)$ is the spatially homogeneous and temporally constant equilibrium Maxwellian velocity distribution and $h_s(\V{R}_s,v_\parallel,v_\perp,t)$ is the perturbed gyrokinetic distribution function at the particle guiding center position $\V{R}_s$ (independent of gyrophase $\theta$ in cylindrical velocity space).

We transform from the perturbed gyrokinetic distribution function $h_s$ to the complementary perturbed gyrokinetic distribution function $g_s$ \citep{schekochihin:2009}, given by 
\begin{equation}
  \label{eq:definegs}
  g_s (\V{R}_s, v_\parallel, v_\perp,t)
  = h_s(\V{R}_s, v_\parallel, v_\perp,t)
  - \frac{q_s F_{0s}}{T_{0s}}
\left\langle  \phi -  \frac{\V{v}_\perp \cdot \V{A}_\perp}{c}\right\rangle_{\V{R}_s},
\end{equation}
substituting for $h_s$ everywhere in the nonlinear gyrokinetic equation \eqref{eq:dhdt}.  After some simplification, the equation can be rearranged to obtain
\begin{eqnarray}
  \lefteqn{   \frac{\partial g_s}{\partial t}
    + v_\parallel  \frac{\partial g_s}{\partial z}
   + \frac{c}{B_0} \left[\langle \chi \rangle_{\V{R}_s},g_s\right]   =  
     - \frac{q_s F_{0s}}{T_{0s}}  \frac{\partial}{\partial t} \left\langle  \frac{v_\parallel A_\parallel}{c} \right\rangle_{\V{R}_s} } &&\\
  &-&  \frac{q_s F_{0s}}{T_{0s}} v_\parallel  \frac{\partial}{\partial z}
   \left\langle  \phi -  \frac{\V{v}_\perp \cdot \V{A}_\perp}{c}\right\rangle_{\V{R}_s}
   +  \frac{q_s F_{0s}}{T_{0s}} \frac{c}{B_0} \left[ \left\langle  \frac{v_\parallel A_\parallel}{c}\right\rangle_{\V{R}_s},  \left\langle  \phi -  \frac{\V{v}_\perp \cdot \V{A}_\perp}{c}\right\rangle_{\V{R}_s}  \right]
   \nonumber
\end{eqnarray}
We can rearrange the terms on the right-hand side to obtain a more physically illuminating form,
\begin{eqnarray}
  \frac{\partial g_s}{\partial t}
  &+&   v_\parallel  \frac{\partial g_s}{\partial z}
   + \frac{c}{B_0} \left[\langle \chi \rangle_{\V{R}_s},g_s\right]\\
   &   = & 
      \frac{q_s F_{0s}}{T_{0s}}  v_\parallel \left[ - \frac{\partial}{\partial z}
        \left\langle  \phi \right\rangle_{\V{R}_s}- \frac{1}{c}   \frac{\partial  \left\langle  A_\parallel \right\rangle_{\V{R}_s}}{\partial t} \right]
      +  \frac{q_s F_{0s}}{T_{0s}}  v_\parallel \frac{1}{B_0} \left[ \left(  \frac{\partial  \left\langle  A_\parallel \right\rangle_{\V{R}_s}}{\partial\V{R}_s }
        \times \zhat \right)        \cdot  \frac{-\partial  \left\langle  \phi \right\rangle_{\V{R}_s}}{\partial\V{R}_s } \right]   \nonumber \\
      & +& \frac{q_s F_{0s}}{T_{0s}}  v_\parallel  \frac{\partial}{\partial z}  \left\langle  \frac{\V{v}_\perp \cdot \V{A}_\perp}{c}\right\rangle_{\V{R}_s}
      +  \frac{q_s F_{0s}}{T_{0s}}  v_\parallel \frac{1}{B_0} \left[ \left(  \frac{\partial  \left\langle  A_\parallel \right\rangle_{\V{R}_s}}{\partial\V{R}_s }
        \times \zhat \right)        \cdot  \frac{1}{c}  \frac{\partial  \left\langle \V{v}_\perp \cdot \V{A}_\perp \right\rangle_{\V{R}_s}}{\partial\V{R}_s } \right]
   \nonumber
\end{eqnarray}

Using the following relations, 
\begin{equation}
 \frac{\partial  \left\langle  A_\parallel \right\rangle_{\V{R}_s}}{\partial\V{R}_s }
        \times \zhat  = \left\langle \delta \V{B}_\perp\right\rangle_{\V{R}_s},
\end{equation}
\begin{equation}
\frac{q_s}{c} \langle  \V{v}_\perp \cdot \V{A}_\perp \rangle_{\V{R}_s}
  = - \frac{1}{2} \frac{q_s}{c}  \frac{v_\perp^2}{\Omega_s} \langle \delta B_\parallel  \rangle_{\V{R}_s}  = -\mu_s \langle \delta B_\parallel  \rangle_{\V{R}_s},
  \label{eq:ex_bz_conversion}
\end{equation}
\begin{equation}
 - \frac{\partial}{\partial z}
        \left\langle  \phi \right\rangle_{\V{R}_s}- \frac{1}{c}   \frac{\partial  \left\langle  A_\parallel \right\rangle_{\V{R}_s}}{\partial t} = \left\langle E_\parallel \right\rangle_{\V{R}_s},
\end{equation}
\begin{equation}
  \frac{-\partial  \left\langle  \phi \right\rangle_{\V{R}_s}}{\partial\V{R}_s }
  = \left\langle \V{E}_\perp \right\rangle_{\V{R}_s},
\end{equation}
we can simplify the result to obtain
\begin{eqnarray}
  \frac{\partial g_s}{\partial t}
  &+&   v_\parallel  \frac{\partial g_s}{\partial z}
   + \frac{c}{B_0} \left[\langle \chi \rangle_{\V{R}_s},g_s\right]\\
   &   = & 
   \frac{q_s F_{0s}}{T_{0s}}  v_\parallel \left( \zhat +  \frac{\left\langle \delta \V{B}_\perp\right\rangle_{\V{R}_s}}{B_0} \right) \cdot  \left\langle \V{E} \right\rangle_{\V{R}_s}
   +  \frac{ F_{0s}}{T_{0s}}  v_\parallel \left[ -\mu_s \left( \zhat +  \frac{\left\langle \delta \V{B}_\perp\right\rangle_{\V{R}_s}}{B_0} \right) \cdot
      \frac{\partial  \left\langle \delta B_\parallel \right\rangle_{\V{R}_s}}{\partial\V{R}_s } \right]\nonumber
\end{eqnarray}
Finally, to put this into a more concise form, we recognize that the direction of the total magnetic field (including the perturbation) to the first order is
$\V{B} = B_0 \zhat +  \left\langle \V{\delta B}_\perp\right\rangle_{\V{R}_s}$, so we can define the unit vector of the total magnetic field direction as $\bhat$
\begin{equation}
\bhat = \frac{ B_0 \zhat +  \left\langle\V{\delta B}_\perp\right\rangle_{\V{R}_s}}{B_0}.
\end{equation}
With this final simplification, we obtain the final result for the nonlinear gyrokinetic equation,
\begin{equation}
 \frac{\partial g_s}{\partial t}+   v_\parallel  \frac{\partial g_s}{\partial z}
   + \frac{c}{B_0} \left[\langle \chi \rangle_{\V{R}_s},g_s\right]
    =    \frac{q_s F_{0s}}{T_{0s}}  v_\parallel \bhat \cdot  \left\langle \V{E} \right\rangle_{\V{R}_s}
   -  \frac{ F_{0s}}{T_{0s}}  v_\parallel  \mu_s \bhat \cdot \nabla _{\V{R}_s}
   \left\langle \delta B_\parallel \right\rangle_{\V{R}_s}
   \label{eq:dgdtphysicalform}
\end{equation}
This equation has a simple physical interpretation with respect to
work done on the distribution functions by the fields: the first term
on the right-hand side is the effect of Landau damping by electric
field parallel to the total magnetic field; and the second term on the
right-hand side is the effect of transit-time damping by the magnetic
mirror force due to the gradient of the magnetic field magnitude along
the total magnetic field, which to lowest order is just due to the
parallel magnetic field perturbations, as shown in  \eqref{eq:deltabpar}. Note also that the nonlinear term involves interactions between the electromagnetic fields and the plasma particles, but when integrated over all guiding-center space $\V{R}_s$, it leads to zero net energy change.

\section{Gyrokinetic Form of the Field-Particle Correlation}
In gyrokinetics, a form of conserved energy inspired from the definition of entropy is calculated by multiplying the 
complementary perturbed gyrokinetic distribution function $g_s$ by  $T_{0s} g_s/F_{0s}$ and integrating over all velocity and physical space \citep{howes:2006,Brizard:2007,schekochihin:2009,TCLi:2016,Howes:2018a}. 
Using a similar approach, we can  obtain an energy equation for the \emph{gyrokinetic phase-space energy density} $w_s(\V{R}_s,v_\parallel,v_\perp,t) = T_{s} g_s^2/(2F_{0s})$ by multiplying \eqref{eq:dgdtphysicalform} by $T_{0s} g_s/F_{0s}$ to obtain
\begin{equation}
  \label{eq:fpcingyro}
  \frac{\partial w_s}{\partial t}
  + v_\parallel  \frac{\partial w_s }{\partial z}
+ \frac{T_{0s} c}{B_0 F_{0s}} \left[\langle \chi \rangle_{\V{R}_s},\frac{g_s^2}{2}\right]
    =     v_\parallel \bhat \cdot  \left\langle q_s \V{E} \right\rangle_{\V{R}_s} g_s
   -   v_\parallel  \mu_s \bhat \cdot \nabla _{\V{R}_s}
   \left\langle \delta B_\parallel \right\rangle_{\V{R}_s} g_s
\end{equation}
In this formulation, the gyrokinetic form of the field-particle correlation for Landau damping would be given by 
\begin{equation}
C_{E_\parallel, s}(\V{R}_{0,s},v_\parallel,v_\perp,t)=  \frac{1}{\tau}\int^{t+\tau/2}_{t-\tau/2}  v_\parallel \bhat \cdot  \left\langle q_s \V{E} \right\rangle_{\V{R}_s} g_s dt',
  \label{eq:corrgk_ld}
\end{equation}
and for transit-time damping would be given by 
\begin{equation}
C_{\delta B_\parallel, s}(\V{R}_{0,s},v_\parallel,v_\perp,t)=  -\frac{1}{\tau}\int^{t+\tau/2}_{t-\tau/2}  v_\parallel  \mu_s \bhat \cdot \nabla _{\V{R}_s}    \left\langle \delta B_\parallel \right\rangle_{\V{R}_s} g_s dt'.
  \label{eq:corrgk_ttd}
\end{equation}

Note that the energy transfer in the nonlinear case is simply
``linear'' collisionless damping occurring along the local total
magnetic field direction (which is a nonlinear correction from the
equilibrium magnetic field direction $\V{B}_0 = B_0 \zhat$). Note also
that, in the gyrokinetic approximation, the nonlinear contribution
(known as the parallel nonlinearity) to the field-particle
interactions is dropped, as it comes in only at higher order.  Thus,
nonlinear saturation of Landau damping does not occur in gyrokinetics,
but the nonlinear correction of ``linear'' collisionless damping due
to the difference in the total magnetic field direction from the
equilibrium magnetic field direction is included in nonlinear
gyrokinetics.

\begin{figure}
    \begin{center}
      \includegraphics[width=0.48\textwidth]{figs_new/ttdb1a_probe001_s1_CB_nc100_t0150_pcolormesh.pdf}
      \includegraphics[width=0.48\textwidth]{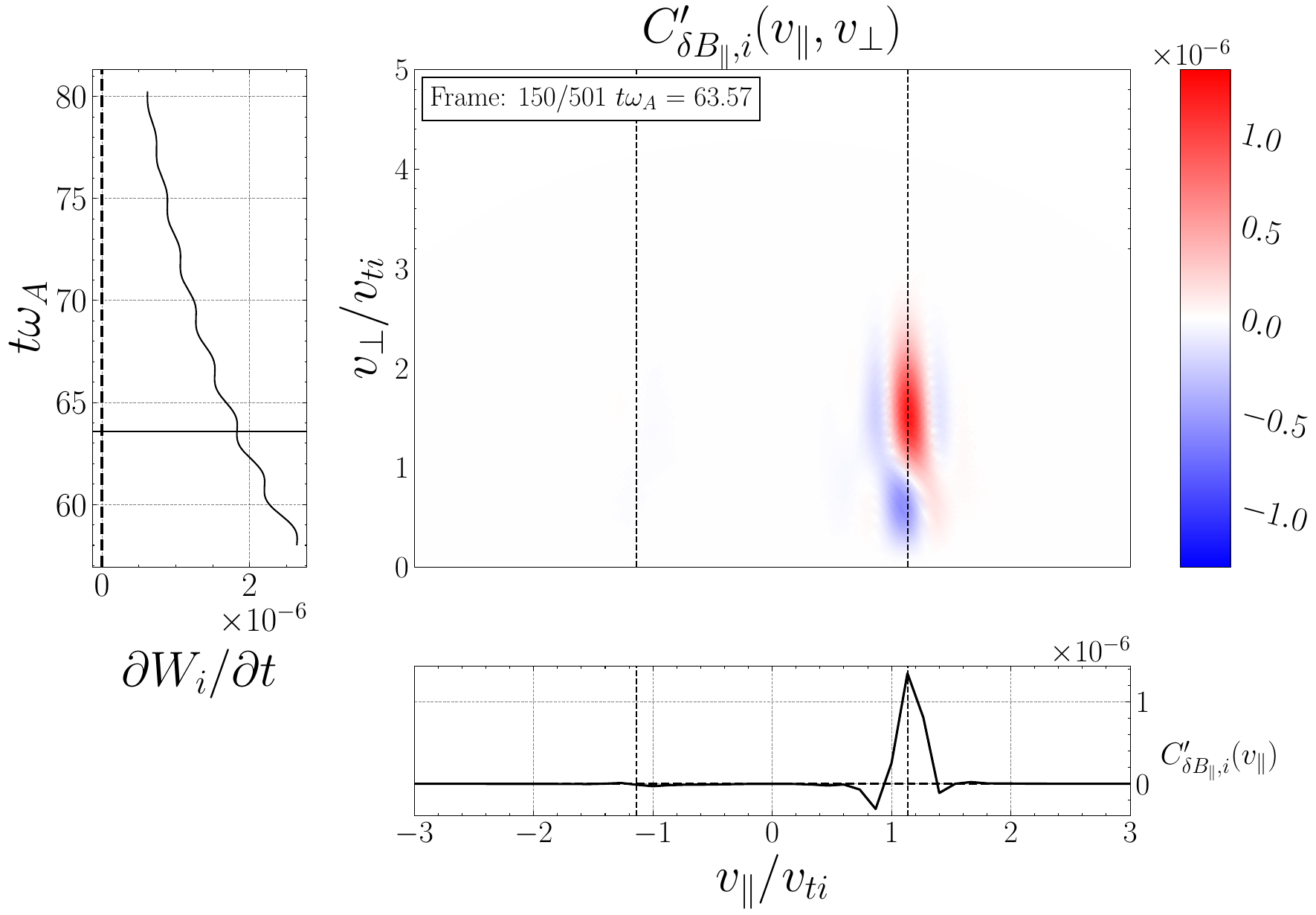}
   \end{center}
    \vskip -1.8in
\hspace*{0.05in} (a)\hspace*{2.3in} (b)
\vskip +1.8in
    \begin{center}
        \includegraphics[width=0.48 \textwidth]{figs_new/ttdb1a_probe001_s1_CE_nc100_t0150_pcolormesh.pdf}
        \includegraphics[width=0.48 \textwidth]{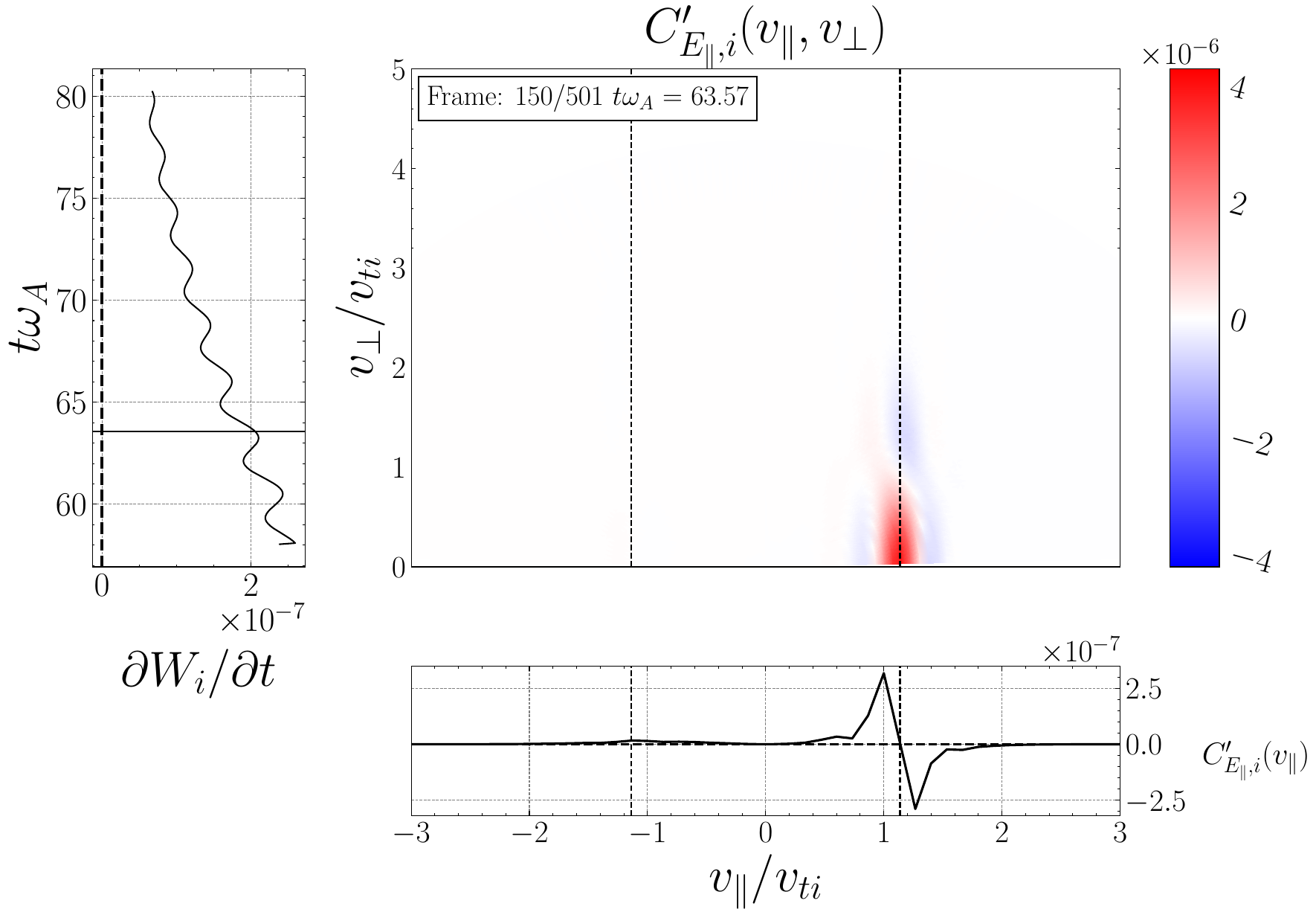}
    \end{center}
    \vskip -1.8in
\hspace*{0.05in} (c)\hspace*{2.3in} (d)
\vskip +1.8in
   \caption{Velocity-space signatures of transit-time damping (top row) and Landau damping (bottom row), plotted from a single kinetic \Alfven wave \T{AstroGK} simulation data with $k_\perp \rho_i = 1, \beta_i = 1, T_i/T_e = 1$. The Vlasov-Maxwell Version correlations, \eqref{eq:corrttd_final} for TTD and the parallel component of \eqref{eq:corrld} for LD, are applied in the left column; and the gyrokinetic version correlations, \eqref{eq:corrgk_ttd} for TTD and \eqref{eq:corrgk_ld} for LD, are applied in the right column. The layout format, simulation data, correlation interval, and normalized parallel phase velocity presented in this figure are identical to those used in \figref{fig:fiducial}.}
    \label{fig:C_CPrime}
\end{figure}

Mathematically, the correlations derived from the Vlasov-Maxwell equations, \eqref{eq:corrttd_final} for TTD and the parallel contribution to the dot product in \eqref{eq:corrld} for LD, and the correlations derived from the gyrokinetics, \eqref{eq:corrgk_ttd} for TTD and \eqref{eq:corrgk_ld} for LD are related by the integration by parts. Physically, the Vlasov-Maxwell version correlation calculates the change in phase-space energy density at fixed positions in velocity space (Eulerian perspective). In contrast, the gyrokinetic version correlation measures the energy changes experienced by particles along their orbits (Lagrangian perspective) \citep{montag:2022}.  The gyrokinetic version is effectively the same as the \emph{alternative} field-particle correlation previously defined \citep{Howes:2017a}, denoted by the symbol $C'$ instead of $C$.
Therefore, as illustrated in \figref{fig:C_CPrime}, where we plot both the Vlasov-Maxwell version correlations and the gyrokinetic version correlations for the fiducial single-KAW simulation, the loss and gain of phase-space energy density in the velocity space are visible in the Vlasov-Maxwell version correlation plots with the zero-crossing centered at the resonant velocity. Conversely, the gyrokinetic version correlation plots highlight the pure energization of particles, with the peak positioned at the resonant velocity. 

We choose not to use the particular forms of the field-particle correlations in \eqref{eq:corrgk_ld} and \eqref{eq:corrgk_ttd} for the study here because the gyrokinetic formulation is the lowest order contribution to the rate of change of phase-space energy density in the gyrokinetic limit of small fluctuations relative to the equilibrium $g_s/F_{0s} \ll 1$ \citep{howes:2006}; we prefer to establish a form that is appropriate for fluctuations of arbitrary amplitude, which is given for TTD by the form in \eqref{eq:corrttd_final}.

\section{Separating Landau Damping from Transit-Time Damping}
As shown by \eqref{eq:dgdtphysicalform}, the nonlinear gyrokinetic equation can be manipulated into a form in which the terms governing Landau damping and transit-time damping are explicitly separated.  Here we explain how one can separate the contributions to the damping rate $\gamma_s$ for a given species $s$ by Landau damping and transit-time damping, both arising from the $n=0$ Landau resonance, in the Vlasov-Maxwell linear dispersion relation \citep{stix:1992,Swanson:2003} in the limit $\gamma_s/\omega\rightarrow 0$.

The rate of work done on a species $s$ by the electric field in a plasma is given by
\begin{equation}
P_s(\V{r},t)= \V{j}_s(\V{r},t) \cdot \V{E}(\V{r},t)
\end{equation}
where the fields $\V{j}_s$ and $\V{E}$ must be real.
For a single plane-wave mode of an arbitrary vector field $\V{A}$, the Fourier transform is given by 
\begin{equation}
  \V{A}(\V{r},t) = \frac{1}{2}\left[ \hat{\V{A}}(\V{k}) e^{i[\V{k} \cdot \V{r} - \omega(\V{k}) t]} e^{\gamma(\V{k}) t} +  \hat{\V{A}}^*(\V{k}) e^{-i[\V{k} \cdot \V{r} - \omega(\V{k}) t]} e^{\gamma(\V{k}) t}\right]
  \label{eq:complex}
\end{equation}
where $\hat{\V{A}}(\V{k})$ is the complex Fourier coefficient for plane-wave vector $\V{k}$ and the complex mode frequency $\omega_c(\V{k}) = \omega(\V{k}) + i \gamma(\V{k})$ is a function of $\V{k}$.  Note that for wave vectors $\V{k} \in [-\infty, \infty]$, the reality condition imposes the constraint  $\hat{\V{A}}(\V{k})= \hat{\V{A}}^*(-\V{k})$.

To determine the average energy transfer rate over one wave period $\langle P_s (\V{r}) \rangle_T$, we compute
\begin{equation}
  \langle P_s (\V{r}) \rangle_T= \frac{1}{T} \int_o^T dt'\V{j}_s(\V{r},t') \cdot \V{E}(\V{r},t')
  \label{eq:ps_je}
\end{equation}
In the case of weak total\footnote{For a fully ionized, single-ion plasma, the total damping or growth rate is the sum of the damping or growth rates due to both species, $\gamma=\gamma_i + \gamma_e$.} damping or growth rate, $|\gamma|/\omega \ll 1$,we can substitute in for $\V{j}_s(\V{x},t)$ and $\V{E}(\V{x},t)$ in terms of the complex Fourier coefficients using \eqref{eq:complex}, which enables us to derive
\begin{equation}
  \lim_{\gamma/\omega \rightarrow 0}\langle P_s (\V{r}) \rangle_T= \frac{1}{4}\left[
    \hat{\V{j}}_s(\V{k}) \cdot  \hat{\V{E}}^*(\V{k})
    +  \hat{\V{j}}_s^*(\V{k}) \cdot  \hat{\V{E}}(\V{k})
    \right]
\end{equation}
as discussed in Sec.~4-2 and eq.~(4-5) in \citet{stix:1992}.

To progress further, we need to determine the contribution to the
plasma current density due to the $n=0$ resonant terms.  In general,
the current density for species $s$ arising from the linear response to an applied
electric field $\hat{\V{E}}(\V{k})$ is given by the linear conductivity
tensor $\bm{\sigma}_s(\omega,\V{k})$,
\begin{equation}
 \hat{\V{j}}_s(\V{k}) =  \bm{\sigma}_s(\omega,\V{k}) \cdot  \hat{\V{E}}(\V{k})
= -\frac{i \omega}{4 \pi}  \bm{\chi}_s(\omega,\V{k}) \cdot  \hat{\V{E}}(\V{k})
\end{equation}
where the linear conductivity tensor can be expressed in terms of the
susceptibility $ \bm{\chi}_s(\omega,\V{k})$ for  species $s$.

Note that the susceptibility tensor  $ \bm{\chi}_s$ involves a sum over all
integers $n$ (the order of the Bessel functions that arise in the integration over the gyrophase of the particles), as shown clearly in  Eq.~(10-57) of \citet{stix:1992}.   The Landau resonance, which gives rise to the collisionless mechanisms of Landau damping and transit-time damping, corresponds to the $n=0$ term in this sum.  Therefore, we need only evaluate the $n=0$ contribution to  $ \bm{\chi}_s$, denoted here by $\bm{\chi}^{(n=0)}_s$.  For an equilibrium magnetic field $\V{B}_0 = B_0 \zhat$ and the choice of a wave vector in the $(x,z)$ plane, $\V{k}= k_\perp \xhat + k_\parallel \zhat$, all elements of the rank 2 tensor $ \bm{\chi}^{(n=0)}_s(\omega,\V{k})$  involving the index $x$ are zero, \emph{i.e.}, $ \chi^{(n=0)}_{xj,s}=\chi^{(n=0)}_{jx,s}=0$, as shown in Stix Eq.~(10-61).
Therefore, the $n=0$ contribution to the current density has  $\xhat \cdot \hat{\V{j}}^{(n=0)}_s(\V{k})=0$.  Thus, we need only concern ourselves with the work done by $\hat{E}_y$ and $\hat{E}_z$.

Physically, Landau damping is governed by the work done by the
$\hat{E}_z$ component of the wave electric field, and transit-time
damping is governed by the work done by the $\hat{E}_y$ component of
the wave electric field \citep{Howes:2024}. Since we know that transit-time damping is
mediated by the magnetic mirror force, it depends on the gradient of
the magnitude of the magnetic field along the magnetic field
direction, $\bhat \cdot \nabla |\V{B}|$. 
In the limit of small amplitudes $|\delta \V{B}| \ll B_0$ appropriate
for linear wave theory with total magnetic field $\V{B} =\V{B}_0 + \delta \V{B}$, the variations of
the magnitude of the total magnetic field $\V{B} =\V{B}_0 + \delta \V{B}$ can be expressed as
$\delta|\V{B}| \simeq \delta B_\parallel$, as shown by \eqref{eq:deltabpar}.
Note that, with variations only in the $x$ and $z$ directions (due to our chosen wavevector orientation), Faraday's Law gives
\begin{equation}
 \frac{\partial B_z}{\partial t} = -c \frac{\partial E_y}{\partial x},
\end{equation}
where $\delta B_z = \delta B_\parallel$.
Therefore, the variations of the magnetic field magnitude that cause
the mirror force arise from the $\hat{E}_y$ component of the electric field.

Since $ \chi^{(n=0)}_{xj,s}=\chi^{(n=0)}_{jx,s}=0$, we can write the components
$\hat{j}_{y,s}$ and $\hat{j}_{z,s}$ in terms of the susceptibility
tensor acting on $\hat{E}_y$ and $\hat{E}_z$.
\begin{equation}
  \hat{\V{j}}^{(n=0)}_{s}(\V{k}) =  -\frac{i \omega(\V{k})}{4 \pi}
  \left( \begin{array}{ccc}
    0 & 0 & 0\\
     0 &\chi^{(n=0)}_{yy,s}(\V{k})  & \chi^{(n=0)}_{yz,s}(\V{k}) \\
     0 &\chi^{(n=0)}_{zy,s}(\V{k})  & \chi^{(n=0)}_{zz,s}(\V{k}) \\
 \end{array}
     \right)
     \left( \begin{array}{c}
    0 \\
    \hat{E}_y(\V{k})\\
     \hat{E}_z(\V{k})\\
     \end{array}
      \right)
 \end{equation}
Thus, we find
\begin{equation}
  \hat{j}^{(n=0)}_{y,s}(\V{k}) =  -\frac{i \omega(\V{k})}{4 \pi}\left[
    \chi^{(n=0)}_{yy,s}(\V{k})\hat{E}_y(\V{k}) +  \chi^{(n=0)}_{yz,s}(\V{k})\hat{E}_z(\V{k})\right]  \label{eq:jy1}    
 \end{equation}
\begin{equation}
  \hat{j}^{(n=0)}_{z,s}(\V{k}) =  -\frac{i \omega(\V{k})}{4 \pi}\left[
    \chi^{(n=0)}_{zy,s}(\V{k})\hat{E}_y(\V{k}) +  \chi^{(n=0)}_{zz,s}(\V{k})\hat{E}_z(\V{k})\right]   \label{eq:jz1}    
\end{equation}

Since transit-time damping is mediated by  $\hat{E}_y$, using \eqref{eq:ps_je} we can compute its energy transfer rate by
\begin{equation}
  \langle P_{TTD,s} (\V{r}) \rangle_T=  \frac{1}{4}\left[
    \hat{j}^{(n=0)}_{y,s}  \hat{E}_{y}^*+  \hat{j}^{(n=0)*}_{y,s}  \hat{E}_{y}
    \right]
  \label{eq:ttd1}    
\end{equation}
where we have suppressed the explicit dependence on $\V{k}$ of the Fourier components for notational simplicity.
Similarly, the energy transfer rate by Landau damping can be computed by
\begin{equation}
  \langle P_{LD,s} (\V{r}) \rangle_T=  \frac{1}{4}\left[
    \hat{j}^{(n=0)}_{z,s}  \hat{E}_{z}^*+  \hat{j}^{(n=0)*}_{z,s} \hat{E}_{z}
    \right]
  \label{eq:ld1}    
\end{equation}

Substituting \eqref{eq:jy1} into \eqref{eq:ttd1} and taking the limit $\gamma/\omega \rightarrow 0$, we obtain the expression for the energy transfer rate due to transit-time damping by species $s$,
\begin{equation}
 \lim_{\gamma/\omega \rightarrow 0}  \langle P_{TTD,s} (\V{r}) \rangle_T=
  -\frac{i \omega}{16 \pi}\left[
    (\chi^{(n=0)}_{yy,s}- \chi^{(n=0)*}_{yy,s})  \hat{E}_{y} \hat{E}_{y}^* +
     \chi^{(n=0)}_{yz,s}\hat{E}_{y}^*\hat{E}_{z} - \chi^{(n=0)*}_{yz,s}\hat{E}_{y}\hat{E}_{z}^*
    \right]
  \label{eq:ttd_fin}    
\end{equation}
Substituting \eqref{eq:jz1} into \eqref{eq:ld1} and taking the limit $\gamma/\omega \rightarrow 0$, we obtain the expression for the energy transfer rate due to Landau damping by species $s$,
\begin{equation}
  \lim_{\gamma/\omega \rightarrow 0}  \langle P_{LD,s} (\V{r}) \rangle_T=
  -\frac{i \omega}{16 \pi}\left[
    (\chi^{(n=0)}_{zz,s}- \chi^{(n=0)*}_{zz,s})  \hat{E}_{z} \hat{E}_{z}^* +
     \chi^{(n=0)}_{zy,s}\hat{E}_{y}\hat{E}_{z}^* - \chi^{(n=0)*}_{zy,s}\hat{E}_{y}^*\hat{E}_{z}
    \right]
  \label{eq:ld_fin}    
\end{equation}

Poynting's theorem integrated over the plasma volume can be used to
connect the period-averaged rate of work done on a plasma species $s$
by a given collisionless damping mechanism $X$ to the linear damping
rate by that mechanism, $\gamma_{X,s}$.  For linear dispersion
relation solutions, the total damping rate $\gamma$ is a linear
combination of the damping rates on each species by each mechanism; if
only the $n=0$ Landau resonant mechanisms contribute to the damping in
a single ion species and electron plasma, this linear combination is
simply
$\gamma=\gamma_{LD,i}+\gamma_{TTD,i}+\gamma_{LD,e}+\gamma_{TTD,e}$. Separating the contributions 
from each mechanism $X$ and each species $s$, 
Poynting's theorem therefore yields the connection $\gamma_{X,s} = \langle
P_{X,s} (\V{x}) \rangle_T/(2 W_{EM})$, where $W_{EM}$ is the total
electromagnetic energy over the integrated volume.

Note that the \citet{stix:1992} representation of the power absorption by species $s$, from his Sec.~(11-8), is of the form
\begin{equation}
 \langle P_{s} \rangle_T=
 \left( \frac{\omega}{8 \pi} \right) 
\left[
 \hat{\V{E}}^* \cdot  \bm{\sigma}^{(a)}_s \cdot  \hat{\V{E}}
  \right]
  \label{eq:stix1}    
\end{equation}
where the anti-Hermitian component of the susceptibility is given by
\begin{equation}
 \bm{\sigma}^{(a)}_s =  -\frac{i}{2} ( \bm{\sigma}_s -  \bm{\sigma}^\dagger_s )
  \label{eq:antichi}    
\end{equation}
In the examples in \citet{stix:1992} of separating the Landau damping and transit-time damping contributions \citep[see also][]{Quataert:1998}, he artificially sets $\hat{E}_{z}=0$ to
determine the power absorption due to transit-time damping, and sets
$\hat{E}_{y}=0$ to determine the power absorption due to Landau
damping.  Such a choice eliminates all of the contributions from the
cross-terms of the susceptibility $\chi_{yz,s}$ and $\chi_{zy,s}$.
However, these terms are not small in general, so their neglect can
lead to an inaccurate estimation of the transit-time damping and
Landau damping rates. Furthermore, by writing the power absorption in
terms of the anti-Hermitian component of the susceptibility $
\bm{\sigma}^{(a)}_s$, the terms due to the cross terms $\chi_{yz,s}$ and
$\chi_{zy,s}$ are exchanged between transit-time damping and Landau
damping, so the separation of the two different mechanisms is not correct.

\subsection{Numerical Results of  Landau Damping and Transit-Time Damping Separation}
\begin{figure}
\begin{center}
            \includegraphics[width=0.75 \textwidth]{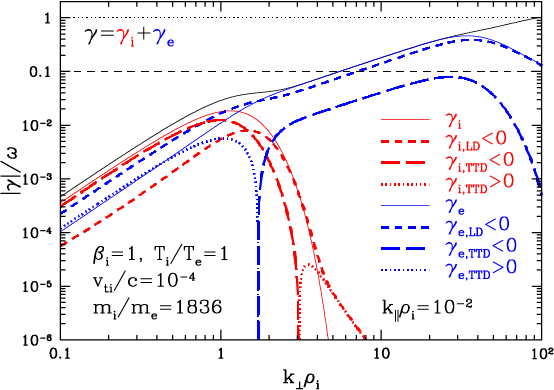} 
\end{center}   
\caption{For a plasma with  $\beta_i=1$, $T_i/T_e=1$,  $m_i/m_e=1836$, $v_{ti}/c=1 \times 10^{-4}$, and $k_\parallel \rho_i =0.01$, plot of total normalized damping or growth rate $\gamma/\omega$ (thin
  black), ion damping or growth rate $\gamma_i/\omega$ (red), and electron
damping or growth rate $\gamma_e/\omega$ (blue). The separate contributions to
the ion damping rate are $\gamma_{i,LD}/\omega<0$ (red short dashed),
$\gamma_{i,TTD}/\omega<0$ (red long dashed), and
$\gamma_{i,TTD}/\omega>0$ (red dotted).  Similarly, The separate
contributions to the electron damping rate are
$\gamma_{e,LD}/\omega<0$ (blue short dashed),
$\gamma_{e,TTD}/\omega<0$ (blue long dashed), and
$\gamma_{e,TTD}/\omega>0$ (blue dotted).}
    \label{fig:ld_ttd}
\end{figure}

As a demonstration of the separation of the Landau damping and transit-time damping rates 
for both ions and electrons, we compute linear Vlasov-Maxwell dispersion relation for the \Alfven wave root using PLUME \citep{klein:2015}
for a proton-electron plasma with Maxwellian equilibrium velocity distributions with $T_i/T_e=1$  and plasma parameters $\beta_i=1$,   $v_{ti}/c=1 \times 10^{-4}$, and $m_i/m_e=1836$. In \figref{fig:ld_ttd}, we plot the normalized damping (or growth) rates $\gamma/\omega$---where damping corresponds to $\gamma<0$ and growth corresponds to  $\gamma>0$, for 
$k_\parallel \rho_i =0.01$ over the range
d $0.1 \le k_\perp \rho_i \le 100$. For this case of isotropic Maxwellian  equilibrium velocity distirbutions, 
there is not source of energy for instabilities, so we obtain a total damping rate  $\gamma/\omega<0$ (thin black).
In the limit  small growth or damping rate limit, $|\gamma|/\omega \ll 1$, we separate the 
ion and electron contributions to the total damping rate, plotting the ion damping or growth rate $\gamma_i/\omega$ (thin solid red) and the 
electron damping or growth rate $\gamma_e/\omega$ (thin solid blue).  Note the breakdown of this separation at $k_\perp \rho_i \gtrsim 30$ when the asymptotic small growth or damping rate limit is violated; the practical limit where the breakdown of this separation by species occurs is 
$|\gamma|/\omega \gtrsim 0.5$, as shown in the figure where sum of the ion and electron damping rates deviates from the total damping rate  $\gamma/\omega$ (thin black) at $k_\perp \rho_i \gtrsim 30$.

Examining the separation of the Landau damping from the transit-time damping rates next, we plot in \figref{fig:ld_ttd}  
the ion Landau damping rate $\gamma_{i,LD}/\omega<0$ (red short dashed), the ion transit-time damping rate
$\gamma_{i,TTD}/\omega<0$ (red long dashed), and the ion transit-time growth rate
$\gamma_{i,TTD}/\omega>0$ (red dotted).  Similarly, we plot the electron Landau damping rate 
$\gamma_{e,LD}/\omega<0$ (blue short dashed), the electron transit-time damping rate
$\gamma_{e,TTD}/\omega<0$ (blue long dashed), and the electron transit-time growth rate
$\gamma_{e,TTD}/\omega>0$ (blue dotted). 
Note that for the ions, the transit-time contribution yields wave growth for $k_\perp\rho_i\ge 3$, and effectively cancels out the contribution from ion Landau damping for  $k_\perp\rho_i\ge 5$, so the total ion damping rate plummets.  For the electrons, the  transit-time contribution is positive yielding growth for $k_\perp\rho_i\lesssim 1.7$.  For $k_\perp\rho_i\lesssim 0.7$, $\gamma_{e,LD} \simeq -2 \gamma_{e,TTD}$, so the total electron damping rate is approximately half of the Landau damping contribution. This plot shows clearly that the contributions  from one mechanism (LD or TTD) can actually yield growth for certain parameters (meaning energy transfer from the particles to the fields via that mechanisms), even when the overall effect is damping.  One can interpret this perhaps puzzling finding with the relative phase between the electric field and the current associated with a particular mechanism.  In some cases---\emph{e.g.}, electron damping at  $k_\perp\rho_i\lesssim 1.7$---this means that energy is transferred from the $E_z$ component of the electric field to the electrons by Landau damping, but at the same time energy can be transferred from the electrons to the field via the $E_y$ component of the electric field by transit-time damping.  The net effect is  damping, but the total damping (or growth) rate is the summation over the Landau and transit-time contributions (which can have either sign). Note also that, for the selection of parameters in this example, the effect of the cyclotron resonance ($n \ne 0$) on the collisionless damping rate is negligible, so the Landau damping and transit-time damping associated with the Landau  ($n =0$) resonance dominates the total wave damping.


\bibliography{abbrev,ttd22}

\begin{thebibliography}{68}
\expandafter\ifx\csname natexlab\endcsname\relax\def\natexlab#1{#1}\fi

\bibitem[{Abel} {\em et~al.\/}(2008){Abel}, {Barnes}, {Cowley}, {Dorland} \& {Schekochihin}]{Abel:2008}
{\sc {Abel}, I.~G., {Barnes}, M., {Cowley}, S.~C., {Dorland}, W. \& {Schekochihin}, A.~A.} 2008 {Linearized model Fokker-Planck collision operators for gyrokinetic simulations. I. Theory}. {\em Phys.~Plasmas\/} {\bf 15}~(12), 122509.

\bibitem[{Afshari} {\em et~al.\/}(2021){Afshari}, {Howes}, {Kletzing}, {Hartley} \& {Boardsen}]{Afshari:2021}
{\sc {Afshari}, A.~S., {Howes}, G.~G., {Kletzing}, C.~A., {Hartley}, D.~P. \& {Boardsen}, S.~A.} 2021 {The Importance of Electron Landau Damping for the Dissipation of Turbulent Energy in Terrestrial Magnetosheath Plasma}. {\em J.~Geophys.~Res.\/} {\bf 126}~(12), e29578.

\bibitem[{Afshari} {\em et~al.\/}(2023){Afshari}, {Howes}, {Shuster}, {Klein}, {McGinnis}, {Martinovic}, {Boardsen}, {Huang}, {Kletzing} \& {Hartley}]{Afshari:2023}
{\sc {Afshari}, A.~S., {Howes}, G.~G., {Shuster}, J.~R., {Klein}, K.~G., {McGinnis}, D., {Martinovic}, M.~M., {Boardsen}, S.~A.~{Brown}, C.~R., {Huang}, R., {Kletzing}, C.~A. \& {Hartley}, D.~P.} 2023 {Direct observation of ion cyclotron damping of turbulence in Earth’s magnetosheath plasma}. {\em Nature Comm.\/} Submitted.

\bibitem[Antonsen~Jr \& Lane(1980)]{antonsen:1980}
{\sc Antonsen~Jr, Thomas~M \& Lane, Barton} 1980 Kinetic equations for low frequency instabilities in inhomogeneous plasmas. {\em The Physics of Fluids\/} {\bf 23}~(6), 1205--1214.

\bibitem[{Arzamasskiy} {\em et~al.\/}(2023){Arzamasskiy}, {Kunz}, {Squire}, {Quataert} \& {Schekochihin}]{Arzamasskiy:2023}
{\sc {Arzamasskiy}, Lev, {Kunz}, Matthew~W., {Squire}, Jonathan, {Quataert}, Eliot \& {Schekochihin}, Alexander~A.} 2023 {Kinetic Turbulence in Collisionless High-{\ensuremath{\beta}} Plasmas}. {\em Phys.~Rev.~X\/} {\bf 13}~(2), 021014.

\bibitem[Barnes(1966)]{Barnes:1966}
{\sc Barnes, Aaron} 1966 Collisionless damping of hydromagnetic waves. {\em The Physics of Fluids\/} {\bf 9}~(8), 1483--1495.

\bibitem[{Barnes} {\em et~al.\/}(2009){Barnes}, {Abel}, {Dorland}, {Ernst}, {Hammett}, {Ricci}, {Rogers}, {Schekochihin} \& {Tatsuno}]{Barnes:2009}
{\sc {Barnes}, M., {Abel}, I.~G., {Dorland}, W., {Ernst}, D.~R., {Hammett}, G.~W., {Ricci}, P., {Rogers}, B.~N., {Schekochihin}, A.~A. \& {Tatsuno}, T.} 2009 {Linearized model Fokker-Planck collision operators for gyrokinetic simulations. II. Numerical implementation and tests}. {\em Phys.~Plasmas\/} {\bf 16}~(7), 072107.

\bibitem[{Brizard} \& {Hahm}(2007)]{Brizard:2007}
{\sc {Brizard}, A.~J. \& {Hahm}, T.~S.} 2007 {Foundations of nonlinear gyrokinetic theory}. {\em Rev. Mod. Phys.\/} {\bf 79}, 421--468.

\bibitem[{Bruno} \& {Carbone}(2013)]{Bruno:2013}
{\sc {Bruno}, R. \& {Carbone}, V.} 2013 {The Solar Wind as a Turbulence Laboratory}. {\em Living Reviews in Solar Physics\/} {\bf 10}, 2.

\bibitem[{Cerri} {\em et~al.\/}(2021){Cerri}, {Arzamasskiy} \& {Kunz}]{Cerri:2021}
{\sc {Cerri}, S.~S., {Arzamasskiy}, L. \& {Kunz}, M.~W.} 2021 {On Stochastic Heating and Its Phase-space Signatures in Low-beta Kinetic Turbulence}. {\em Astrophys.~J.\/} {\bf 916}~(2), 120.

\bibitem[{Chandran} {\em et~al.\/}(2010){Chandran}, {Li}, {Rogers}, {Quataert} \& {Germaschewski}]{Chandran:2010}
{\sc {Chandran}, Benjamin D.~G., {Li}, Bo, {Rogers}, Barrett~N., {Quataert}, Eliot \& {Germaschewski}, Kai} 2010 {Perpendicular Ion Heating by Low-frequency Alfv{\'e}n-wave Turbulence in the Solar Wind}. {\em Astrophysical Journal Letters\/} {\bf 720}~(1), 503--515.

\bibitem[{Chandran} {\em et~al.\/}(2013){Chandran}, {Verscharen}, {Quataert}, {Kasper}, {Isenberg} \& {Bourouaine}]{Chandran:2013}
{\sc {Chandran}, B.~D.~G., {Verscharen}, D., {Quataert}, E., {Kasper}, J.~C., {Isenberg}, P.~A. \& {Bourouaine}, S.} 2013 {Stochastic Heating, Differential Flow, and the Alpha-to-proton Temperature Ratio in the Solar Wind}. {\em Astrophys.~J.\/} {\bf 776}~(1), 45.

\bibitem[Chen {\em et~al.\/}(2019)Chen, Klein \& Howes]{Chen:2019}
{\sc Chen, C. H.~K., Klein, K.~G. \& Howes, G.~G.} 2019 Evidence for electron {Landau} damping in space plasma turbulence. {\em Nature Comm.\/} {\bf 10}~(1), 740.

\bibitem[{Coburn} {\em et~al.\/}(2015){Coburn}, {Forman}, {Smith}, {Vasquez} \& {Stawarz}]{Coburn:2015}
{\sc {Coburn}, J.~T., {Forman}, M.~A., {Smith}, C.~W., {Vasquez}, B.~J. \& {Stawarz}, J.~E.} 2015 {Third-moment descriptions of the interplanetary turbulent cascade, intermittency and back transfer}. {\em {Phil. Trans. Roy. Soc. A}\/} {\bf 373}~(2041), 20140150--20140150.

\bibitem[{Coburn} {\em et~al.\/}(2014){Coburn}, {Smith}, {Vasquez}, {Forman} \& {Stawarz}]{Coburn:2014}
{\sc {Coburn}, Jesse~T., {Smith}, Charles~W., {Vasquez}, Bernard~J., {Forman}, Miriam~A. \& {Stawarz}, Julia~E.} 2014 {Variable Cascade Dynamics and Intermittency in the Solar Wind at 1 AU}. {\em Astrophys.~J.\/} {\bf 786}~(1), 52.

\bibitem[{Conley} {\em et~al.\/}(2023){Conley}, {Howes} \& {McCubbin}]{Conley:2023}
{\sc {Conley}, Sarah~A., {Howes}, Gregory~G. \& {McCubbin}, Andrew~J.} 2023 {Characterizing the velocity-space signature of electron Landau damping}. {\em J.~Plasma Phys.\/} {\bf 89}~(5), 905890514.

\bibitem[{Drake} {\em et~al.\/}(2013){Drake}, {Schroeder}, {Howes}, {Kletzing}, {Skiff}, {Carter} \& {Auerbach}]{Drake:2013}
{\sc {Drake}, D.~J., {Schroeder}, J.~W.~R., {Howes}, G.~G., {Kletzing}, C.~A., {Skiff}, F., {Carter}, T.~A. \& {Auerbach}, D.~W.} 2013 {Alfv{\'e}n wave collisions, the fundamental building block of plasma turbulence. IV. Laboratory experiment}. {\em Phys.~Plasmas\/} {\bf 20}~(7), 072901.

\bibitem[Frieman \& Chen(1982)]{frieman:1982}
{\sc Frieman, EA \& Chen, Liu} 1982 Nonlinear gyrokinetic equations for low-frequency electromagnetic waves in general plasma equilibria. {\em The Physics of Fluids\/} {\bf 25}~(3), 502--508.

\bibitem[Goldreich \& Sridhar(1995)]{goldreich:1995}
{\sc Goldreich, P \& Sridhar, S} 1995 Toward a theory of interstellar turbulence. 2: Strong alfvenic turbulence. {\em The Astrophysical Journal\/} {\bf 438}, 763--775.

\bibitem[{Horvath} {\em et~al.\/}(2020){Horvath}, {Howes} \& {McCubbin}]{Horvath:2020}
{\sc {Horvath}, S.~A., {Howes}, G.~G. \& {McCubbin}, A.~J.} 2020 Electron landau damping of kinetic alfv\'en waves in simulated magnetosheath turbulence. {\em Phys.~Plasmas\/} {\bf 27}~(10), 102901.

\bibitem[{Howes}(2016)]{Howes:2016b}
{\sc {Howes}, G.~G.} 2016 {The Dynamical Generation of Current Sheets in Astrophysical Plasma Turbulence}. {\em Astrophys.~J.~Lett.\/} {\bf 82}, L28.

\bibitem[{Howes}(2017)]{Howes:2017c}
{\sc {Howes}, G.~G.} 2017 A prospectus on kinetic heliophysics. {\em Phys.~Plasmas\/} {\bf 24}~(5), 055907.

\bibitem[Howes {\em et~al.\/}(2006)Howes, Cowley, Dorland, Hammett, Quataert \& Schekochihin]{howes:2006}
{\sc Howes, Gregory~G, Cowley, Steven~C, Dorland, William, Hammett, Gregory~W, Quataert, Eliot \& Schekochihin, Alexander~A} 2006 Astrophysical gyrokinetics: basic equations and linear theory. {\em The Astrophysical Journal\/} {\bf 651}~(1), 590.

\bibitem[{Howes} {\em et~al.\/}(2008{\natexlab{{\em a\/}}}){Howes}, {Cowley}, {Dorland}, {Hammett}, {Quataert} \& {Schekochihin}]{Howes:2008b}
{\sc {Howes}, G.~G., {Cowley}, S.~C., {Dorland}, W., {Hammett}, G.~W., {Quataert}, E. \& {Schekochihin}, A.~A.} 2008{\natexlab{{\em a\/}}} {A model of turbulence in magnetized plasmas: Implications for the dissipation range in the solar wind}. {\em J.~Geophys.~Res.\/} {\bf 113}~(A12), A05103.

\bibitem[{Howes} {\em et~al.\/}(2008{\natexlab{{\em b\/}}}){Howes}, {Dorland}, {Cowley}, {Hammett}, {Quataert}, {Schekochihin} \& {Tatsuno}]{Howes:2008a}
{\sc {Howes}, G.~G., {Dorland}, W., {Cowley}, S.~C., {Hammett}, G.~W., {Quataert}, E., {Schekochihin}, A.~A. \& {Tatsuno}, T.} 2008{\natexlab{{\em b\/}}} {Kinetic Simulations of Magnetized Turbulence in Astrophysical Plasmas}. {\em Phys.~Rev.~Lett.\/} {\bf 100}~(6), 065004.

\bibitem[{Howes} {\em et~al.\/}(2012){Howes}, {Drake}, {Nielson}, {Carter}, {Kletzing} \& {Skiff}]{Howes:2012b}
{\sc {Howes}, G.~G., {Drake}, D.~J., {Nielson}, K.~D., {Carter}, T.~A., {Kletzing}, C.~A. \& {Skiff}, F.} 2012 {Toward Astrophysical Turbulence in the Laboratory}. {\em Phys.~Rev.~Lett.\/} {\bf 109}~(25), 255001.

\bibitem[{Howes} {\em et~al.\/}(2024){Howes}, {Huang} \& {Felix}]{Howes:2024}
{\sc {Howes}, G.~G., {Huang}, R. \& {Felix}, A.~A.} 2024 {The Physics of the Magnetic Mirror Force and Transit-Time Damping: Asymptotic Solution}. {\em J.~Plasma Phys.\/} In preparation.

\bibitem[{Howes} {\em et~al.\/}(2017){Howes}, {Klein} \& {Li}]{Howes:2017a}
{\sc {Howes}, G.~G., {Klein}, K.~G. \& {Li}, T.~C.} 2017 {Diagnosing collisionless energy transfer using field-particle correlations: Vlasov-Poisson plasmas}. {\em J.~Plasma Phys.\/} {\bf 83}~(1), 705830102.

\bibitem[{Howes} {\em et~al.\/}(2014){Howes}, {Klein} \& {TenBarge}]{Howes:2014a}
{\sc {Howes}, G.~G., {Klein}, K.~G. \& {TenBarge}, J.~M.} 2014 {Validity of the Taylor Hypothesis for Linear Kinetic Waves in the Weakly Collisional Solar Wind}. {\em Astrophys.~J.\/} {\bf 789}, 106.

\bibitem[{Howes} {\em et~al.\/}(2018){Howes}, {McCubbin} \& {Klein}]{Howes:2018a}
{\sc {Howes}, G.~G., {McCubbin}, A.~J. \& {Klein}, K.~G.} 2018 {Spatially localized particle energization by Landau damping in current sheets produced by strong Alfv{\'e}n wave collisions}. {\em J.~Plasma Phys.\/} {\bf 84}~(1), 905840105.

\bibitem[{Howes} \& {Nielson}(2013)]{Howes:2013a}
{\sc {Howes}, G.~G. \& {Nielson}, K.~D.} 2013 {Alfv{\'e}n wave collisions, the fundamental building block of plasma turbulence. I. Asymptotic solution}. {\em Phys.~Plasmas\/} {\bf 20}~(7), 072302.

\bibitem[{Howes} {\em et~al.\/}(2013){Howes}, {Nielson}, {Drake}, {Schroeder}, {Skiff}, {Kletzing} \& {Carter}]{Howes:2013b}
{\sc {Howes}, G.~G., {Nielson}, K.~D., {Drake}, D.~J., {Schroeder}, J.~W.~R., {Skiff}, F., {Kletzing}, C.~A. \& {Carter}, T.~A.} 2013 {Alfv{\'e}n wave collisions, the fundamental building block of plasma turbulence. III. Theory for experimental design}. {\em Phys.~Plasmas\/} {\bf 20}~(7), 072304.

\bibitem[{Howes} {\em et~al.\/}(2011){Howes}, {Tenbarge} \& {Dorland}]{Howes:2011b}
{\sc {Howes}, G.~G., {Tenbarge}, J.~M. \& {Dorland}, W.} 2011 {A weakened cascade model for turbulence in astrophysical plasmas}. {\em Phys.~Plasmas\/} {\bf 18}, 102305.

\bibitem[Howes {\em et~al.\/}(2011)Howes, TenBarge, Dorland, Quataert, Schekochihin, Numata \& Tatsuno]{Howes:2011a}
{\sc Howes, G.~G., TenBarge, J.~M., Dorland, W., Quataert, E., Schekochihin, A.~A., Numata, R. \& Tatsuno, T.} 2011 Gyrokinetic simulations of solar wind turbulence from ion to electron scales. {\em Phys.~Rev.~Lett.\/} {\bf 107}, 035004.

\bibitem[{Isenberg} \& {Hollweg}(1983)]{Isenberg:1983}
{\sc {Isenberg}, P.~A. \& {Hollweg}, J.~V.} 1983 {On the preferential acceleration and heating of solar wind heavy ions}. {\em J.~Geophys.~Res.\/} {\bf 88}, 3923--3935.

\bibitem[{Isenberg} \& {Vasquez}(2019)]{Isenberg:2019}
{\sc {Isenberg}, Philip~A. \& {Vasquez}, Bernard~J.} 2019 {Perpendicular Ion Heating by Cyclotron Resonant Dissipation of Turbulently Generated Kinetic Alfv{\'e}n Waves in the Solar Wind}. {\em Astrophys.~J.\/} {\bf 887}~(1), 63.

\bibitem[Klein \& Howes(2015)]{klein:2015}
{\sc Klein, Kristopher~G \& Howes, Gregory~G} 2015 Predicted impacts of proton temperature anisotropy on solar wind turbulence. {\em Physics of Plasmas\/} {\bf 22}~(3), 032903.

\bibitem[{Klein} \& {Howes}(2016)]{Klein:2016a}
{\sc {Klein}, K.~G. \& {Howes}, G.~G.} 2016 {Measuring Collisionless Damping in Heliospheric Plasmas using Field-Particle Correlations}. {\em Astrophys.~J.~Lett.\/} {\bf 826}, L30.

\bibitem[{Klein} {\em et~al.\/}(2017){Klein}, {Howes} \& {TenBarge}]{Klein:2017b}
{\sc {Klein}, K.~G., {Howes}, G.~G. \& {TenBarge}, J.~M.} 2017 {Diagnosing collisionless enegy transfer using field-particle correlations: gyrokinetic turbulence}. {\em J.~Plasma Phys.\/} {\bf 83}~(4), 535830401.

\bibitem[{Klein} {\em et~al.\/}(2020){Klein}, {Howes}, {TenBarge} \& {Valentini}]{Klein:2020}
{\sc {Klein}, Kristopher~G., {Howes}, Gregory~G., {TenBarge}, Jason~M. \& {Valentini}, Francesco} 2020 {Diagnosing collisionless energy transfer using field-particle correlations: Alfv{\'e}n-ion cyclotron turbulence}. {\em J.~Plasma Phys.\/} {\bf 86}~(4), 905860402.

\bibitem[{Kolmogorov}(1941)]{Kolmogorov:1941b}
{\sc {Kolmogorov}, A.~N.} 1941 {Dissipation of Energy in Locally Isotropic Turbulence}. {\em Dokl. Akad. Nauk SSSR\/} {\bf 32}, 16.

\bibitem[{Landau}(1946)]{Landau:1946}
{\sc {Landau}, L.~D.} 1946 {On the Vibrations of the Electronic Plasma}. {\em Journal of Physics\/} {\bf 10}, 25.

\bibitem[{Li} {\em et~al.\/}(2019){Li}, {Howes}, {Klein}, {Liu} \& {TenBarge}]{TCLi:2019}
{\sc {Li}, Tak~Chu, {Howes}, Gregory~G., {Klein}, Kristopher~G., {Liu}, Yi-Hsin \& {TenBarge}, Jason~M.} 2019 Collisionless energy transfer in kinetic turbulence: field–particle correlations in fourier space. {\em Journal of Plasma Physics\/} {\bf 85}~(4), 905850406.

\bibitem[{Li} {\em et~al.\/}(2016){Li}, {Howes}, {Klein} \& {TenBarge}]{TCLi:2016}
{\sc {Li}, T.~C., {Howes}, G.~G., {Klein}, K.~G. \& {TenBarge}, J.~M.} 2016 {Energy Dissipation and Landau Damping in Two- and Three-dimensional Plasma Turbulence}. {\em Astrophys.~J.~Lett.\/} {\bf 832}, L24.

\bibitem[{Lichko} \& {Egedal}(2020)]{Lichko:2020}
{\sc {Lichko}, E. \& {Egedal}, J.} 2020 {Magnetic pumping model for energizing superthermal particles applied to observations of the Earth's bow shock}. {\em Nature Comm.\/} {\bf 11}, 2942.

\bibitem[{Loureiro} \& {Boldyrev}(2017)]{Loureiro:2017a}
{\sc {Loureiro}, N.~F. \& {Boldyrev}, S.} 2017 {Role of Magnetic Reconnection in Magnetohydrodynamic Turbulence}. {\em Phys.~Rev.~Lett.\/} {\bf 118}~(24), 245101.

\bibitem[{Mallet} {\em et~al.\/}(2017){Mallet}, {Schekochihin} \& {Chandran}]{Mallet:2017}
{\sc {Mallet}, Alfred, {Schekochihin}, Alexander~A. \& {Chandran}, Benjamin D.~G.} 2017 {Disruption of Alfv{\'e}nic turbulence by magnetic reconnection in a collisionless plasma}. {\em J.~Plasma Phys.\/} {\bf 83}~(6), 905830609.

\bibitem[{Martinovi{\'c}} {\em et~al.\/}(2020){Martinovi{\'c}}, {Klein}, {Kasper}, {Case}, {Korreck}, {Larson}, {Livi}, {Stevens}, {Whittlesey}, {Chandran}, {Alterman}, {Huang}, {Chen}, {Bale}, {Pulupa}, {Malaspina}, {Bonnell}, {Harvey}, {Goetz}, {Dudok de Wit} \& {MacDowall}]{Martinovic:2020}
{\sc {Martinovi{\'c}}, Mihailo~M., {Klein}, Kristopher~G., {Kasper}, Justin~C., {Case}, Anthony~W., {Korreck}, Kelly~E., {Larson}, Davin, {Livi}, Roberto, {Stevens}, Michael, {Whittlesey}, Phyllis, {Chandran}, Benjamin D.~G., {Alterman}, Ben~L., {Huang}, Jia, {Chen}, Christopher H.~K., {Bale}, Stuart~D., {Pulupa}, Marc, {Malaspina}, David~M., {Bonnell}, John~W., {Harvey}, Peter~R., {Goetz}, Keith, {Dudok de Wit}, Thierry \& {MacDowall}, Robert~J.} 2020 {The Enhancement of Proton Stochastic Heating in the Near-Sun Solar Wind}. {\em Astrophys.~J.~Supp.\/} {\bf 246}~(2), 30.

\bibitem[Melrose(1980)]{melrose:1980}
{\sc Melrose, Donald~B} 1980 {\em Plasma astrophysics\/}, , vol.~1. CRC Press.

\bibitem[Montag \& Howes(2022)]{montag:2022}
{\sc Montag, P \& Howes, Gregory~G} 2022 A field-particle correlation analysis of magnetic pumping. {\em Physics of plasmas\/} {\bf 29}~(3).

\bibitem[{Nielson} {\em et~al.\/}(2013){Nielson}, {Howes} \& {Dorland}]{Nielson:2013a}
{\sc {Nielson}, K.~D., {Howes}, G.~G. \& {Dorland}, W.} 2013 {Alfv{\'e}n wave collisions, the fundamental building block of plasma turbulence. II. Numerical solution}. {\em Phys.~Plasmas\/} {\bf 20}~(7), 072303.

\bibitem[Numata {\em et~al.\/}(2010)Numata, Howes, Tatsuno, Barnes \& Dorland]{numata:2010}
{\sc Numata, Ryusuke, Howes, Gregory~G, Tatsuno, Tomoya, Barnes, Michael \& Dorland, William} 2010 Astrogk: Astrophysical gyrokinetics code. {\em Journal of Computational Physics\/} {\bf 229}~(24), 9347--9372.

\bibitem[{Osman} {\em et~al.\/}(2011){Osman}, {Matthaeus}, {Greco} \& {Servidio}]{Osman:2011}
{\sc {Osman}, K.~T., {Matthaeus}, W.~H., {Greco}, A. \& {Servidio}, S.} 2011 {Evidence for Inhomogeneous Heating in the Solar Wind}. {\em Astrophys.~J.~Lett.\/} {\bf 727}, L11.

\bibitem[{Politano} \& {Pouquet}(1998)]{Politano:1998a}
{\sc {Politano}, H. \& {Pouquet}, A.} 1998 {von K{\'a}rm{\'a}n-Howarth equation for magnetohydrodynamics and its consequences on third-order longitudinal structure and correlation functions}. {\em Phys.~Rev.~E\/} {\bf 57}, 21.

\bibitem[{Quataert}(1998)]{Quataert:1998}
{\sc {Quataert}, E.} 1998 {Particle Heating by Alfv\'enic Turbulence in Hot Accretion Flows}. {\em Astrophys.~J.\/} {\bf 500}, 978--991.

\bibitem[{Sahraoui} {\em et~al.\/}(2013){Sahraoui}, {Huang}, {Belmont}, {Goldstein}, {R{\'e}tino}, {Robert} \& {De Patoul}]{Sahraoui:2013b}
{\sc {Sahraoui}, F., {Huang}, S.~Y., {Belmont}, G., {Goldstein}, M.~L., {R{\'e}tino}, A., {Robert}, P. \& {De Patoul}, J.} 2013 {Scaling of the Electron Dissipation Range of Solar Wind Turbulence}. {\em Astrophys.~J.\/} {\bf 777}, 15.

\bibitem[Schekochihin {\em et~al.\/}(2009)Schekochihin, Cowley, Dorland, Hammett, Howes, Quataert \& Tatsuno]{schekochihin:2009}
{\sc Schekochihin, AA, Cowley, SC, Dorland, W, Hammett, GW, Howes, Gregory~G, Quataert, E \& Tatsuno, T} 2009 Astrophysical gyrokinetics: kinetic and fluid turbulent cascades in magnetized weakly collisional plasmas. {\em The Astrophysical Journal Supplement Series\/} {\bf 182}~(1), 310.

\bibitem[Spitzer~Jr \& Witten(1953)]{spitzer:1953}
{\sc Spitzer~Jr, L \& Witten, L} 1953 On the ionization and heating of plasma. {\em Tech. Rep.\/}. U.S. Atomic Energy Commission Report No. NYO-999 (PM-S-6).

\bibitem[Stix(1992)]{stix:1992}
{\sc Stix, Thomas~H} 1992 {\em Waves in plasmas\/}. Springer Science \& Business Media.

\bibitem[{Swanson}(2003)]{Swanson:2003}
{\sc {Swanson}, D.~Gary} 2003 {\em {Plasma Waves, 2nd Edition}\/}.

\bibitem[{TenBarge} \& {Howes}(2013)]{TenBarge:2013a}
{\sc {TenBarge}, J.~M. \& {Howes}, G.~G.} 2013 {Current Sheets and Collisionless Damping in Kinetic Plasma Turbulence}. {\em Astrophys.~J.~Lett.\/} {\bf 771}, L27.

\bibitem[{TenBarge} {\em et~al.\/}(2013){TenBarge}, {Howes} \& {Dorland}]{TenBarge:2013b}
{\sc {TenBarge}, J.~M., {Howes}, G.~G. \& {Dorland}, W.} 2013 {Collisionless Damping at Electron Scales in Solar Wind Turbulence}. {\em Astrophys.~J.\/} {\bf 774}, 139.

\bibitem[{TenBarge} {\em et~al.\/}(2014){TenBarge}, {Howes}, {Dorland} \& {Hammett}]{TenBarge:2014a}
{\sc {TenBarge}, J.~M., {Howes}, G.~G., {Dorland}, W. \& {Hammett}, G.~W.} 2014 {An oscillating Langevin antenna for driving plasma turbulence simulations}. {\em Comp.~Phys.~Comm.\/} {\bf 185}, 578--589.

\bibitem[{Tu} \& {Marsch}(1995)]{Tu:1995}
{\sc {Tu}, C.-Y. \& {Marsch}, E.} 1995 {MHD structures, waves and turbulence in the solar wind: Observations and theories}. {\em Space Sci.~Rev.\/} {\bf 73}, 1--2.

\bibitem[{Verniero} \& {Howes}(2018)]{Verniero:2018b}
{\sc {Verniero}, J.~L. \& {Howes}, G.~G.} 2018 {The Alfv{\'e}nic nature of energy transfer mediation in localized, strongly nonlinear Alfv{\'e}n wavepacket collisions}. {\em J.~Plasma Phys.\/} {\bf 84}~(1), 905840109.

\bibitem[{Verniero} {\em et~al.\/}(2018){Verniero}, {Howes} \& {Klein}]{Verniero:2018a}
{\sc {Verniero}, J.~L., {Howes}, G.~G. \& {Klein}, K.~G.} 2018 {Nonlinear energy transfer and current sheet development in localized Alfv{\'e}n wavepacket collisions in the strong turbulence limit}. {\em J.~Plasma Phys.\/} {\bf 84}~(1), 905840103.

\bibitem[{Villani}(2014)]{Villani:2014}
{\sc {Villani}, C.} 2014 {Particle systems and nonlinear Landau damping}. {\em Physics of Plasmas\/} {\bf 21}~(3), 030901.

\bibitem[{Zhdankin} {\em et~al.\/}(2015){Zhdankin}, {Uzdensky} \& {Boldyrev}]{Zhdankin:2015}
{\sc {Zhdankin}, V., {Uzdensky}, D.~A. \& {Boldyrev}, S.} 2015 {Temporal Intermittency of Energy Dissipation in Magnetohydrodynamic Turbulence}. {\em Phys.~Rev.~Lett.\/} {\bf 114}~(6), 065002.

\end{thebibliography}

\end{document}